\documentclass[aps, prd, twocolumn, lengthcheck, superscriptaddress, 
nofootinbib]{revtex4-1}

\usepackage{setspace}
\usepackage{textcomp}
\usepackage{diagbox}
\usepackage{titlesec}
\usepackage[titletoc,toc,page,header]{appendix}
\usepackage{tabularx}
\usepackage{booktabs}
\usepackage{epstopdf}
\usepackage{epsfig}
\usepackage{graphicx}
\usepackage{amsmath}
\usepackage{geometry}
\usepackage{amssymb}
\usepackage{mathrsfs}
\usepackage{float}
\usepackage{makecell}
\usepackage{multirow}
\usepackage{tikz}
\usepackage{gensymb}
\usepackage{amstext}
\usepackage{bm}
\usepackage{indentfirst}
\usepackage{latexsym}
\usepackage{color}
\usepackage{subfigure}
\usepackage{slashed}
\usepackage{hyperref}
\usepackage{lipsum}
\usepackage{ulem}
\usepackage{flushend}
\usepackage{threeparttable}

\def\be{\begin{eqnarray}}
\def\ee{\end{eqnarray}}

\begin{document}
\title{Quark structure of isoscalar- and isovector-scalar mesons and nuclear matter property}
\author{Yao Ma}
\email{mayao@ucas.ac.cn}
\affiliation{School of Fundamental Physics and Mathematical Sciences, Hangzhou Institute for Advanced Study, UCAS, Hangzhou, 310024, China}
\affiliation{University of Chinese Academy of Sciences, Beijing 100049, China}

\author{Yong-Liang Ma}
\email{ylma@nju.edu.cn}
\affiliation{Nanjing University, Suzhou, 215163, China}
\affiliation{School of Fundamental Physics and Mathematical Sciences, Hangzhou Institute for Advanced Study, UCAS, Hangzhou, 310024, China}
\affiliation{International Center for Theoretical Physics Asia-Pacific (ICTP-AP) , UCAS, Beijing, 100190, China}

\begin{abstract}
It was found that isoscalar-scalar and isovector-scalar mesons play significant roles in nuclear matter physics. However, the underlying structures of these resonances are not yet well understood. We construct a three-flavor baryonic extended linear sigma model including both the two- and four-quark constitutes of scalar mesons with respect to the axial transformation and study the nuclear matter properties with relativistic mean field method. The nuclear matter properties at saturation density and hadron spectra are well reproduced with this model simultaneously, and only the two-quark component of the scalar mesons couples to baryon fields and dominates the nuclear force. A plateaulike structure is found in the symmetry energy of nuclear matter due to the multimeson couplings, and it is crucial for understanding the neutron skin thickness of $^{208}\rm Pb$ and the tidal deformation of neutron star from GW170817. At high-density regions, the vector mesons are found to be rather crucial, especially the ones coupling with four-quark configurations. This model can be easily extended to study neutron stars, even with hyperons in their interior.
\end{abstract}

\maketitle

\allowdisplaybreaks{}

\section{Introduction}

Nuclear matter (NM) is a kind of strongly interacting system which is mainly composed of nucleons and mesons and, maybe other novel structures (see, e.g., Refs.~\cite{Serot:1984ey,Holt:2014hma,Baym:2017whm,Ma:2019ery} for reviews). Since NM is in the nonperturbative region of QCD which is not yet fully understood so far, many properties are still under debate.

Finite nuclei as well as infinite NM at low densities can be fairly well accessed by nuclear chiral effective field theories, pionless or pionful, anchored on some symmetries and invariances of QCD. However, these theories break at high densities relevant to the interiors of massive stars~\cite{Holt:2014hma}, and the higher-order expansion of the power counting and hadron resonances such as the lightest isoscalar-scalar meson ($\sigma$) and lowest-lying vector mesons are indispensable; see, e.g., Refs.~\cite{Paeng:2015noa,Paeng:2017qvp,Ma:2019kiq}, where the $\sigma$ meson is regarded as the Nambu-Goldstone boson of scale symmetry-breaking~\cite{Crewther:2013vea,Crewther:2020tgd} and the vector mesons are included through the hidden local symmetry approach~\cite{Bando:1984ej,Bando:1987br,Harada:2003jx}.

In the relativistic mean field (RMF) approach to NM, it has long been recognized that both the isoscalar-scalar mesons and the isovector-scalar mesons are crucial for obtaining the NM properties around the saturation density $n_0\approx 0.16\ \rm fm^{-3}$, for example, the binding energy of nucleon $e_0$ due to the competition between the attractive force from $\sigma$ meson and repulsive force from the $\omega$ meson~\cite{Walecka:1974qa,Serot:1984ey,Serot:1997xg}. The three- and four-point interactions of the sigma meson reduce the nuclear incompressibility~\cite{Boguta:1977xi} and the six-point interaction reproduces the properties of NM~\cite{Motohiro:2015taa}. The interaction between sigma and vector mesons is also significant: The $\sigma$ and $\omega$ meson coupling has been introduced to obtain a reasonable value of incompressibility coefficient $K_0$~\cite{Boguta:1977xi,Lalazissis:1996rd}; the interactions between \(\sigma\) and $\rho$ meson are crucial for describing the isospin asymmetric NM~\cite{Walecka:1974qa,Serot:1984ey,Mueller:1996pm,Serot:1997xg} and the neutron skin thickness of heavy nuclei $^{208}$Pb~\cite{Horowitz:2000xj,Horowitz:2001ya}.

In addition to the isoscalar-scalar meson, the isovector-scalar mesons $\delta$ [denoted as $a_0(980)$ in particle physics] should also be included in the RMF approach to NM, especially for asymmetric NM at high densities~\cite{Kubis:1997ew,Hofmann:2000vz,Liu:2001iz,Chen:2007ih}. Recently, it has found that isoscalar- and isovector-scalar meson couplings are closely related to the stiffness of the equation of state (EOS) of NM and, therefore, the properties of neutron stars~\cite{Zabari:2018tjk,Zabari:2019clk,Miyatsu:2022wuy,Li:2022okx}.

The existing studies have already shown that meson resonances heavier than the pion should be seriously considered in the study of nuclear matter, especially for asymmetric matter at high densities. However, these studies are considered separately, and the results are highly model dependent. So it is very interesting to construct a self-consistent model including $\sigma, \delta$ [following the convention of particle physics, we denote $\delta$ as $a_0(980)$ in the rest of this work] and $\omega, \rho$, in addition to $\pi$, to study these resonance effects in a systematical way.

A systematic study of the scalar meson effects on NM is not so easy if one wants to concern more underlying dynamics of QCD, since, unlike pions, scalar mesons cannot be regarded as Nambu-Goldstone bosons of chiral symmetry-breaking, and their quark contents still puzzle physicists~\cite{Close:2002zu}. Considering their mass ordering, it is very difficult to arrange scalar mesons below 1 GeV as pure $q\bar{q}$ states, and the four-quark components of scalar mesons should be taken into account. The purpose of this work is to investigate the effects of the quark contents of scalar mesons on NM properties.

To consider the two- and four-quark contents in the model, we follow the idea of Refs.~\cite{Napsuciale:2004au,Fariborz:2005gm,Fariborz:2007ai,Fariborz:2009cq}, where an extended linear sigma model (ELSM) with both the two-quark and four-quark configurations was included and the physical states---the mixing states of the two- and four-quark configurations---were obtained by fitting the data~\cite{Harada:2004mh,Fariborz:2011in,Harada:2012km,Fariborz:2014gsa}. We couple the scalar meson states to baryons with respect to the axial symmetry~\cite{thesis2015Olbrich,Olbrich:2015gln}. Since the two- and four-quark components have different axial transformations, the quark contents of the scalar mesons affect the properties of NM. In addition to the scalar mesons, the vector mesons are introduced as chiral representations with two-quark configurations~\cite{Parganlija:2012fy}.

By taking the RMF approximation using such constructed baryonic ELSM (bELSM), we found that both scalar mesons and vector mesons contribute to the EOS and the quark contents of the scalar mesons also affect the EOS in a sizable way, especially at high-density regions. A plateaulike structure of symmetry energy is found around saturation density $n_0\approx 0.16\ \rm{fm^{-3}}$, which is the key to understand neutron skin of $\rm {}^{208}Pb$ and tidal deformability of neutron stars. Another interesting finding is $|g_{\sigma NN}|\sim 10$ and $|g_{\omega NN}|\sim 13$, which is consistent with the results from other approaches~\cite{Sugahara:1993wz,Shen:1998gq,Li:2022okx}.

This paper is organized as follows: In Sec.~\ref{sec:model}, we discuss the theoretical framework that will be used. The numerical results are discussed in detail in Sec.~\ref{sec:num}. Our summary and perspective are given in the last section and the Lagrangian of the three- and four-meson couplings are listed in the Appendix.

\section{Extended linear sigma model with baryon fields}

\label{sec:model}

We consider  two three-flavor meson nonets $\Phi$ and $\hat{\Phi}$, which represent two-quark configuration $q\bar{q}$ and four-quark configuration $(qq)(\bar{q}\bar{q})$, respectively. Under chiral transformation, they transform as~\cite{Fariborz:2005gm,Fariborz:2007ai,Fariborz:2009cq}
\begin{equation}
	\Phi \rightarrow g_{\rm L} \Phi g_{\rm R}^{\dagger}, \quad \hat{\Phi} \rightarrow g_{\rm L} \hat{\Phi} g_{\rm R}^{\dagger}
\end{equation}
where $g_{\rm L, R} \in\rm S U(3)_{\rm L, R}$. In addition, they transform under \(\rm U(1)_A\) axial transformation as
\begin{equation}
	\Phi \rightarrow e^{2 i \nu} \Phi, \quad \hat{\Phi} \rightarrow e^{-4 i \nu} \hat{\Phi}
\end{equation}
with $\nu$ being the phase angle.

In the ELSM, we introduce the vector meson fields as current operators $L_{\mu}$ and $R_{\mu}$ with $\bar{q}_{L}\gamma_{\mu}q_{\rm L}$ and $\bar{q}_{R}\gamma_{\mu}q_{\rm R}$ configurations, respectively~\cite{Parganlija:2012fy}. Under chiral transformation, they transform as
\be
L_\mu \rightarrow g_{\rm L} L_\mu g_{\rm L}^{\dagger}, \quad R_\mu \rightarrow g_{\rm R} R_\mu g_{\rm R}^{\dagger}.
\label{eq:transV}
\ee

Considering the baryon as a three-quark configuration, we apply the diquark model for  simplicity~\cite{thesis2015Olbrich,Olbrich:2015gln}. Then the baryon states are donated as $N_{\rm R}^{\rm (R R)}, N_{\rm L}^{\rm (R R)}, N_{\rm L}^{\rm (L L)}$, and $N_{\rm R}^{\rm (L L)}$, where $(\rm R R)$ and $(\rm L L)$ denote right- and left-hand diquark configurations, respectively, and $\rm R$ and $\rm L$ represent the quark configurations. Their transformation behaviors are given by
\be
N_{\rm R}^{\rm (R R)} & \rightarrow & g_{\rm R} N_{\rm R}^{\rm (R R)} g_{\rm R}^{\dagger}, N_{\rm L}^{\rm (R R)} \rightarrow g_{\rm L} N_{\rm L}^{\rm (R R)} g_{\rm R}^{\dagger}, \nonumber\\
N_{\rm R}^{\rm (L L)} & \rightarrow & g_{\rm R} N_{\rm R}^{\rm (L L)} g_{\rm L}^{\dagger}, N_{\rm L}^{\rm (L L)} \rightarrow g_{\rm L} N_{\rm L}^{\rm (L L)} g_{\rm L}^{\dagger},
\ee
and
\be
N_{\rm R}^{\rm (R R)} & \rightarrow & e^{-3 i v} N_{\rm R}^{\rm (R R)}, \quad N_{\rm L}^{\rm (R R)} \rightarrow e^{-i v} N_{\rm L}^{\rm (R R)}\ , \nonumber\\
N_{\rm R}^{\rm (L L)} & \rightarrow & e^{i v} N_{\rm R}^{\rm (L L)}, \quad N_{\rm L}^{\rm (L L)} \rightarrow e^{3 i v} N_{\rm L}^{\rm (L L)}\ .
\ee

Since in the effective model based on the linear realization of the chiral symmetry to be used here, there is no self-consistent power-counting mechanism, like the derivative expansion in chiral perturbation theory, we set the following counting rules to truncate our model:
\begin{itemize}
	\item[I.] The order of an effective term is determined by the number of valence quarks $N_{q}$.
	\item[II.] Double trace terms are neglected because of large $N_c$ suppression~\cite{Parganlija:2010fz}.
	\item[III.] The terms containing vector mesons should be given priority considering rule-I.
\end{itemize}

\subsection{The Lagrangian}

After the above discussion on the transformation properties of the meson and baryon fields, we are ready to construct the effective model. Formally, we decompose the Lagrangian of the bELSM as
\be
\mathcal{L} & = & \mathcal{L}_{\mathrm{M}} + \mathcal{L}_{\mathrm{V}} + \mathcal{L}_{\mathrm{B}}.
\ee

The $\mathcal{L}_{\mathrm{M}}$ is the gauged scalar and pseudoscalar part of ELSM which takes the form
\be
\mathcal{L}_{\mathrm{M}} & = &{} \frac{1}{2} \operatorname{Tr}\left(D_\mu \Phi D^\mu \Phi^{\dagger}\right)+\frac{1}{2} \operatorname{Tr}\left(D_\mu \hat{\Phi} D^\mu \hat{\Phi}^{\dagger}\right) \nonumber\\
& & {} - V_0(\Phi, \hat{\Phi})-V_{\mathrm{SB}},
\ee
where $D_\mu \Phi=\partial_\mu \Phi-i g_1\left(L_\mu \Phi- \Phi R_\mu\right)$. $V_{\rm{SB}}$ stands for the effect of the explicit symmetry-breaking which will not be considered in this work. The interaction potential $V_0$ is given by
\begin{widetext}
\be
V_0 & = &{} -c_2 \operatorname{Tr}\left(\Phi \Phi^{\dagger}\right)+c_4 \operatorname{Tr}\left(\Phi \Phi^{\dagger} \Phi \Phi^{\dagger}\right) + d_2 \operatorname{Tr}\left(\hat{\Phi} \hat{\Phi}^{\dagger}\right) + e_3\left(\epsilon_{a b c} \epsilon^{d e f} \Phi_d^a \Phi_e^b \hat{\Phi}_f^c+\text { H.c. }\right) \nonumber\\
& &{} +c_3\left[\gamma_1 \ln \left(\frac{\operatorname{det} \Phi}{\operatorname{det} \Phi^{\dagger}}\right) +\left(1-\gamma_1\right) \ln \left(\frac{\operatorname{Tr}\left(\Phi \hat{\Phi}^{\dagger}\right)}{\operatorname{Tr}\left(\hat{\Phi} \Phi^{\dagger}\right)}\right)\right]^2 ,
\ee
\end{widetext}
where the $c_3$ term accounts for the $\rm U(1)_A$ anomaly of QCD~\cite{Rosenzweig:1979ay,Hsu:1998jd,Fariborz:2005gm} and the other terms are chiral invariant. It was found that the $e_3$ terms are crucial for scalar meson mass split between octet and singlet~\cite{Black:1998wt}.

By considering the transformation properties~\eqref{eq:transV}, the chiral and axial invariant Lagrangian for the vector meson part at the lowest order takes the form
\begin{widetext}
\be
\mathcal{L}_{\rm V} & = & -\frac{1}{4} \operatorname{Tr}\left(R_{\mu \nu}^2+L_{\mu \nu}^2\right) + i \frac{g_2}{2}\left\{\operatorname{Tr}\left(L_{\mu \nu}\left[L^\mu, L^\nu\right]\right)+\operatorname{Tr}\left(R_{\mu \nu}\left[R^\mu, R^\nu\right]\right)\right\} \nonumber\\
& &{} + h_2 \operatorname{Tr}\left(\left|L_\mu \Phi\right|^2+\left|\Phi R_\mu\right|^2\right) +\hat{h}_2 \operatorname{Tr}\left(\left|L_\mu \hat{\Phi}\right|^2+\left|\hat{\Phi} R_\mu\right|^2\right) \nonumber\\
& &{} +2 h_3 \operatorname{Tr}\left(L_\mu \Phi R^\mu \Phi^{\dagger}\right)+2 \hat{h}_3 \operatorname{Tr}\left(L_\mu \hat{\Phi} R^\mu \hat{\Phi}^{\dagger}\right) \nonumber\\
& &{} +g_3\left[\operatorname{Tr}\left(L_\mu L_\nu L^\mu L^\nu\right)+\operatorname{Tr}\left(R_\mu R_\nu R^\mu R^\nu\right)\right] +g_4\left[\operatorname{Tr}\left(L_\mu L^\mu L_\nu L^\nu\right)+\operatorname{Tr}\left(R_\mu R^\mu R_\nu R^\nu\right)\right],
\ee
\end{widetext}
where $R_{\mu \nu}=\partial_\mu R_\nu-\partial_\nu R_\mu-i\left[R_{\mu},R_{\nu}\right]$ and $L_{\mu \nu}=\partial_\mu L_\nu-\partial_\nu L_\mu-i\left[L_{\mu},L_{\nu}\right]$. In this Lagrangian, all the terms have $N_{q}=8$ at most except the $\hat{h}_2$ and $\hat{h}_3$ terms which are allowed by power-counting rule III.

Next, let us turn to the baryon sector. According to the above discussion, with respect to the chiral and axial invariances, we can write down the effective Lagrangian in term of the basis $N_{\rm R}^{\rm (R R)}, N_{\rm L}^{\rm (R R)}, N_{\rm L}^{\rm (L L)}$, and $N_{\rm R}^{\rm (L L)}$ as 
\begin{widetext}
\be
\mathcal{L}_{\rm B} & = & {\rm Tr}\left\{ \bar{N}_{\rm R}^{\rm (RR)} i\gamma_\mu D_{1R}^\mu N_{\rm R}^{\rm (RR)} + \bar{N}_{\rm L}^{\rm (RR)} i\gamma_\mu D_{2L}^\mu N_{\rm L}^{\rm (RR)} + \bar{N}_{\rm R}^{\rm (LL)} i\gamma_\mu D_{2R}^\mu N_{\rm R}^{\rm (LL)} + \bar{N}_{\rm L}^{\rm (LL)} i\gamma_\mu D_{1L}^\mu N_{\rm L}^{\rm (LL)} \right\} \nonumber\\
& & {} + 2 c^\prime{\rm Tr}\left\{R^\mu \bar{N}_{\rm R}^{\rm (RR)}\gamma_\mu N_{\rm R}^{\rm (RR)} + L^\mu \bar{N}_{\rm L}^{\rm (LL)} \gamma_\mu N_{\rm L}^{\rm (LL)}\right\} \nonumber\\
& & {} - g{\rm Tr}\left\{ \bar{N}_{\rm L}^{\rm (RR)} \Phi N_{\rm R}^{\rm (RR)} + \bar{N}_{\rm R}^{\rm (RR)} \Phi^\dagger N_{\rm L}^{\rm (RR)} + \bar{N}_{\rm L}^{\rm (LL)} \Phi N_{\rm R}^{\rm (LL)} + \bar{N}_{\rm R}^{\rm (LL)} \Phi^\dagger N_{\rm L}^{\rm (LL)} \right\},
\ee
\end{widetext}
where $D_{i R}^\mu = \partial^\mu - i c_{i} R^\mu$ and $D_{i L}^\mu = \partial^\mu - i c_{i} L^\mu$ with $c_{i}$ being constants. An interesting and important observation from this Lagrangian is that, due to the invariance under axial transformation, only the two-quark component of the scalar fields couples to baryon fields at the lowest order. This is one of the main conclusions of this work.

Without considering the negative parity baryons, one can make the identification
\be
N_{\rm R,L}^{\rm (R R)} & = & \frac{1}{\sqrt{2}}\frac{1\pm\gamma_5}{2}B, \nonumber\\
N_{\rm R,L}^{\rm (L L)} & = &{} - \frac{1}{\sqrt{2}}\frac{1\pm\gamma_5}{2}B,
\ee
with $B$ being the baryon octet. Then, the final Lagrangian for baryons to be used in the following is
\be
\mathcal{L}_{\mathrm{B}} & = & \frac{1}{2} \operatorname{Tr}\left\{\bar{B} i \gamma_\mu\left[\left(D_{\rm R}^\mu+D_{\rm L}^\mu\right) \right.\right. \nonumber\\
& & \left.\left. \;\;\;\;\;\;\;\;\;\;\;\;\;\;\;\;\;\; {} +\gamma_5\left(D_{\rm R}^\mu-D_{\rm L}^\mu\right)\right] B\right\} \nonumber\\
& &{} -\frac{g}{2} \operatorname{Tr}\left\{\bar{B}\left[\left(\Phi+\Phi^{\dagger}\right)+\gamma_5\left(\Phi-\Phi^{\dagger}\right)\right] B\right\},\nonumber\\
\ee
where  
$D_{\rm R}^\mu B=\partial^\mu B-i c R^\mu B - i c^\prime B R^\mu$ and $D_{\rm L}^\mu B =\partial^\mu B -i c L^\mu B - i c^\prime B L^\mu $ with $c$ and $c^\prime$ being coupling constants.

\subsection{Physical states}

There are two meson multiplates $\Phi$ and $\hat{\Phi}$ in the bELSM discussed above. The physical states are the admixture of them. 
The two-quark nonets can be organized in terms of the scalar and pseudoscalar states as 
\begin{widetext}
\be
\Phi & = & S+i P=\left(
	\begin{array}{ccc}
		\frac{\left(\sigma_{N}+a_{0}^{0}\right)+i\left(\eta_{N}+\pi^{0}\right)}{\sqrt{2}} & a_{0}^{+}+i \pi^{+} & K_{S}^{+}+i K^{+} \\
		a_{0}^{-}+i \pi^{-} & \frac{\left(\sigma_{N}-a_{0}^{0}\right)+i\left(\eta_{N}-\pi^{0}\right)}{\sqrt{2}} & K_{S}^{0}+i K^{0} \\
		K_{S}^{-}+i K^{-} & \bar{K}_{S}^{0}+i \bar{K}^{0} & \sigma_{S}+i \eta_{S}
	\end{array}
	\right).
\ee
The octet and singlet states can be obtained via
	\begin{equation}
		\Phi=\frac{1}{\sqrt{2}}\sum_{i=0}^{8}\Phi_{i}\lambda_{i} = \frac{1}{\sqrt{2}}\left(\sum_{i=0}^{8}S_{i}\lambda_{i} + i\sum_{i=0}^{8}P_{i}\lambda_{i}\right) ,
	\end{equation}
where $\lambda_0$ is $\sqrt{\frac{2}{3}}I_{3\times3}$ and the others are Gell-Mann matrices. The four-quark configurations can be written in the similar way and will be denoted with an overcaret in the following.

Moreover, vector and axial-vector mesons can be obtained through the combination of left- and right-hand currents:
\be
(L, R)^{\mu} & = & V^{\mu}\pm A^{\mu}=\frac{1}{\sqrt{2}}\left(
		\begin{array}{ccc}
			\frac{\omega_{N}^{\mu}+\rho^{\mu 0}}{\sqrt{2}}\pm\frac{f_{1 N}^{\mu}+a_{1}^{\mu 0}}{\sqrt{2}} & \rho^{\mu+}\pm a_{1}^{\mu+} & K^{* \mu+}\pm K_{1}^{\mu+} \\
			\rho^{\mu-}\pm a_{1}^{\mu-} & \frac{\omega_{N}^{\mu}-\rho^{\mu 0}}{\sqrt{2}}\pm \frac{f_{1 N}^{\mu}-a_{1}^{\mu 0}}{\sqrt{2}} & K^{* \mu 0} \pm  K_{1}^{\mu 0} \\
			K^{* \mu-}\pm K_{1}^{\mu-} & \bar{K}^{* \mu 0}\pm \bar{K}_{1}^{\mu 0} & \omega_{S}^{\mu}\pm f_{1 S}^{\mu}
		\end{array}
		\right).
\ee
\end{widetext}
Finally, the baryon octet takes the form 
\begin{equation}~\label{eq:baryonm}
	B \equiv\left(
	\begin{array}{ccc}
		\frac{\Lambda}{\sqrt{6}}+\frac{\Sigma^{0}}{\sqrt{2}} & \Sigma^{+} & p \\
		\Sigma^{-} & \frac{\Lambda}{\sqrt{6}}-\frac{\Sigma^{0}}{\sqrt{2}} & n \\
		\Xi^{-} & \Xi^{0} & -\frac{2 \Lambda}{\sqrt{6}}
	\end{array}
	\right).
\end{equation}

At low energies, the chiral symmetry of QCD breaks to $\rm U(3)_V$ spontaneously, so it is more reasonable to let the scalar components which lie on the diagonal position of the $\Phi$ matrix to have nonzero vacuum expectation values
\begin{equation}
	\left\langle S_{a}^{b}\right\rangle=\alpha_{a} \delta_{a}^{b}, \quad\left\langle \hat{S}_{a}^{b}\right\rangle=\beta_{a} \delta_{a}^{b},
\end{equation}
where $ S_{a}^{b}$ and $\hat{S}_{a}^{b}$ stand for the two-quark and four-quark scalar components, respectively. Certainly, $\alpha_{1}=\alpha_{2}=\alpha_{3}=\alpha$ and $\beta_{1}=\beta_{2}=\beta_{3}=\beta$ if $\rm U(3)_V$ symmetry is exact.  Then the fluctuations of the two- and four-quark fields are defined as $\tilde{S}_a^b=S_a^b-\alpha \delta_a^b$ and $\tilde{\hat{S}}_a^b=\hat{S}_a^b-\beta \delta_a^b$, respectively.

The existence of $c_3$ and $e_3$ terms in $V_0$ leads to the mixing of two-quark and four-quark states. The mixing angles can be obtained by solving two-point vertex functions, and mixing matrix can be defined as
\be
\label{eq:rot}
\left[\begin{array}{c}
		\Phi_{i,j} \\
		\hat{\Phi}_{i,j}
	\end{array}\right] & = & R\left[\begin{array}{c}
		\Phi_{i,j}^{\prime} \\
		\hat{\Phi}_{i,j}^{\prime}
	\end{array}\right] \nonumber\\
& = & \left[\begin{array}{cc}
		\cos \theta_{i,j} & -\sin \theta_{i,j} \\
		\sin \theta_{i,j} & \cos \theta_{i,j}
	\end{array}\right]\left[\begin{array}{c}
		\Phi_{i,j}^{\prime} \\
		\hat{\Phi}_{i,j}^{\prime}
	\end{array}\right],
\ee
where superscript the superscript prime denotes the physical states. Currently, it leads to four mixing sets of (pseudo)scalar sectors (octets and singlets).

The four-quark configurations follow the same terminology.
Then, the minimum potential conditions should also be considered in order to pin down vacuum expectation values of isoscalar scalar states, $\alpha$ and $\beta$:
\be
\label{eq:mvp}
\left\langle\frac{\partial V_{0}}{\partial S_{a,a}}\right\rangle & = & 2 \alpha\left(-c_{2}+2 c_{4} \alpha^{2}+4 e_{3} \beta\right)=0, \nonumber\\
\left\langle\frac{\partial V_{0}}{\partial \hat{S}_{a,a}}\right\rangle & = & 2\left(d_{2} \beta+2 e_{3} \alpha^{2}\right)=0.
\ee

In this work, we will not consider strange hadrons or mesons heavier than $1\ \mathrm{GeV}$. And, since EOS of nuclear matter is obtained via RMF approximation, the axial-vector and pseudoscalar mesons will also be dropped out.
Then the baryon field $B$ can be rewritten as 
\begin{equation*}
	\Psi=\left(
	\begin{array}{c}
		p \\
		n
	\end{array}
	\right),
\end{equation*}
and the relevant mesons should be scalar and vector mesons including $f_0(500)$ (denoted as $\sigma$), $f_0(980),\ a_0(980),\ \omega$, and $\rho$.

\subsection{Physical parameters}

The physical parameters can be defined in terms of the parameters in the above discussed Lagrangian and the corresponding mixing angles. By diagonalizing of two-point effective potentials, one can obtain mass terms for scalar singlets and scalar octets, respectively, as
\begin{widetext}
\be
m_{s,0,\pm} & = &{} -c_2+6\alpha^2c_4+d_2+4\beta e_3 \nonumber\\
& &{} \pm \sqrt{\left(-c_2+6 \alpha^{2}c_4+d_2+4 \beta e_3\right)^{2}+4\left(c_2 d_2-6 \alpha^{2} c_4 d_2+4 e_3\left(4 \alpha^{2} e_3-\beta d_2\right)\right)},\nonumber\\
m_{s,8,\pm} & = &{} -c_2+6\alpha^2c_4+d_2-2\beta e_3 \nonumber\\
& &{} \pm \sqrt{\left(c_2-6 \alpha^{2}c_4-d_2+2 \beta e_3\right)^{2}+4\left(c_2 d_2-6 \alpha^{2} c_4 d_2+2 e_3\left(2 \alpha^{2} e_3+\beta d_2\right)\right)}.
\ee
\end{widetext}
It is obvious that one stands for the light states and the other one stands for heavier ones and the mixing angle in Eq.~\eqref{eq:rot} can be also obtained as
\be
\label{eq:mxa}
\theta_{s,0} & = & {} -\frac{1}{2}\arctan\left(
		\frac{8\alpha e_3}{c_2-6\alpha^2 c_4-4\beta e_3+d_2}
		\right),\nonumber\\
\theta_{s,8} & = &{} -\frac{1}{2}\arctan\left(
		\frac{4\alpha e_3}{c_2-6\alpha^2 c_4+2\beta e_3+d_2}
		\right).\nonumber\\
\ee

With the above arguments, the Lagrangian contributing to the mean field approach can be written as
\be
\mathcal{L}_{RMF} & = & \bar{\Psi}\left(i\slashed{\partial}-m_N\right)\Psi \nonumber\\
& &{} +\mathcal{L}_{\sigma}+\mathcal{L}_{\omega}+\mathcal{L}_{\rho}+\mathcal{L}_{a_0}+\mathcal{L}_{f_0}+\mathcal{L}_I,\nonumber\\
\ee
where $m_N=g\alpha$. The Lagrangian for the meson part can be organized as
\be
\mathcal{L}_{\sigma} & = & \frac{1}{2}\left(\partial_{\mu}\sigma\partial^{\mu}\sigma-m_{\sigma}^2\sigma^2\right),\nonumber\\
\mathcal{L}_{\omega} & = &{} -\frac{1}{4}\omega_{\mu\nu}\omega^{\mu\nu}+\frac{1}{2}m_{\omega}^2\omega_{\mu}\omega^{\mu},\nonumber\\
\mathcal{L}_{\rho} & = &{} -\frac{1}{4}\vec{\rho}_{\mu\nu}\cdot\vec{\rho}^{\mu\nu}+\frac{1}{2}m_{\rho}^2\vec{\rho}_{\mu}\cdot\vec{\rho}^{\mu},\nonumber\\
\mathcal{L}_{a_0} & = & \frac{1}{2}\left(\partial_{\mu}\vec{a}_{0}\cdot\partial^{\mu}\vec{a}_0-m_{a_0}^2\vec{a}_{0}^2\right),\nonumber\\
\mathcal{L}_{f_0} & = & \frac{1}{2}\left(\partial_{\mu}f_{0}\partial^{\mu}f_0-m_{f_0}^2f_{0}^2\right)\ ,
\ee
where $ a_0=a^i_{0}\tau^i$, $\rho_{\mu}=\rho^i_{\mu}\tau^i $ with $\tau^i$ being the Pauli matrices, and
\be  
m_{\rho} & = & m_{\omega} \nonumber\\
& = &\sqrt{2h_2\alpha^2+2\hat{h}_2\beta^2+2h_3\alpha^2+2\hat{h}_3\beta^2}.\nonumber\\
\ee
The interaction terms can be written as
\be
\mathcal{L}_I & = & \bar{\Psi}\left(g_{a_0NN}a_0+g_{f_0NN}f_0 + g_{\sigma NN}\sigma_0\right. \nonumber\\
& & \left.\;\;\;\;\; + g_{\rho NN}\gamma_{\mu}\rho^{\mu}+g_{\omega NN}\gamma_{\mu}\omega^{\mu}\right)\Psi \nonumber\\
& &{} +\mathcal{L}_3+\mathcal{L}_4\ ,
\ee
where $g_{a_0NN}=-\frac{1}{\sqrt{2}}g\cos\tilde{\theta}_{s,8},\ g_{f_0NN}=-\frac{g}{\sqrt{6}}\cos\tilde{\theta}_{s,8},\ g_{\sigma NN}=-\frac{g}{\sqrt{3}}\cos\tilde{\theta}_{s,0},\text{and} g_{\rho NN}=g_{\omega NN}=\frac{c}{2}$. $\tilde{\theta}$ can be $\theta$ or $\frac{\pi}{2}+\theta$ defined in Eq.~\eqref{eq:mxa} corresponding to the choice of light set between $\Phi_{i,j}^{\prime}$ and $\hat{\Phi}_{i,j}^{\prime}$.  $\mathcal{L}_{3(4)}$ refers to three-(four)-meson coupling terms which are listed in the Appendix.

Note that with RMF approximation only zero components in isospin space and nonderivative terms in Lorentz space survive for meson fields. Therefore, $g_1$ and $g_2$ make no contribution and the effects of $g_4,\ h_3$ and $\hat{h}_3$ can be absorbed by $g_3,\ h_2$ and $\hat{h}_2$, respectively.

\section{Phenomenological analysis and numerical results}
\label{sec:num}

\subsection{Parameter values}

The parameters interrelated in this work are $\alpha,\ \beta,\ c_2,\ c_4,\ d_2,\ e_3,\ h_2,\ \hat{h}_2,\ g_3,$ and $c$. By using  Eq.~\eqref{eq:mvp}, two parameters, e.g., $d_2$ and $c_2$ can be reduced. The other parameters are determined by pinning the empirical values of meson spectra and nuclear matter properties. Our results are listed in Table~\ref{tab:parameters-fit}. In the following, this set of parameters is denoted as ``opt."

\begin{table}[htb]\small
	\centering
	\caption{
		Values of the parameters interrelated in this work. $\alpha$ and $\beta$ are in units of MeV and $e_3$ is in units of GeV. 
	}
	\label{tab:parameters-fit}
	\begin{threeparttable}   
		\begin{tabular}{@{}cccccccc}
			\hline
			\hline
			$\alpha$ & $\beta$& $e_3$ &   $c_4$  & $h_2$ & $\hat{h}_2$ & $g_3$ & $c$\\
			\hline
			$32.3$  & $9.55$  & $-3.58$ &  $154$ & $370$ & $-915$ & $630$ & $25.8$\\
			\hline
			\hline
		\end{tabular}
	\end{threeparttable}
\end{table}

The light isosinglet scalar meson $\sigma$ takes the choice of $\tilde{\theta}=\theta$ and the light mesons $a_0$ and $f_0$ in the original octet take the choice of $\tilde{\theta}=\theta+\frac{\pi}{2}$, where $\theta_{s,0}=41.0\degree$ and $\theta_{s,8}=40.7\degree$ which indicates the two-quark component in the singlet is $56.9\%$ and that in the octet is $42.5\%$. This leads to $g_{\sigma NN}=-13.0,\ g_{a_0NN}=13.8,\ g_{f_0NN}=7.98,$ and $g_{\rho(\omega)NN}=12.7$. 

\begin{widetext}
The three-(four)-meson couplings $\mathcal{L}_{3(4)}$ after RMF simplification are organized as
\be
\mathcal{L}_3 & = &{} -m_N\left(-13.2 a_0^2f_0+4.41f_0^3+1.84f_0\rho^2+9.72a_0^2\sigma+9.72f_0^2\sigma-4.06\rho^2\sigma+1.96\sigma^3 \right.\nonumber\\
& &\left.\;\;\;\;\;\;\;\;\;\;\;\;{} +6.36a_0\rho\omega +1.84f_0\omega^2-4.06\omega^2\sigma\right)
\ee
and
\be
\mathcal{L}_4 & = &{} -13.9 a_0^4-27.8 a_0^2f_0^2-13.9 f_0^4+116a_0^2\rho^2-38.6f_0^2\rho^2+158\rho^4+90.9a_0^2f_0\sigma-30.3f_0^3\sigma \nonumber\\
& &{}  +98.7f_0\rho^2\sigma -74.4a_0^2\sigma^2-74.4f_0^2\sigma^2 -61.2\rho^2\sigma^2-16.6\sigma^4-267a_0f_0\rho\omega+342a_0\rho\sigma\omega\nonumber\\
& &{} -116a_0^2\omega^2-38.6f_0^2\omega^2 +946\rho^2\omega^2+98.7f_0\sigma\omega^2-61.2\sigma^2\omega^2+158\omega^4\ .
\ee
where, to clearly show the magnitudes of the contributions, we explicitly write down the values of the coupling constants. The above parameter space leads to the spectra of mesons and nuclear matter quantities listed in Table~\ref{tab:mass-spectra} and Table~\ref{tab:nuclear-matter}, respectively.

\begin{table*}[htb]\small
	\centering
	\caption{
		The spectra of hadrons in units of $\mathrm{MeV}$. The empirical values are chosen as the real parts of the corresponding resonance T-matrix poles from Ref.~\cite{ParticleDataGroup:2022pth}.
		The excited state $f_0'$ is chosen to be $f_0(1370)$ with $a_0'$ being $a_0(1450)$ and $\sigma'$ being $f_0(1500)$.
		Since $\rm U(3)_V$ is exact in current work, the empirical values for $\rho$ and $\omega$ or $f_0(1370)$ and $a_0(1450)$ are combined together.
		The nucleon mass is fixed as $938 \rm MeV$ for simplicity.
	}
	\begin{threeparttable}~\label{tab:mass-spectra}
		\begin{tabular}{@{}ccccccc}
			\hline
			\hline
			& $m_N$ & $m_{\sigma}$ & $m_{f_0(a_0)}$ & $m_{f_0'(a_0')}$ & $m_{\sigma'}$ & $m_{\rho(\omega)}$ \\
			\hline
			Empirical & $938$-$940$ & $400$-$800$ & $960$-$1010$ & $1250$-$1500$ & $1430$-$1530$ & $761$-$783$ \\
			\hline
			opt & $938$ & $701$ & $965$ & $1359$ & $1526$ & $779$ \\
			\hline
			\hline
		\end{tabular}
	\end{threeparttable}
\end{table*}

\begin{table*}[htb]\small
	\centering
	\caption{Nuclear matter properties at saturation density $n_0$. $e_0$ is the binding energy of nucleon, $E_{\mathrm{sym}}(n)=\left.\frac{1}{2} \frac{\partial^2 E(n, a)}{\partial a^2}\right|_{a=0}$ is the symmetry energy, $K_0=\left.9 n_0^2 \frac{\partial^2 E(n, 0)}{\partial n^2}\right|_{n=n_0}$ is the incompressibility coefficient, $J_0=\left.27 n_0^3 \frac{\partial^{3} E(n, 0)}{\partial n^3}\right|_{n=n_0}$ is the skewness coefficient and $L_0=\left.3 n_0 \frac{\partial E_{\mathrm{sym}}(n_0)}{\partial n}\right|_{n=n_0}$ is the symmetry energy density slope.}
	\begin{threeparttable}~\label{tab:nuclear-matter}
		\begin{tabular}{@{}ccccccc}
			\hline
			\hline
			& $n_0\ (\mathrm{fm^{-3}})$ & \(e_0\ (\mathrm{MeV})\) & $E_{\mathrm{sym}}(n_0)\ (\mathrm{MeV})$ & $J_0\ (\mathrm{MeV})$ & $L_0\ (\mathrm{MeV})$ & $K_0\ (\mathrm{MeV})$ \\
			\hline
			Empirical & $0.155\pm0.050$\cite{Sedrakian:2022ata} & $-15.0\pm1.0$\cite{Sedrakian:2022ata} & $30.9\pm1.9$\cite{Lattimer:2012xj} & $-700\pm500$\cite{Farine:1997vuz} & $52.5\pm17.5$\cite{Lattimer:2012xj} & $230\pm30$\cite{Dutra:2012mb} \\
			\hline
			opt & $0.155$ & $-16.0$ & $31.9$ & $-449$ & $62.7$ & $225$ \\
			\hline
			\hline
		\end{tabular}
	\end{threeparttable}
\end{table*}
\end{widetext}

It can be seen that the choice of parameter space opt leads to reasonable physical results. In the numerical calculation, it is found $\alpha \simeq 30\ \rm{MeV}$, which makes $|g_{\sigma NN}|\sim 10$, in order to obtain the physical quantities of NM.
The $g_{\sigma NN}$ is consistent with Refs.~\cite{Sugahara:1993wz,Lalazissis:1996rd,Shen:1998gq,Fattoyev:2010mx,Li:2022okx}, where the couplings between sigma and nucleon are parametrized as one-boson-exchange (OBE) form. It provides evidence that this model is suited for the description of NM properties and also makes a suggestion that $m_{\sigma}$ may be around $700\ \rm MeV$ in the nuclear medium.

\subsection{Parameter dependence of nuclear matter properties}

\begin{figure*}[tbh]
	\centering
	\subfigure[]{\includegraphics[scale=0.15]{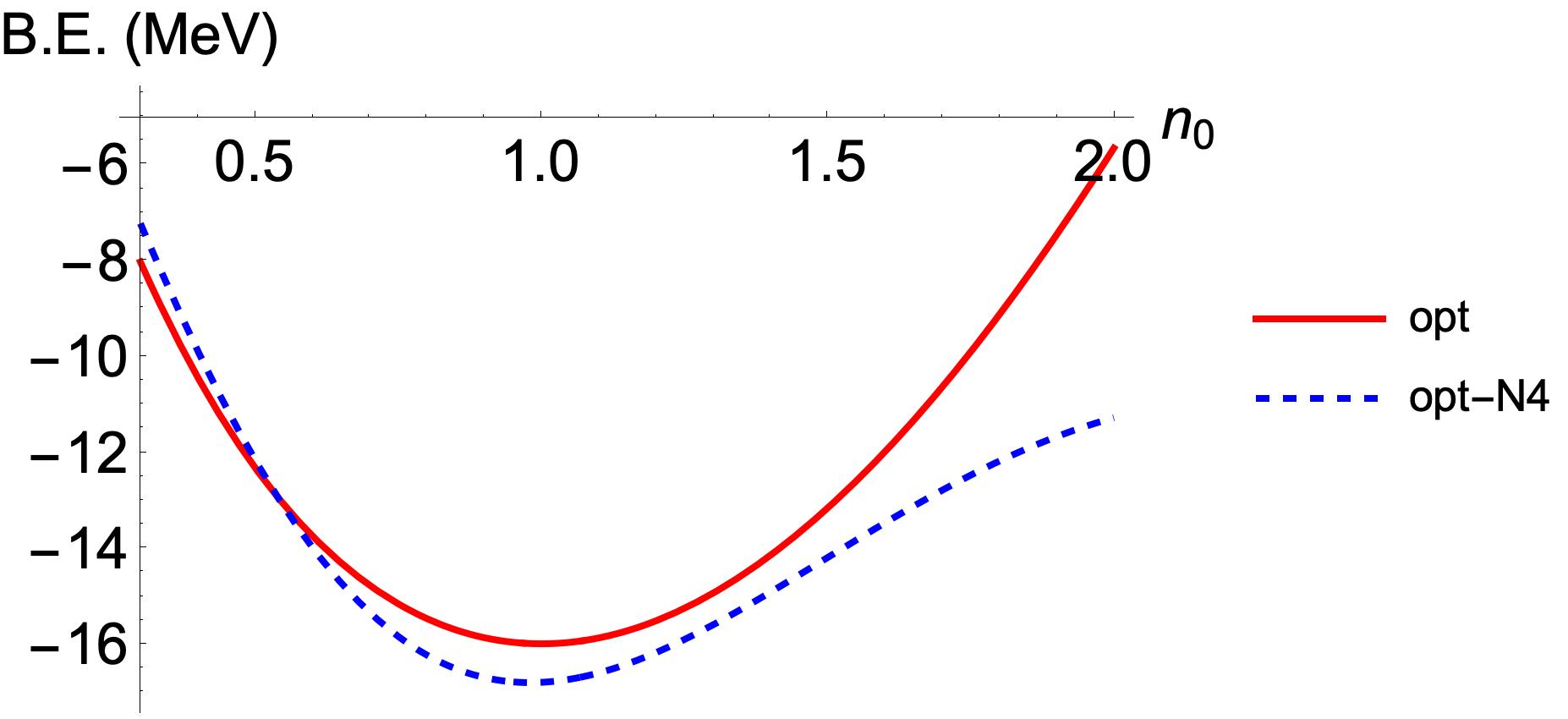}} 
	\subfigure[]{\includegraphics[scale=0.15]{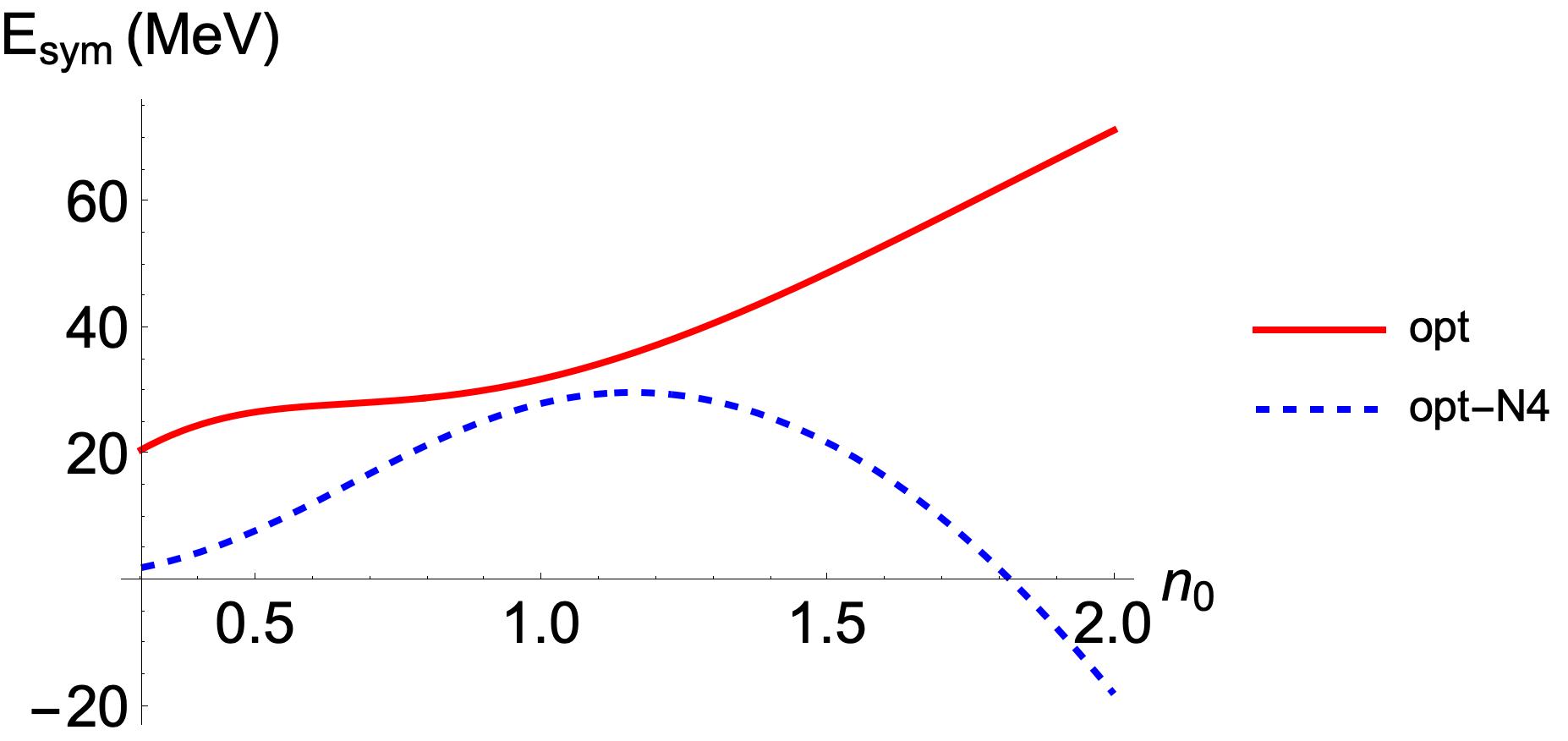}}
	\subfigure[]{\includegraphics[scale=0.15]{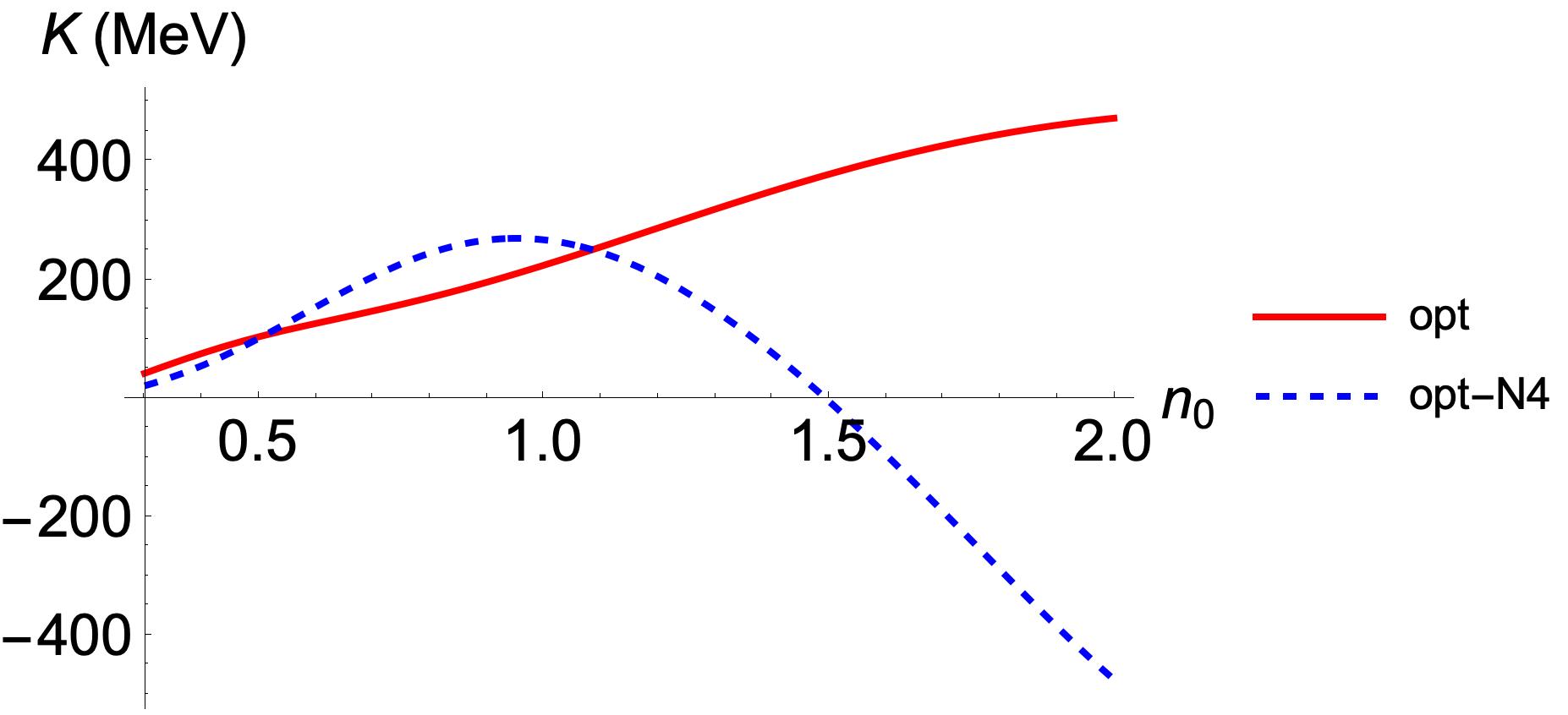}}
	\subfigure[]{\includegraphics[scale=0.15]{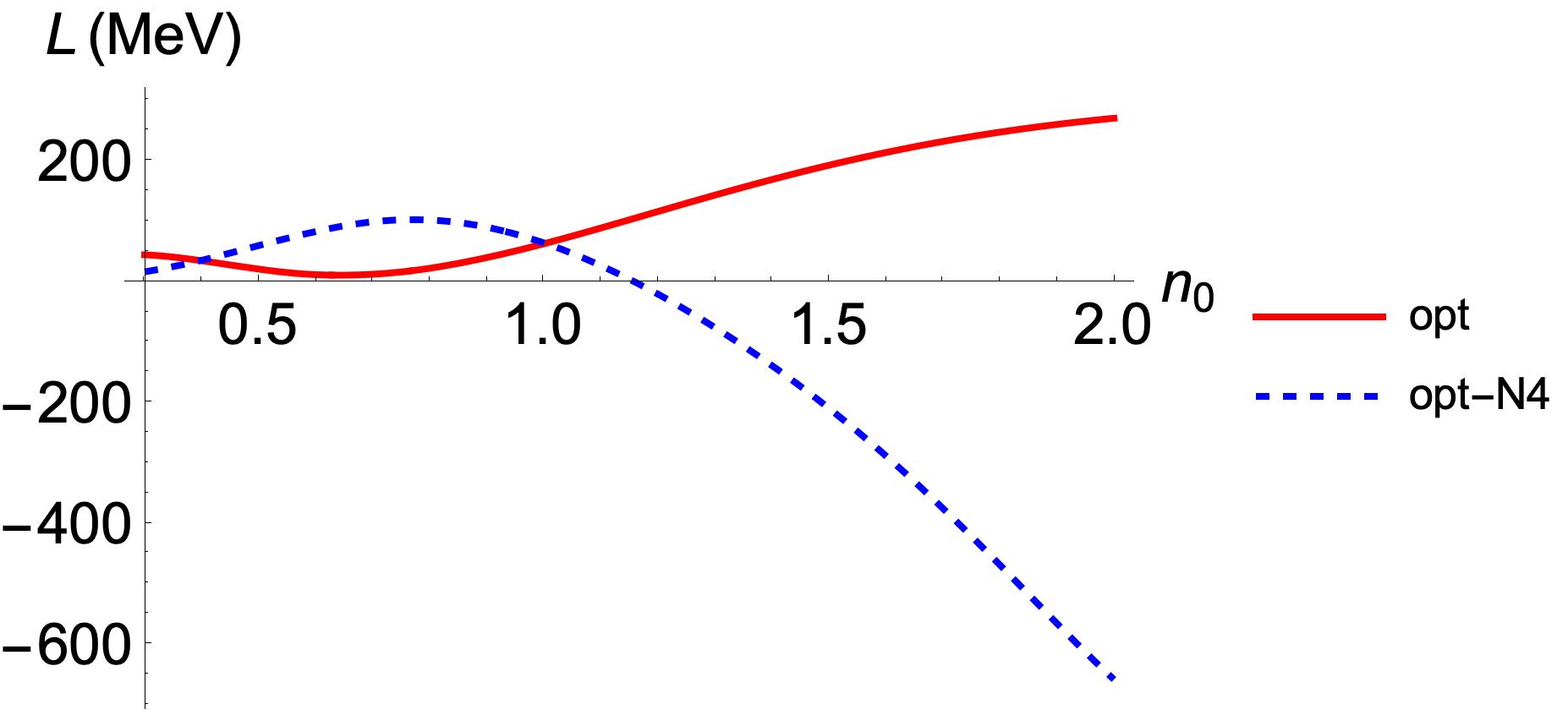}}
	\subfigure[]{\includegraphics[scale=0.15]{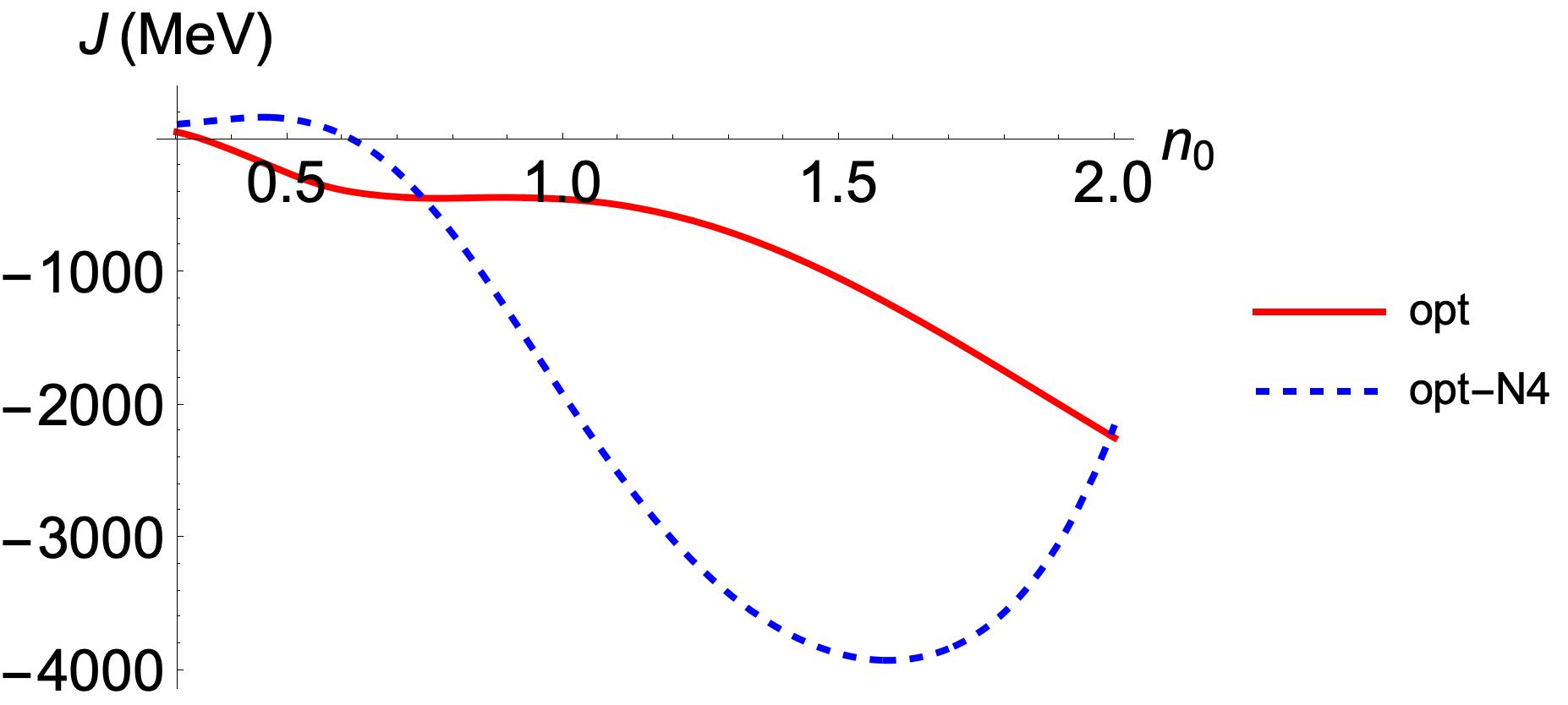}}
	\caption{
		Comparison between models with (opt) and without $\hat{h}_2$ (opt-N4) terms.
		(a) Binding energy difference.
		(b) Symmetry energy difference.
		(c) Incompressibility coefficient difference.
		(d) Symmetry energy slope difference.
		(e) Skewness coefficient difference.
	}
	\label{fig:n4}
\end{figure*}

The couplings between scalar mesons and nucleons contribute to attractive potential, while those between vector mesons and nucleons contribute to repulsive potential, and the competition between them leads to the saturation of NM and the value of $e_0$. After obtaining the reasonable $e_0$ in current bELSM, the $K_0$ and $E_{\rm sym}(n_0)$ are found to be too large compared to experimental data. Then, the four-vector meson self-interaction term $g_3$ is found to be crucial to suppress these higher-order density-dependent quantities. But when pinning down all these quantities, the choice of $g_3$ makes the $E_{\rm sym}$ and $K$ to be negative, which may lead to a too soft neutron star structure in the current scheme. With this, the importance of the four-quark configuration $\hat{h}_2$ term comes in and compensates the effects brought by $g_3$ terms at high-density regions while keeping higher-order quantities at $n_0$ reasonable. In addition, the $\hat{h}_2$ terms yield $|g_{\omega NN}|\simeq 13$, which is also consistent with the models in Refs.~\cite{Sugahara:1993wz,Lalazissis:1996rd,Shen:1998gq,Fattoyev:2010mx,Li:2022okx}. These effects can be seen in Fig.~\ref{fig:n4}, where opt and opt-N4 which refers to the parameter set without $\hat{h}_2$ terms are compared. In opt-N4, the parameters are chosen as $\alpha= 35.8\ \rm{MeV},\ \beta= 8.03\ \rm{MeV},\ c_4=146,\ e_3=-2640\ \rm{MeV},\ h_2=232,\ c=14.6,$ and $g_3=123$ in order to pin down the NM properties and this choice yields $g_{\omega NN}=7.28$ with $m_{\sigma}=910~\rm MeV$.

\begin{figure*}[htb]
	\centering
	\subfigure[\(\alpha\) variarion.]{\includegraphics[scale=0.2]{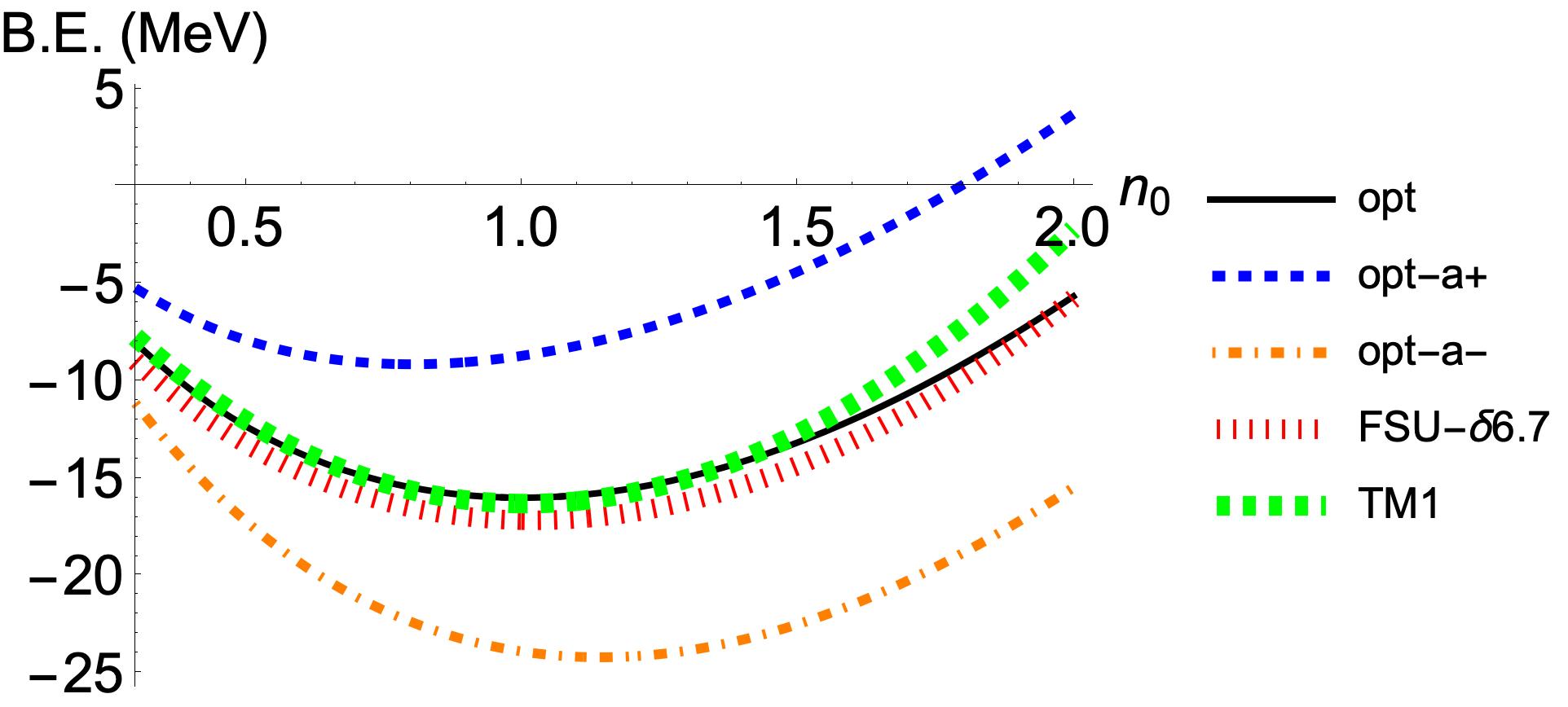}}
	\subfigure[\(\beta\) variarion.]{\includegraphics[scale=0.2]{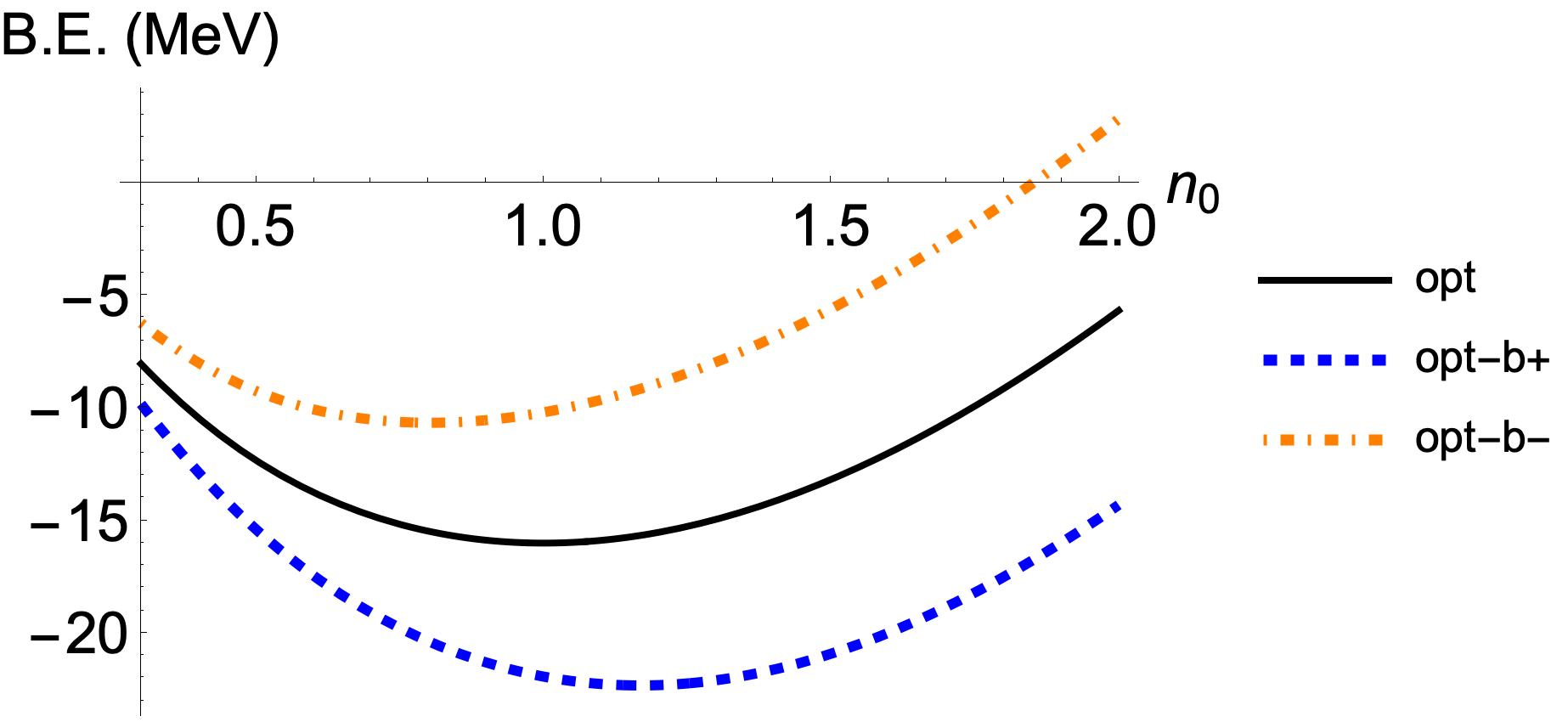}}
	\subfigure[\(c_4\) variarion.]{\includegraphics[scale=0.2]{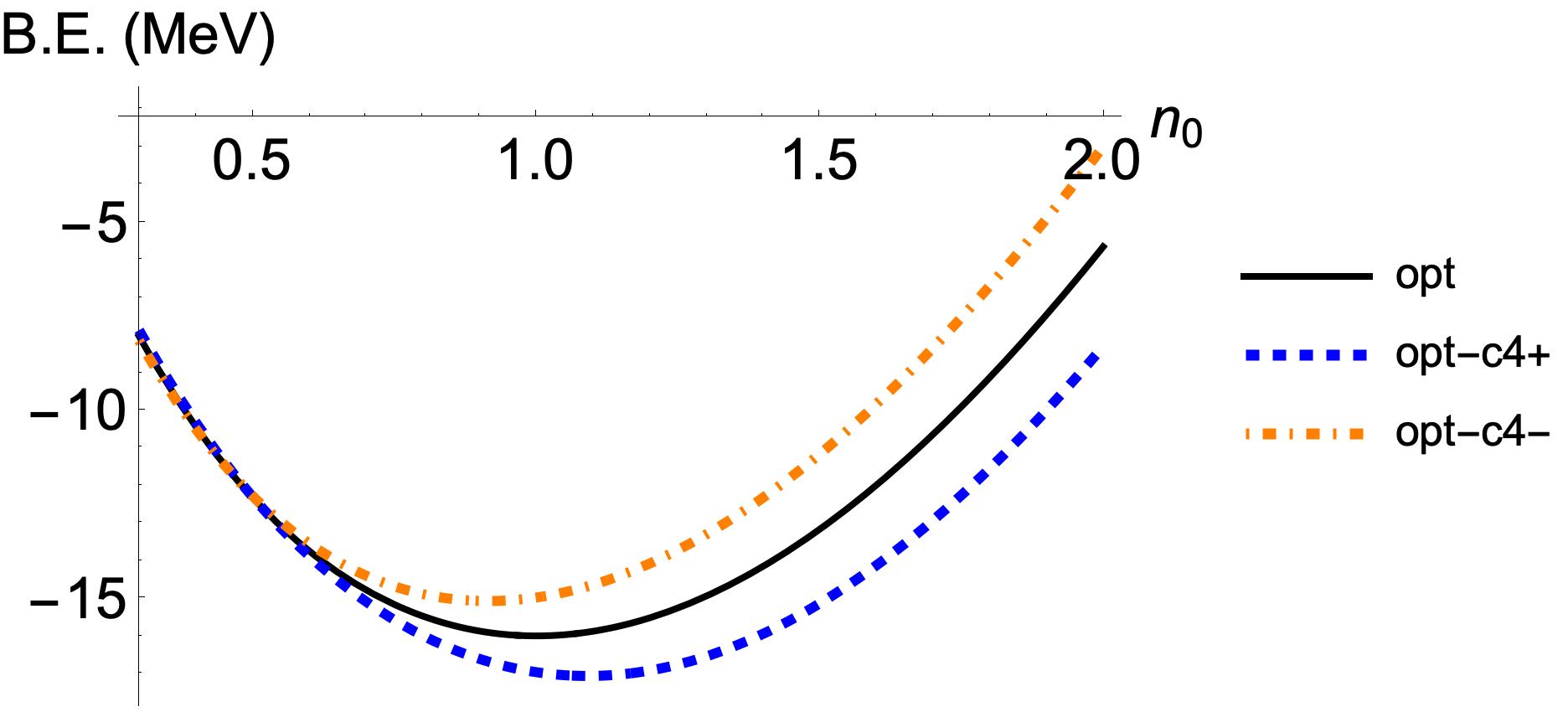}}
	\subfigure[\(e_3\) variarion.]{\includegraphics[scale=0.2]{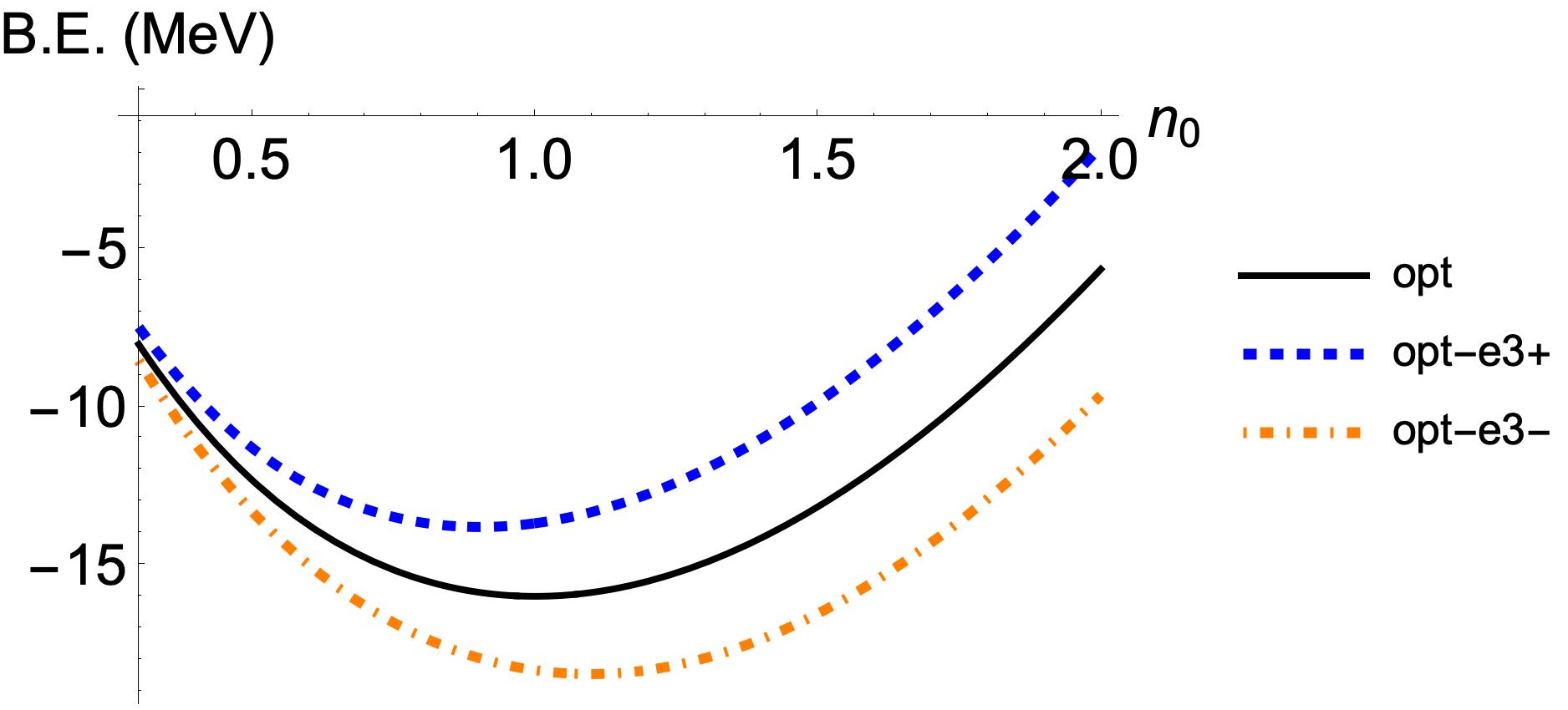}}
	\subfigure[\(c\) variarion.]{\includegraphics[scale=0.2]{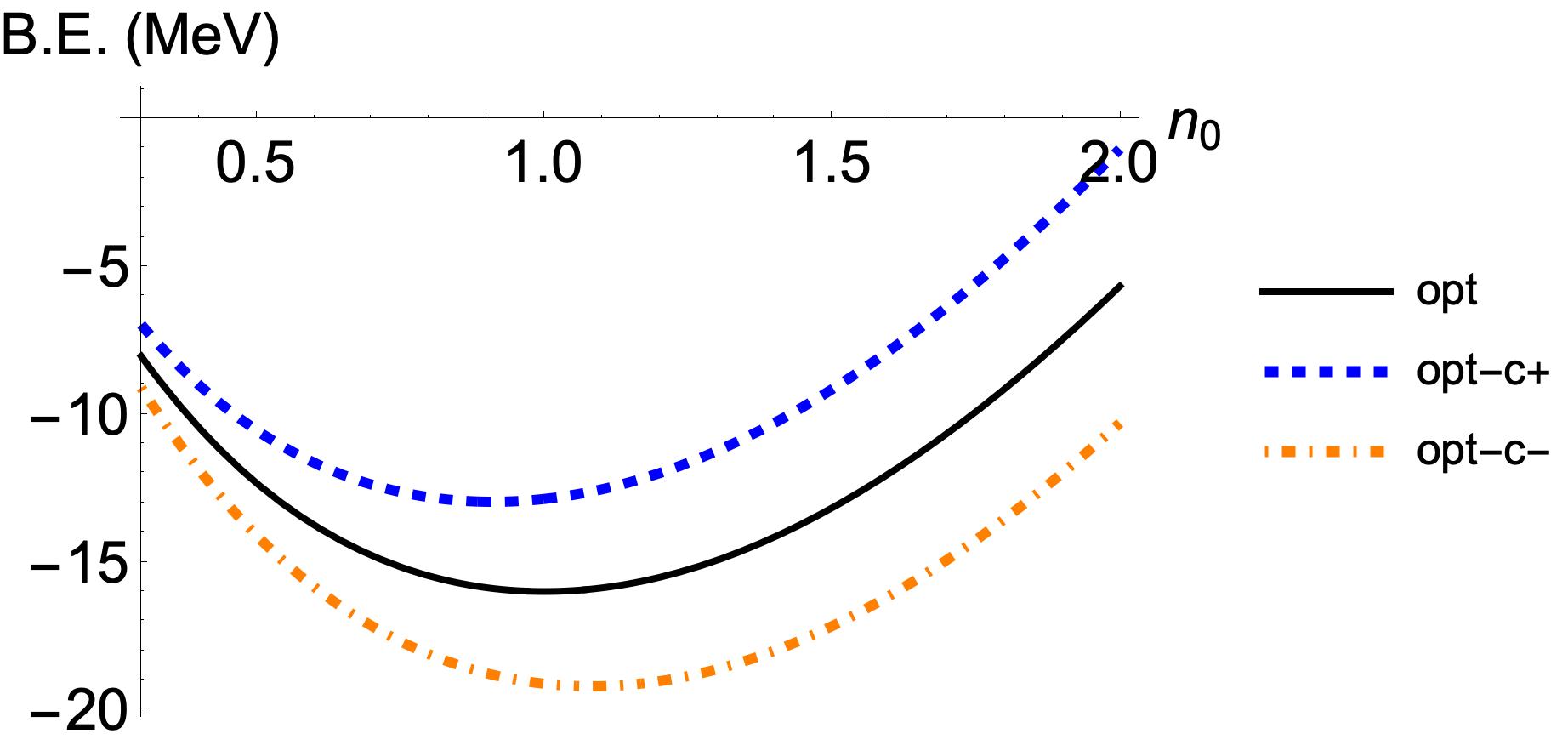}}
	\subfigure[\(g_3\) variarion.]{\includegraphics[scale=0.2]{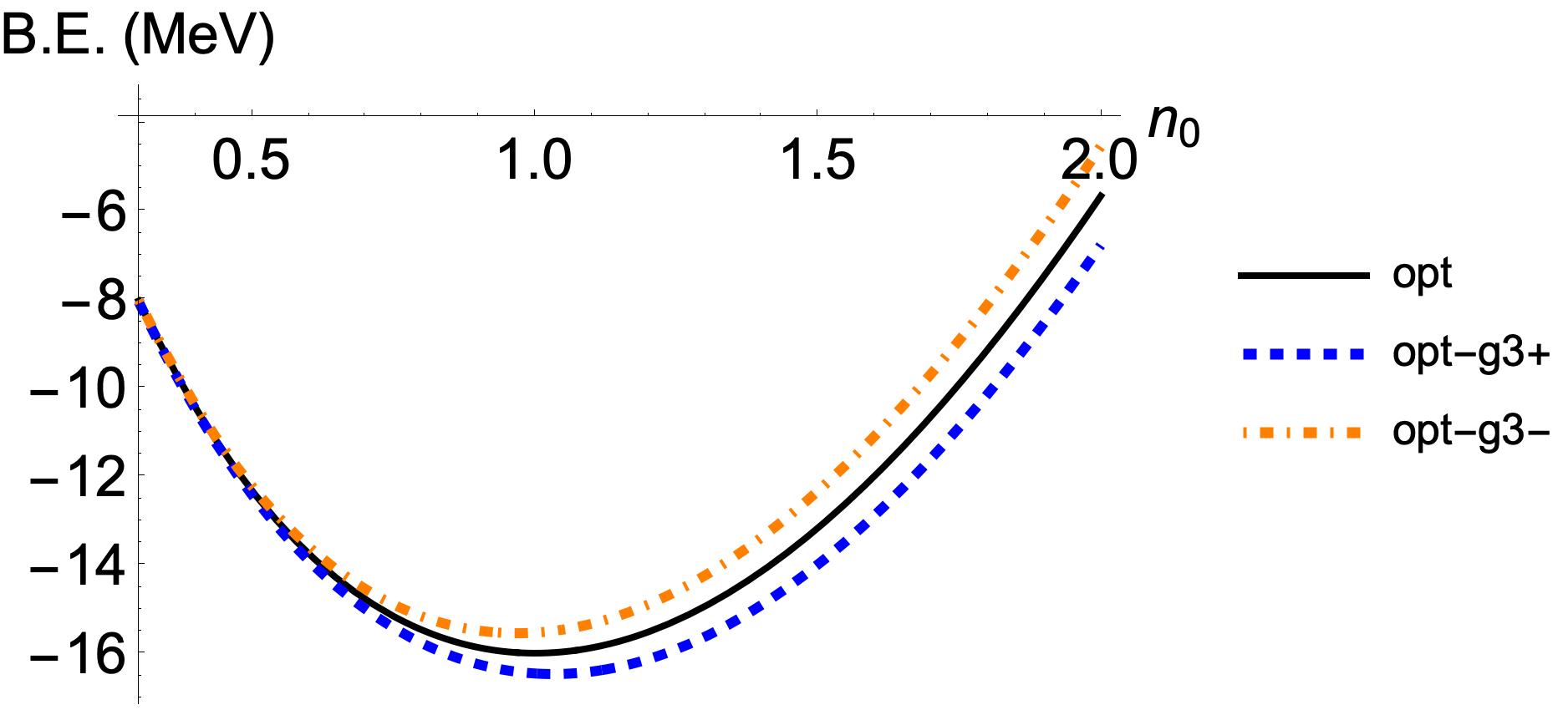}}
	\subfigure[\(h_2\) variarion.]{\includegraphics[scale=0.2]{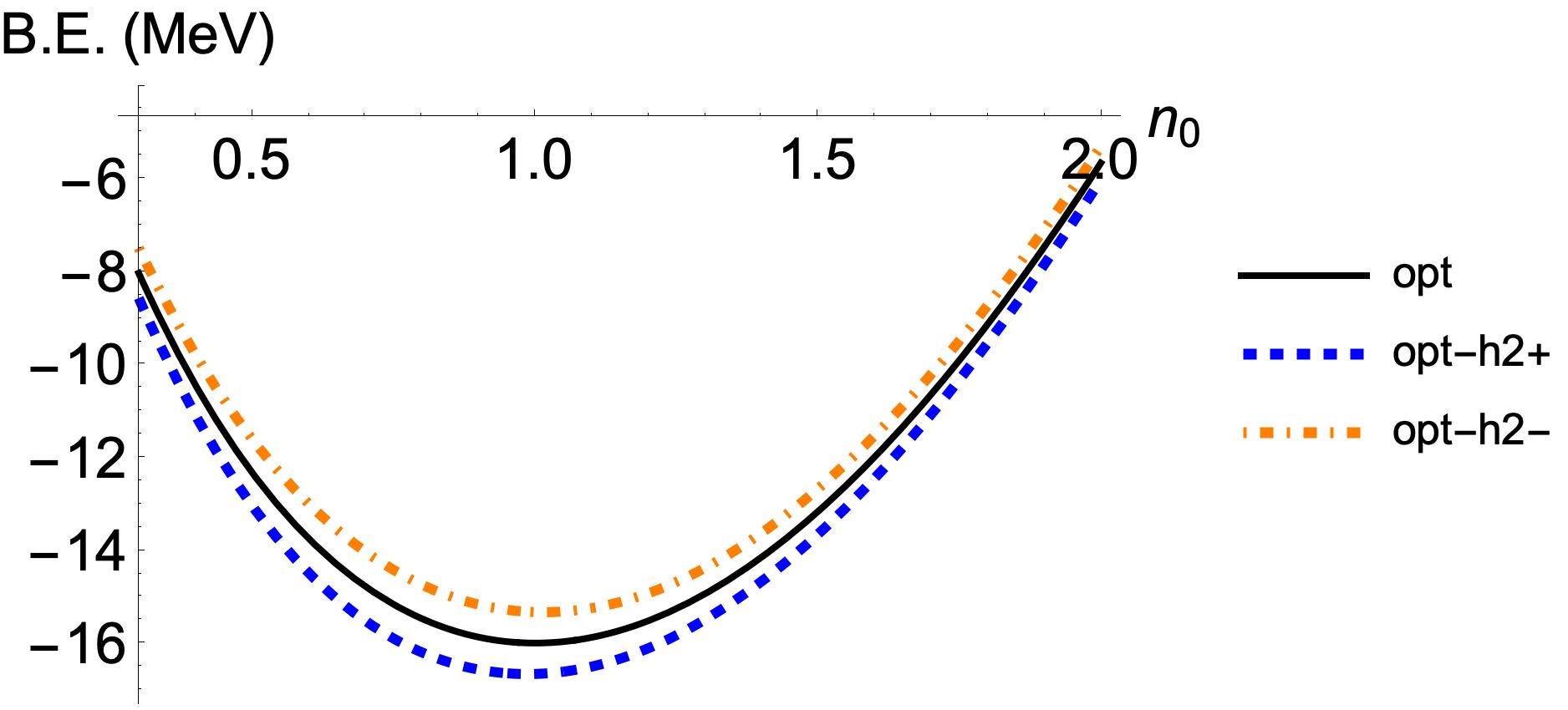}}
	\subfigure[\(\hat{h}_2\) variarion.]{\includegraphics[scale=0.2]{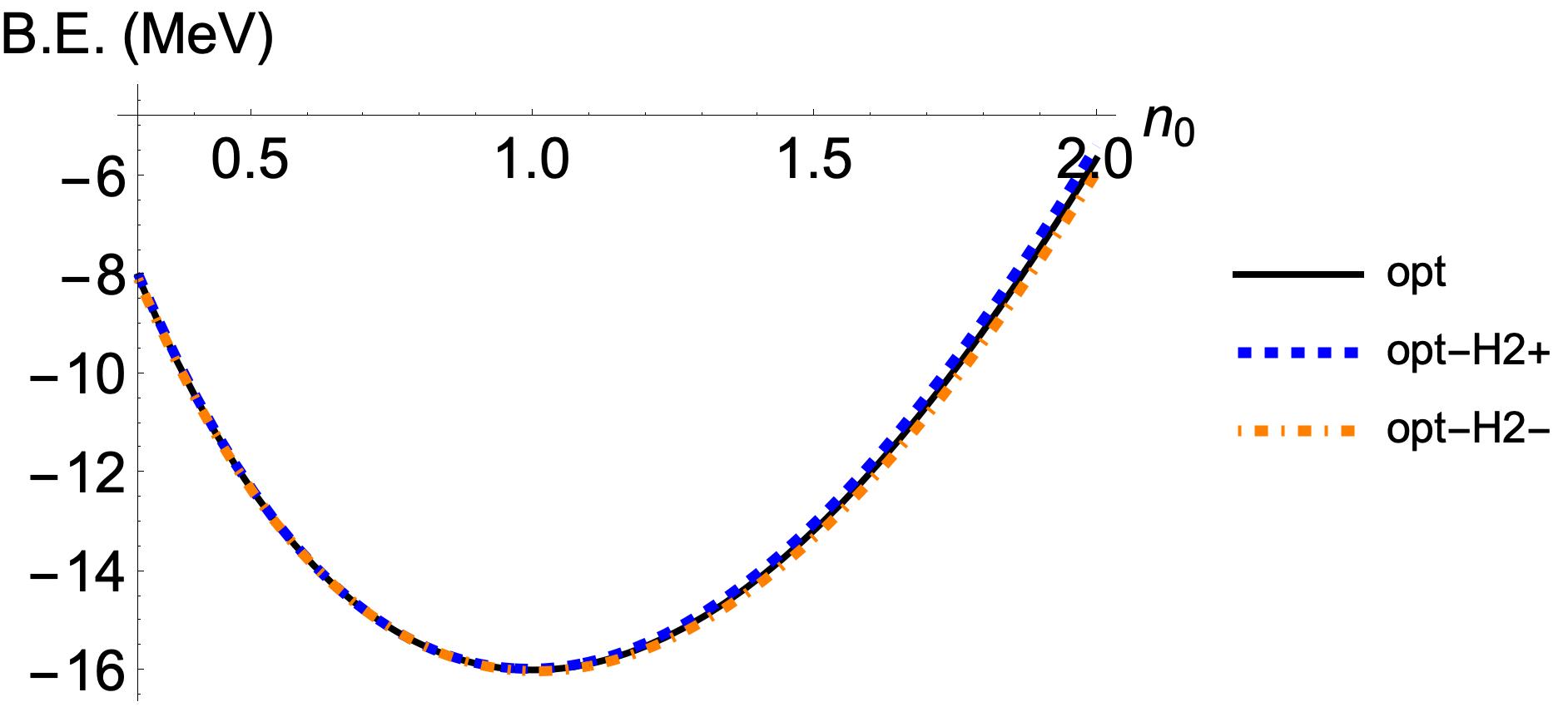}}
	\caption{
		Parameter effects on the binding energy. TM1 is the result from Ref.~\cite{Sugahara:1993wz} where $\rho$, $\omega$ and $\sigma$ mesons are considered with $n_0=0.145\ \rm fm^{-3}$. FSU-$\rm \delta$6.7 is the result from Ref.~\cite{Li:2022okx} where $\delta$ [here $a_0(980)$] meson is added with $n_0=0.148\ \rm fm^{-3}$.
	}
	\label{fig:bevar}
\end{figure*}

\begin{figure*}[htb]
	\centering
	\subfigure[\(\alpha\) variarion.]{\includegraphics[scale=0.2]{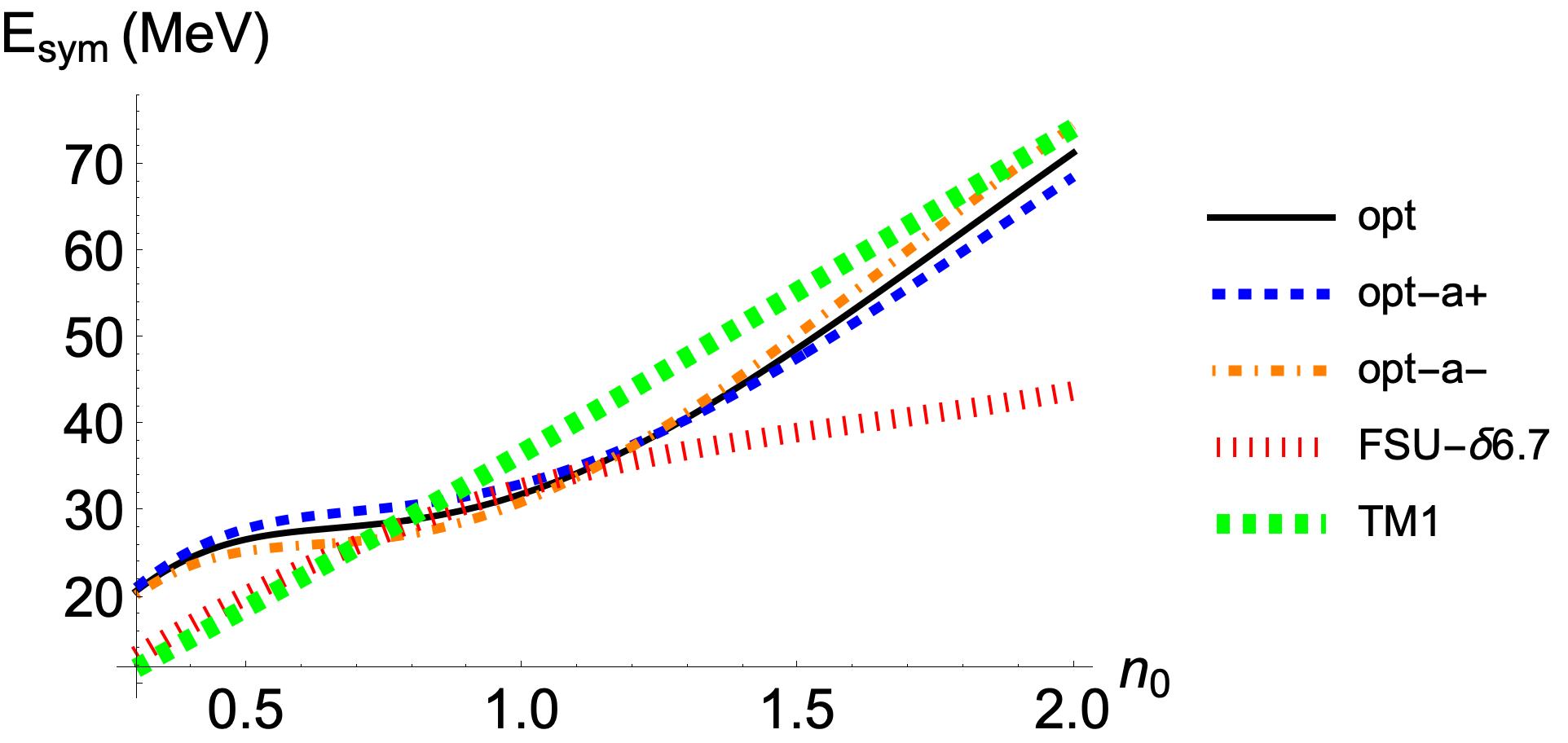}}
	\subfigure[\(\beta\) variarion.]{\includegraphics[scale=0.2]{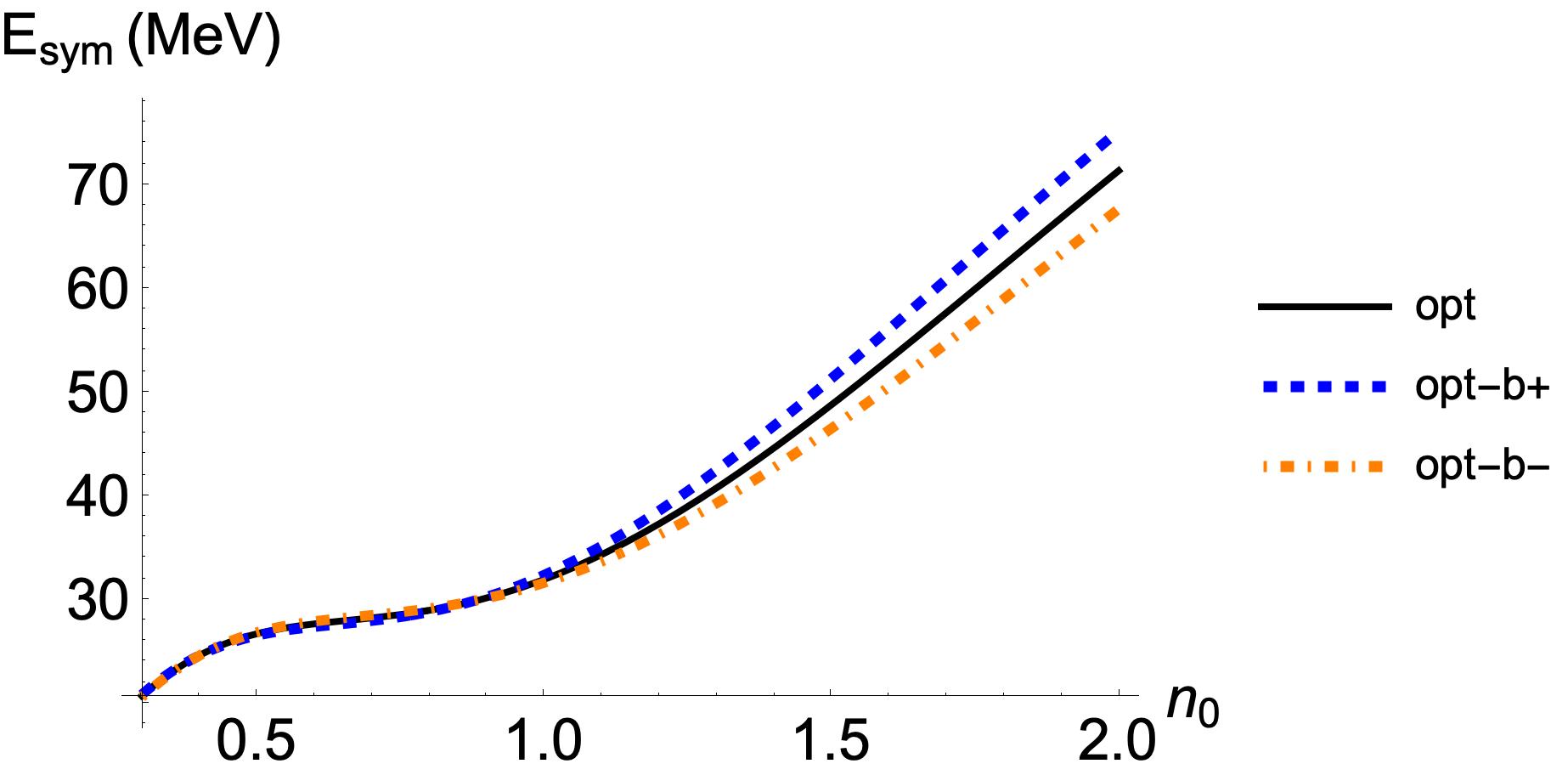}}
	\subfigure[\(c_4\) variarion.]{\includegraphics[scale=0.2]{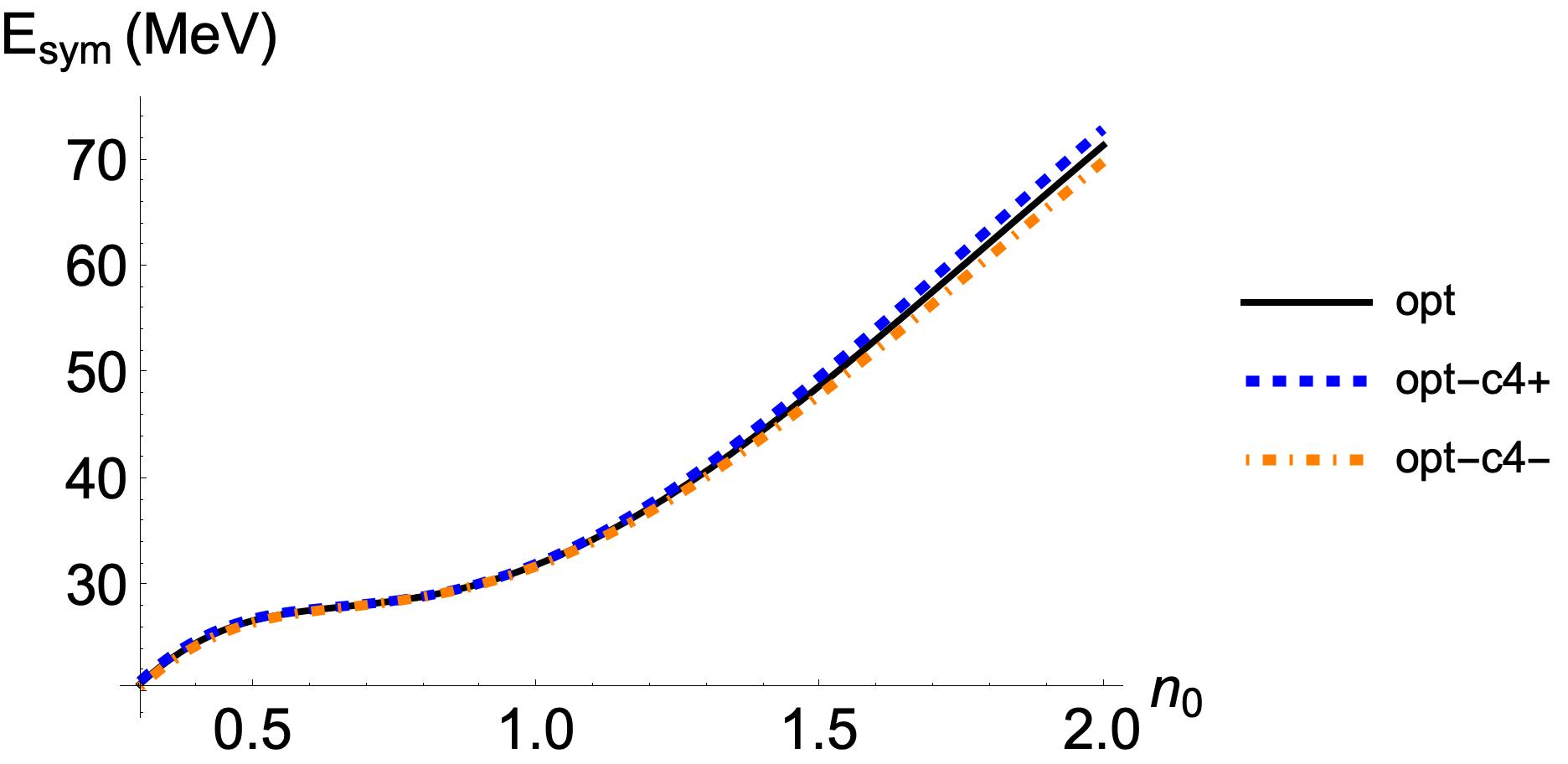}}
	\subfigure[\(e_3\) variarion.]{\includegraphics[scale=0.2]{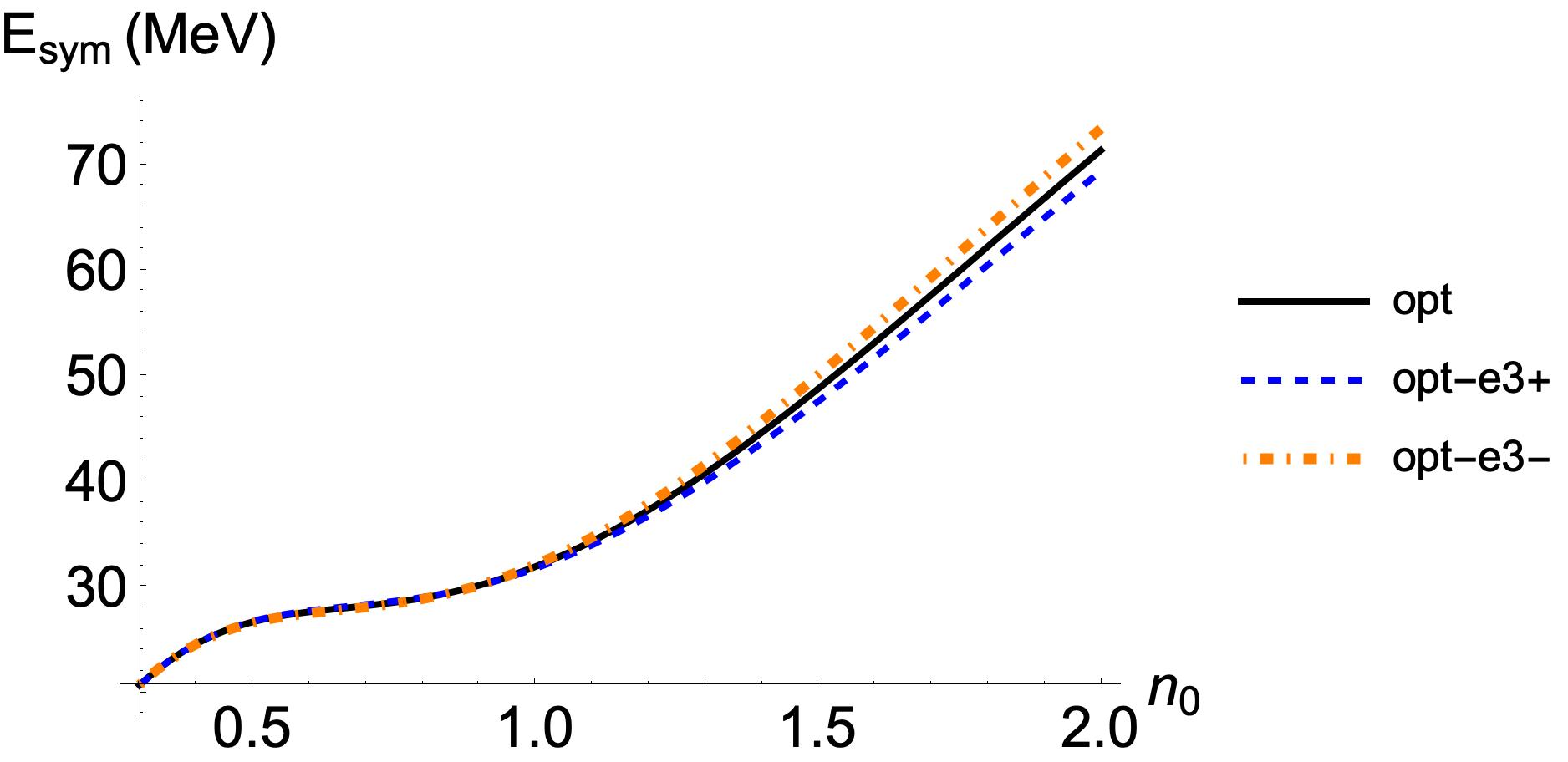}}
	\subfigure[\(c\) variarion.]{\includegraphics[scale=0.2]{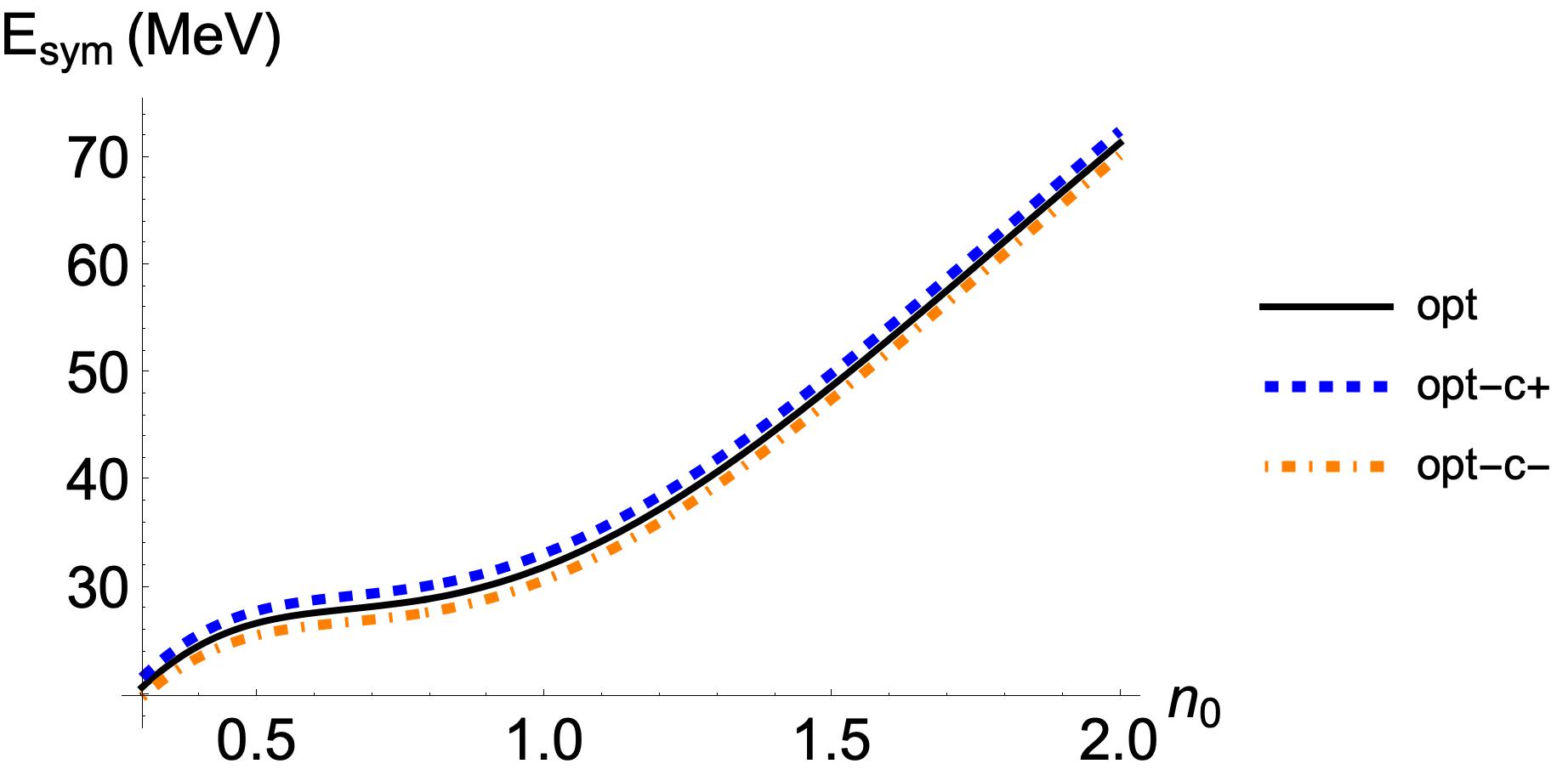}}
	\subfigure[\(g_3\) variarion.]{\includegraphics[scale=0.2]{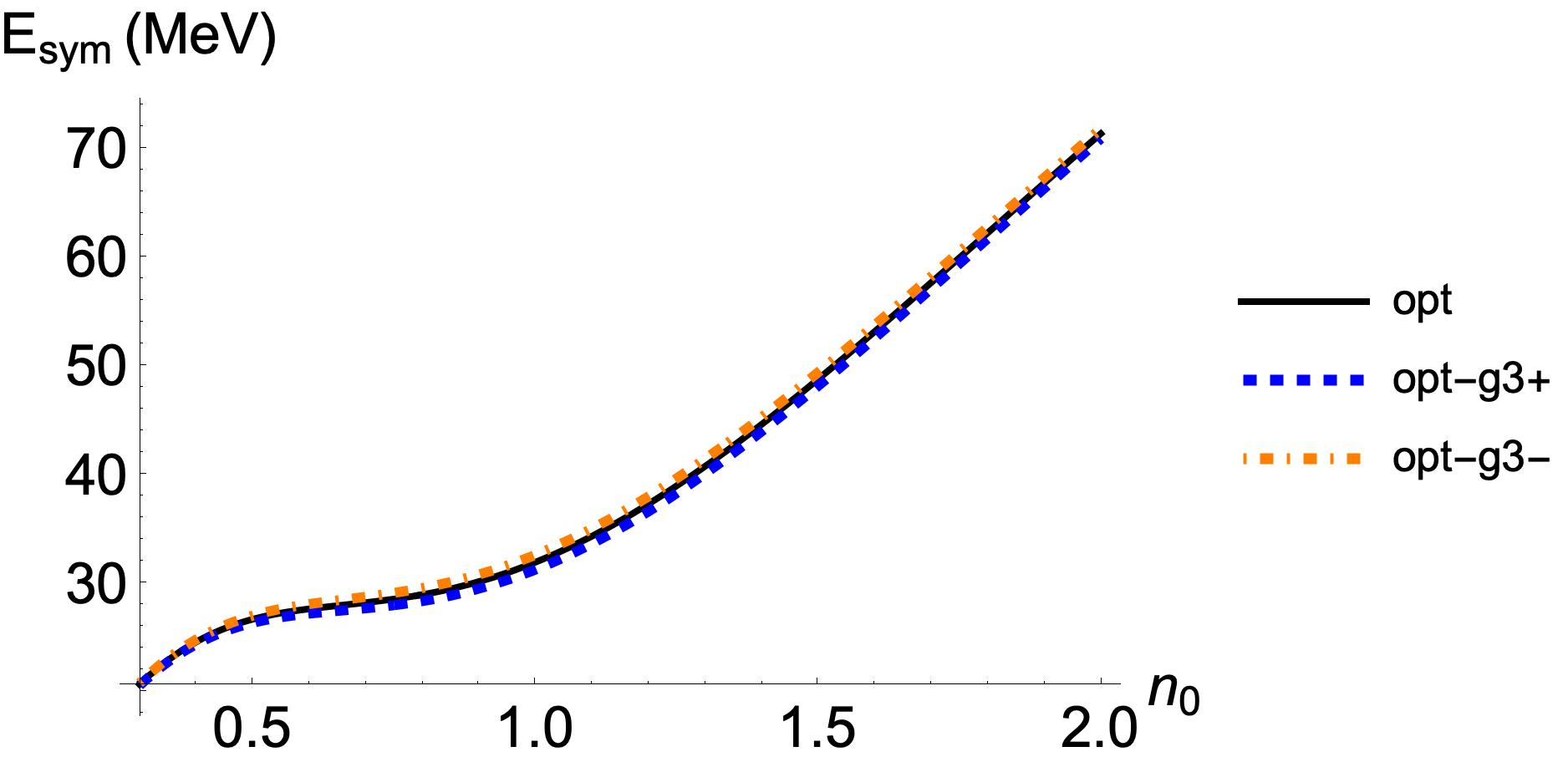}}
	\subfigure[\(h_2\) variarion.]{\includegraphics[scale=0.2]{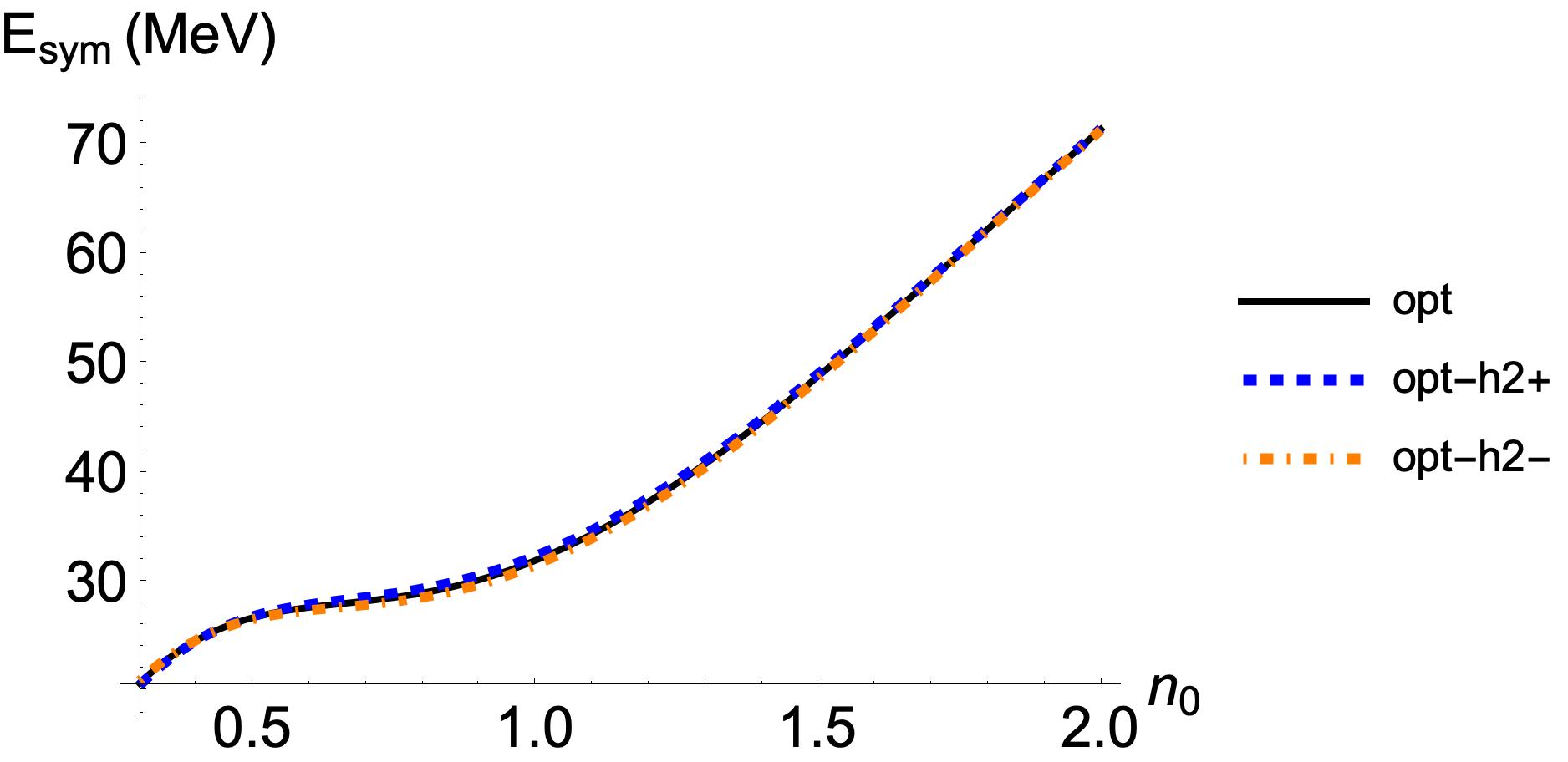}}
	\subfigure[\(\hat{h}_2\) variarion.]{\includegraphics[scale=0.2]{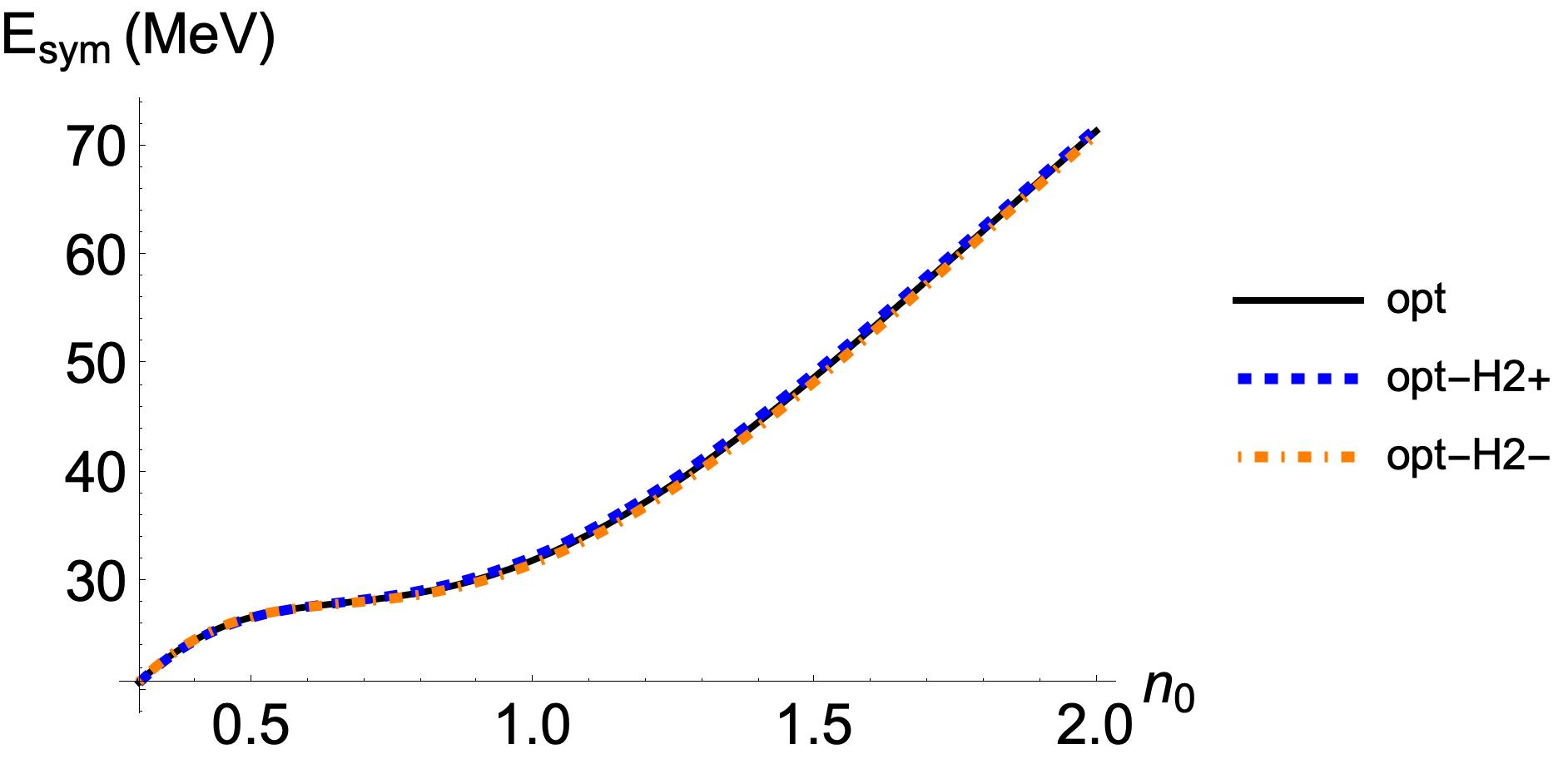}}
	\caption{Parameter effects on  symmetry energy $E_{\rm sym}$.}
	\label{fig:esymvar}
\end{figure*}

\begin{figure*}[htb]
	\centering
	\subfigure[\(\alpha\) variarion.]{\includegraphics[scale=0.2]{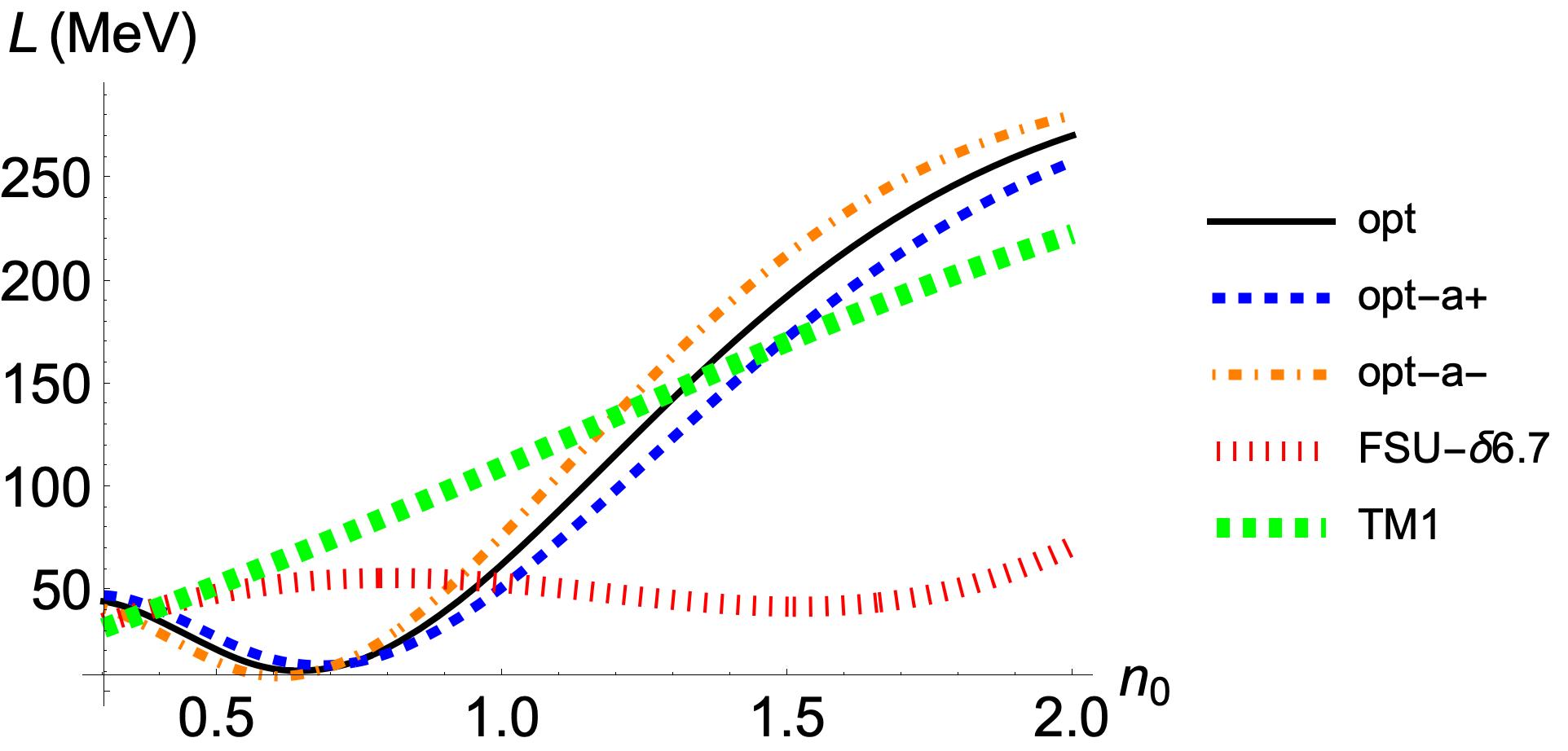}}
	\subfigure[\(\beta\) variarion.]{\includegraphics[scale=0.2]{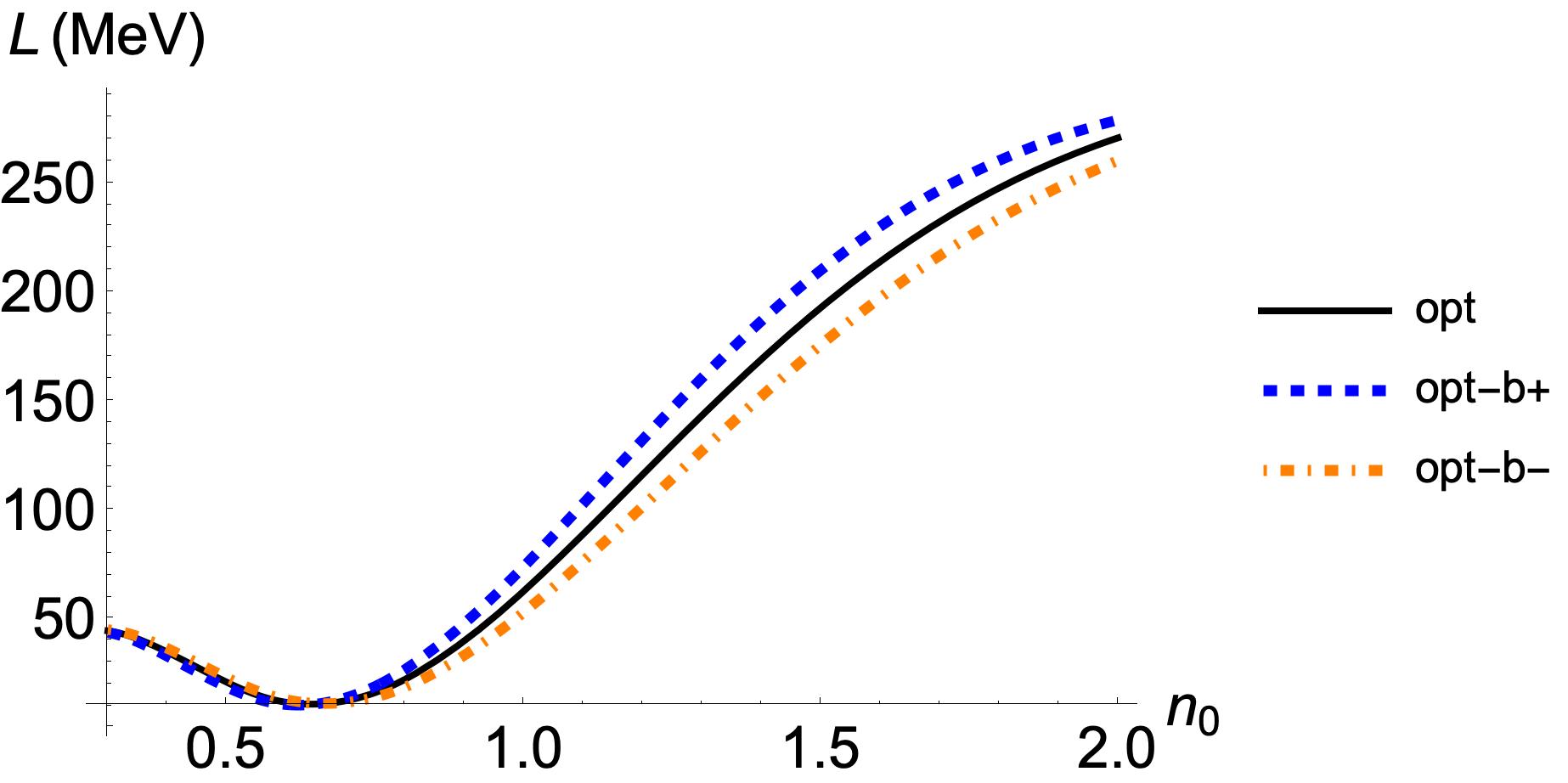}}
	\subfigure[\(c_4\) variarion.]{\includegraphics[scale=0.2]{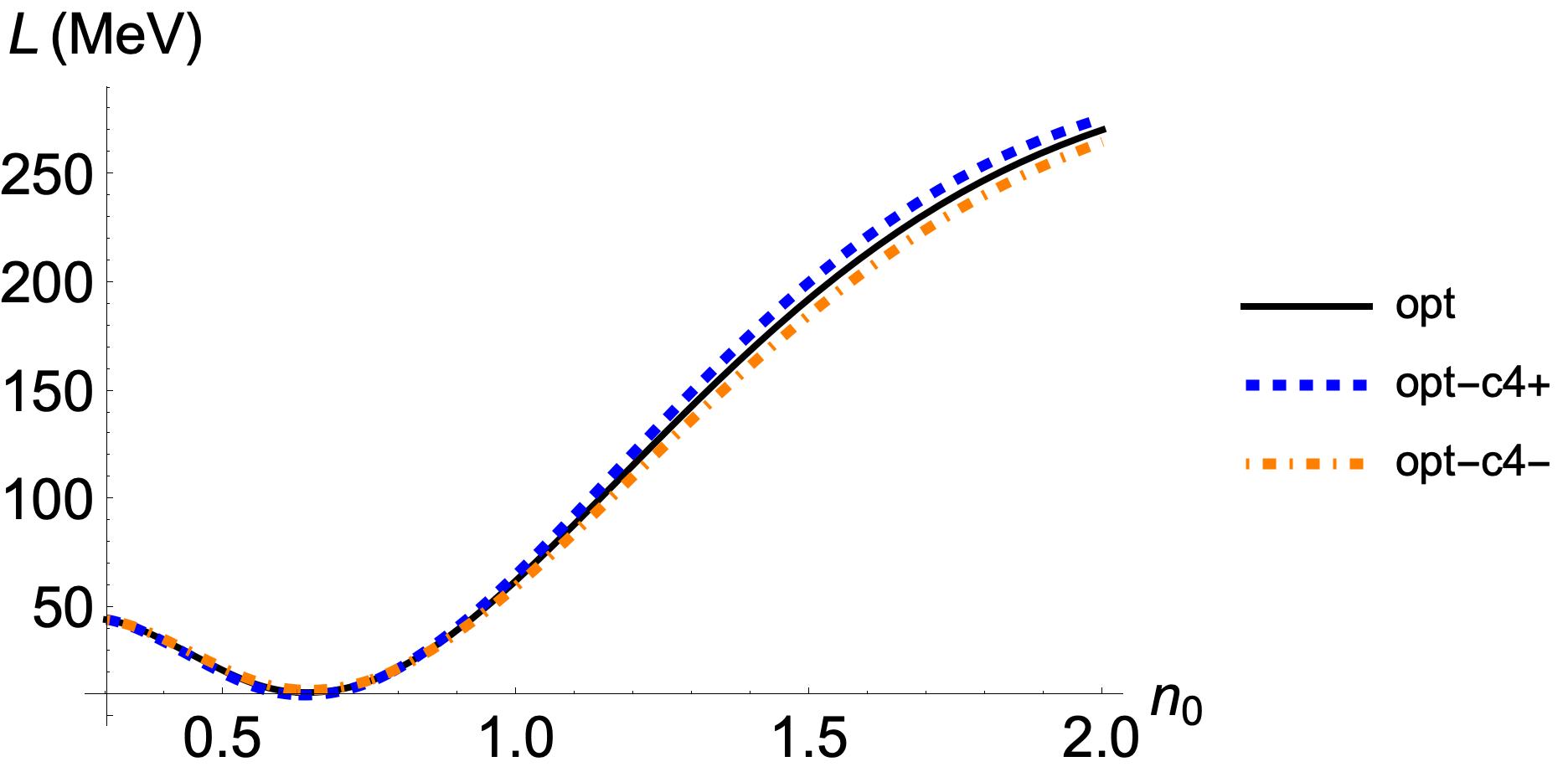}}
	\subfigure[\(e_3\) variarion.]{\includegraphics[scale=0.2]{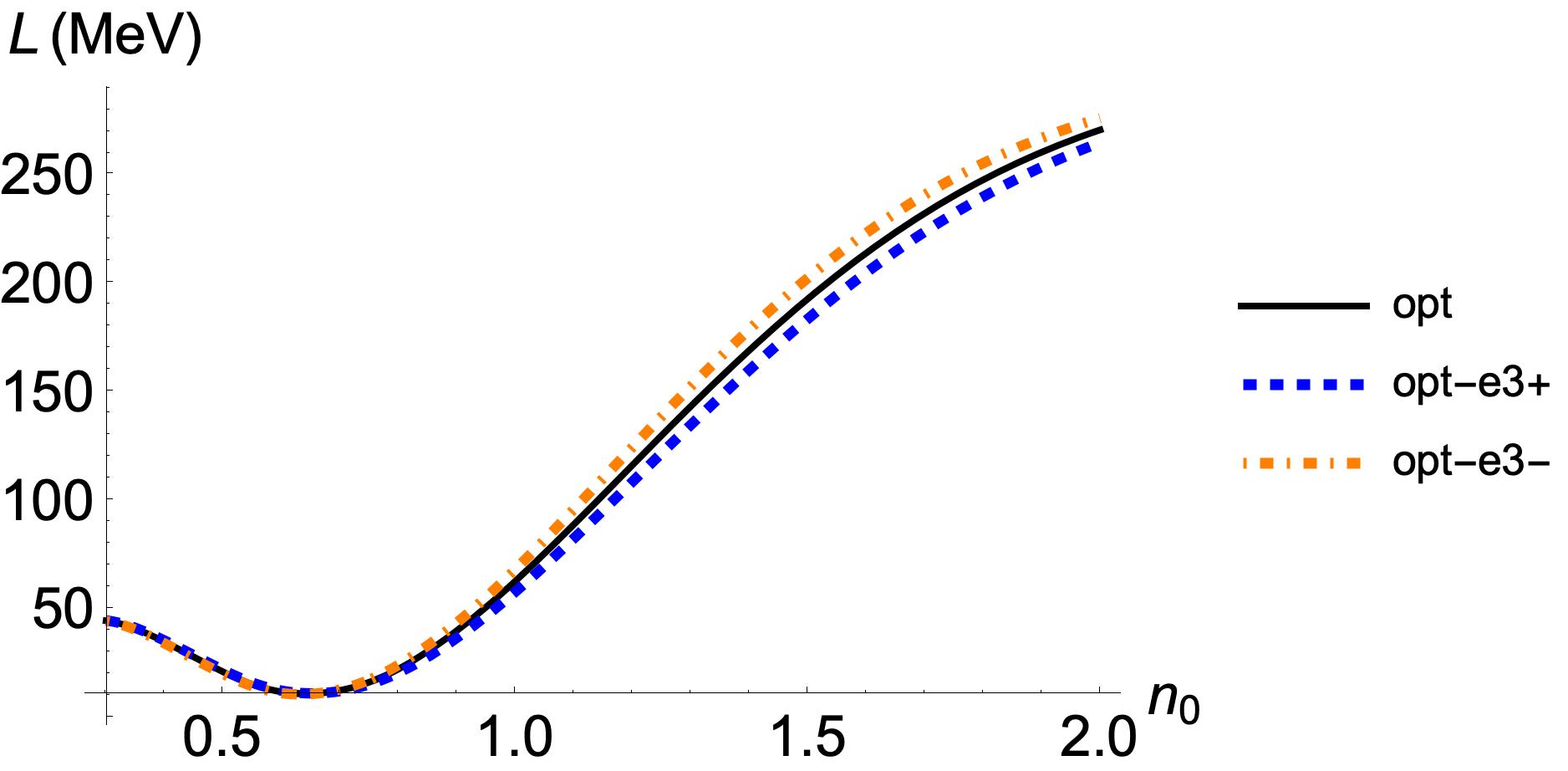}}
	\subfigure[\(c\) variarion.]{\includegraphics[scale=0.2]{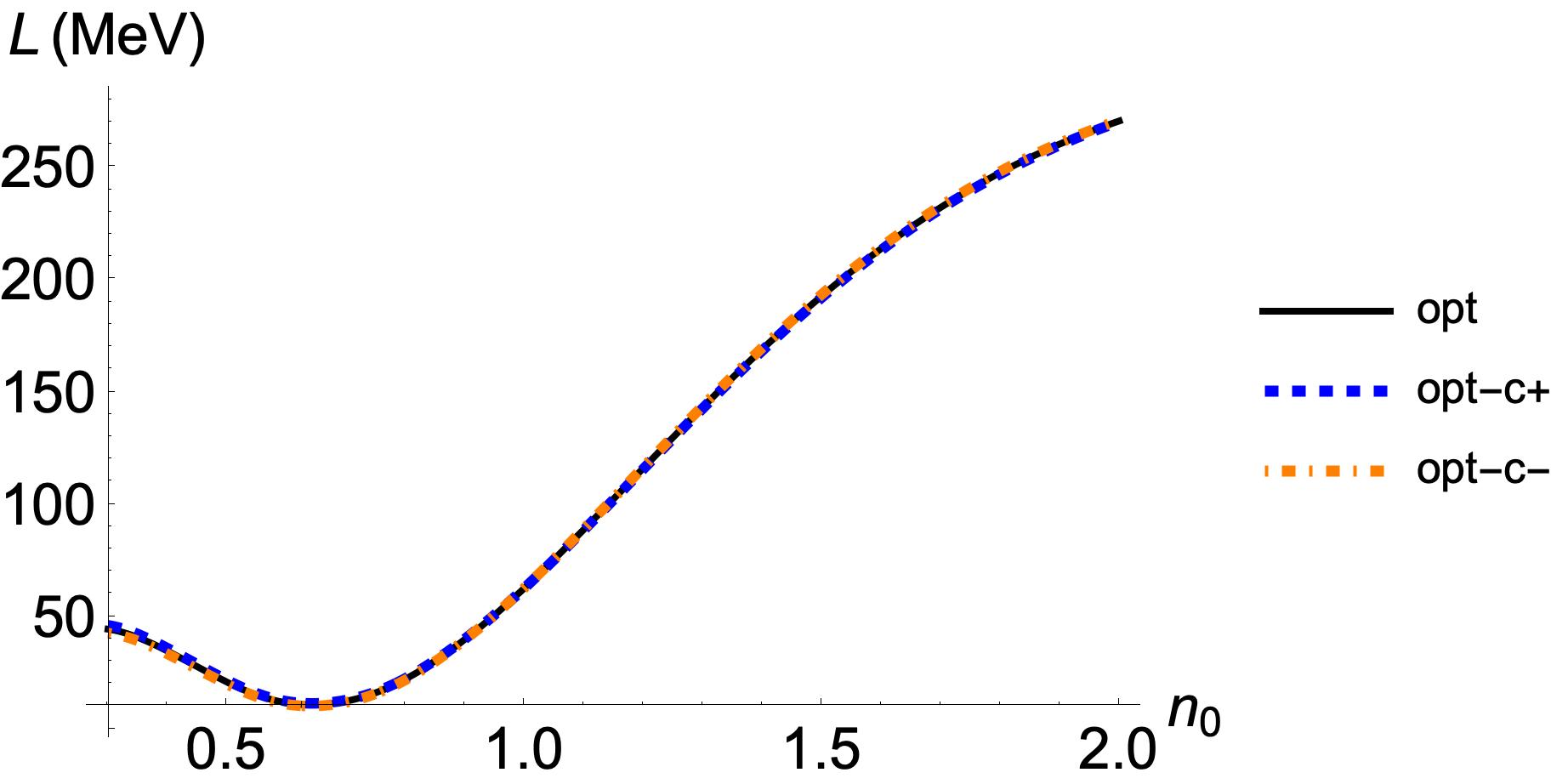}}
	\subfigure[\(g_3\) variarion.]{\includegraphics[scale=0.2]{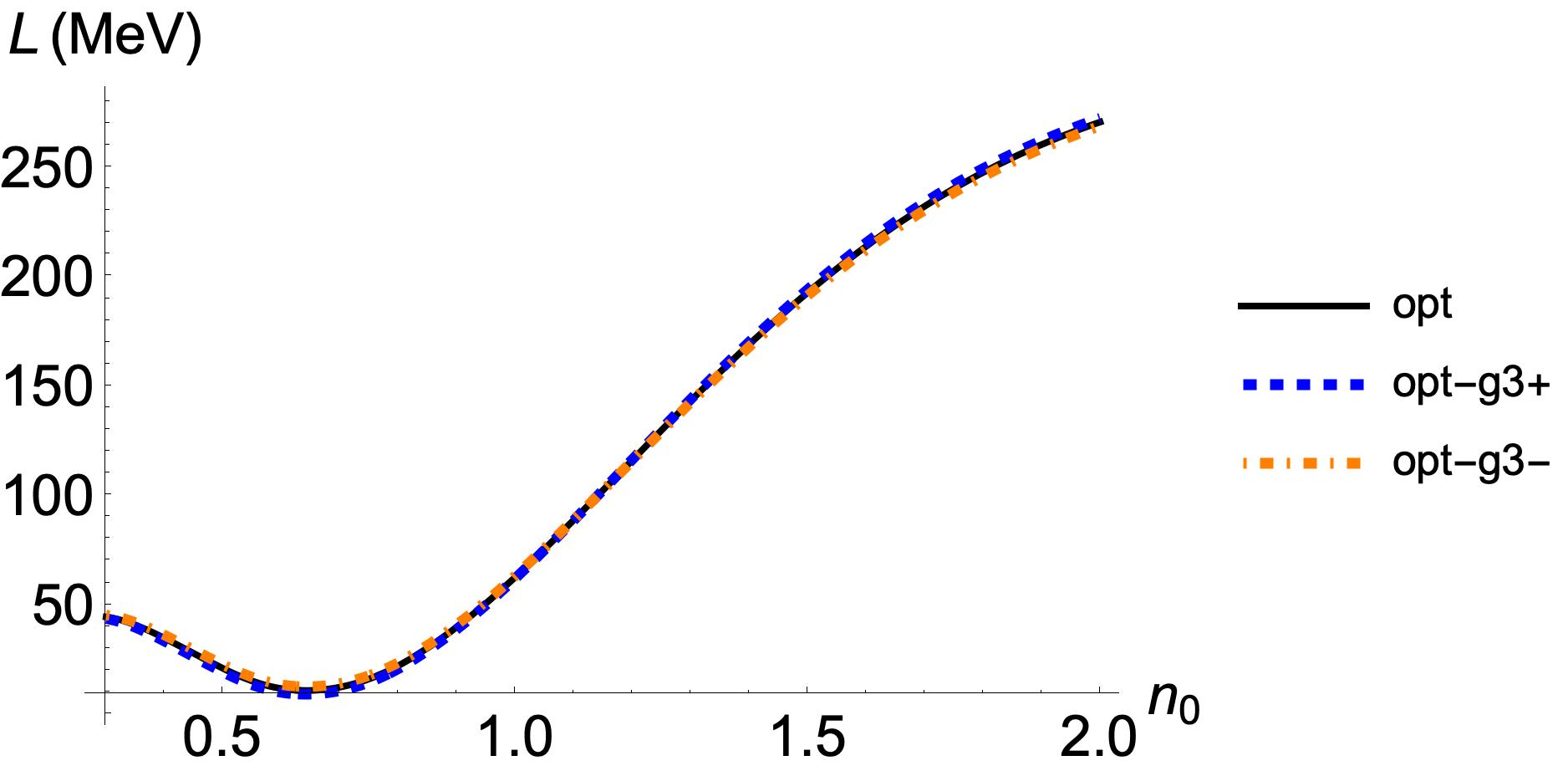}}
	\subfigure[\(h_2\) variarion.]{\includegraphics[scale=0.2]{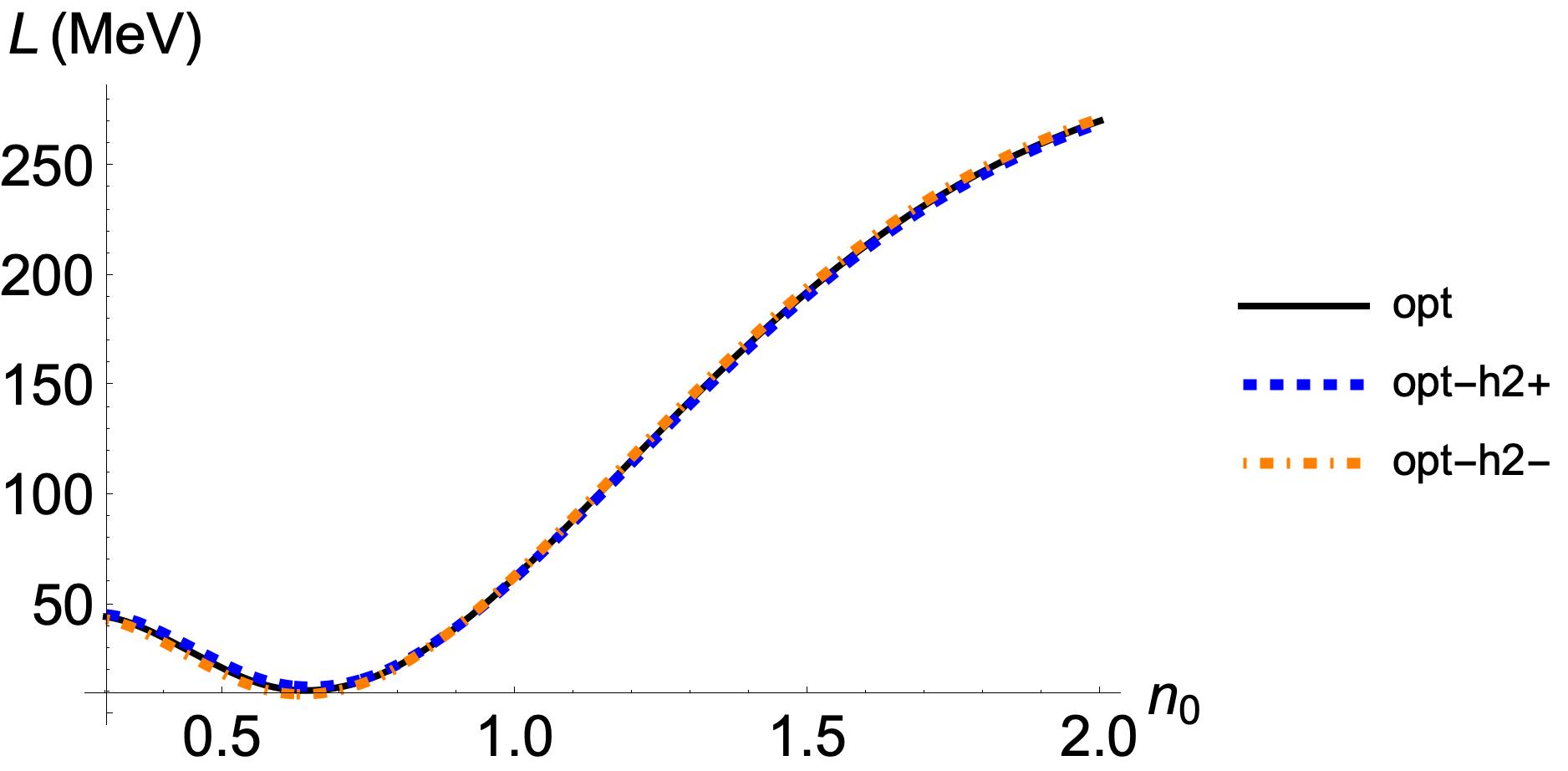}}
	\subfigure[\(\hat{h}_2\) variarion.]{\includegraphics[scale=0.2]{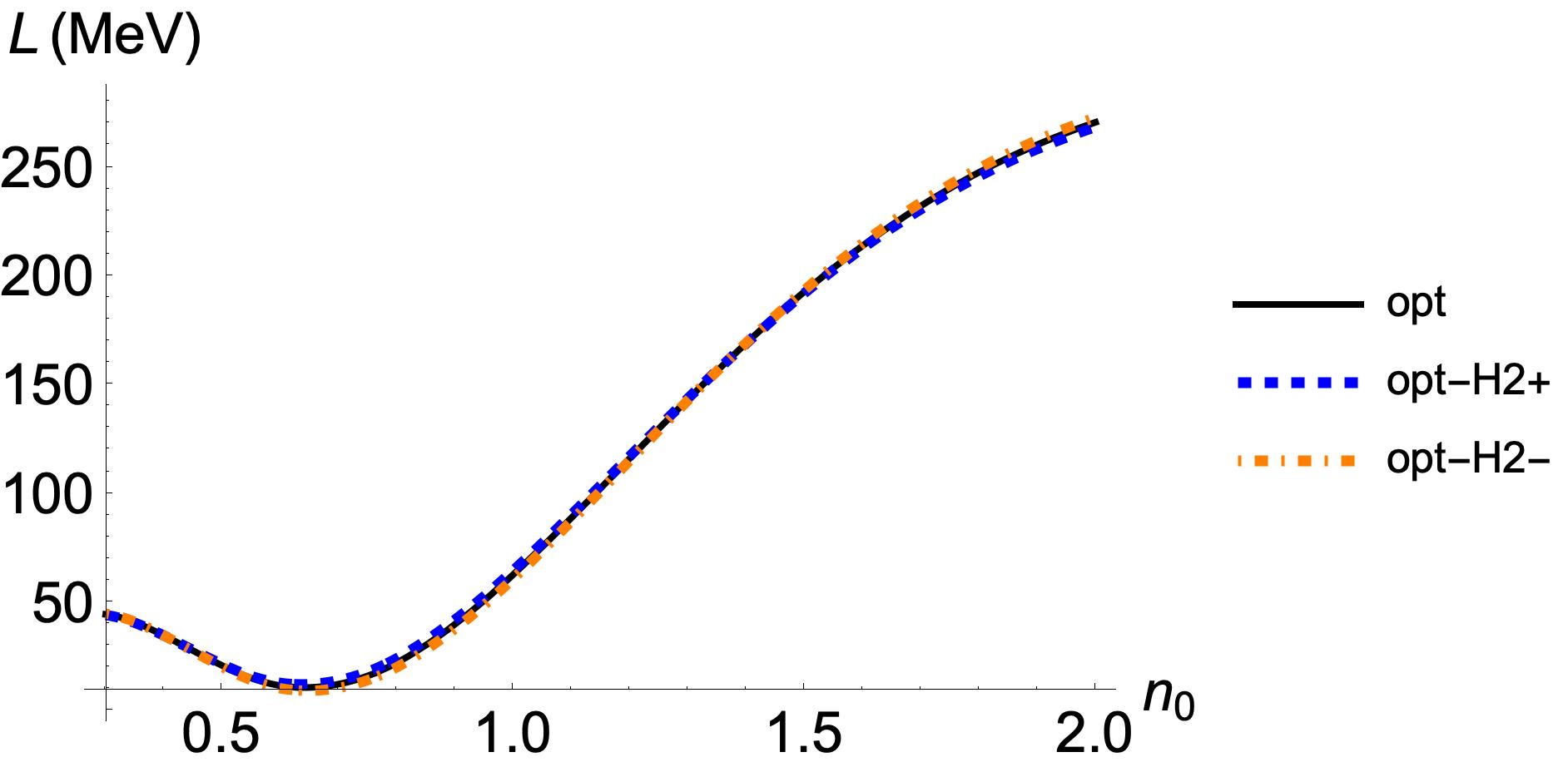}}
	\caption{Parameter effects on the slope of the symmetry energy $L$.}
	\label{fig:lvar}
\end{figure*}

\begin{figure*}[htb]
	\centering
	\subfigure[\(\alpha\) variarion.]{\includegraphics[scale=0.2]{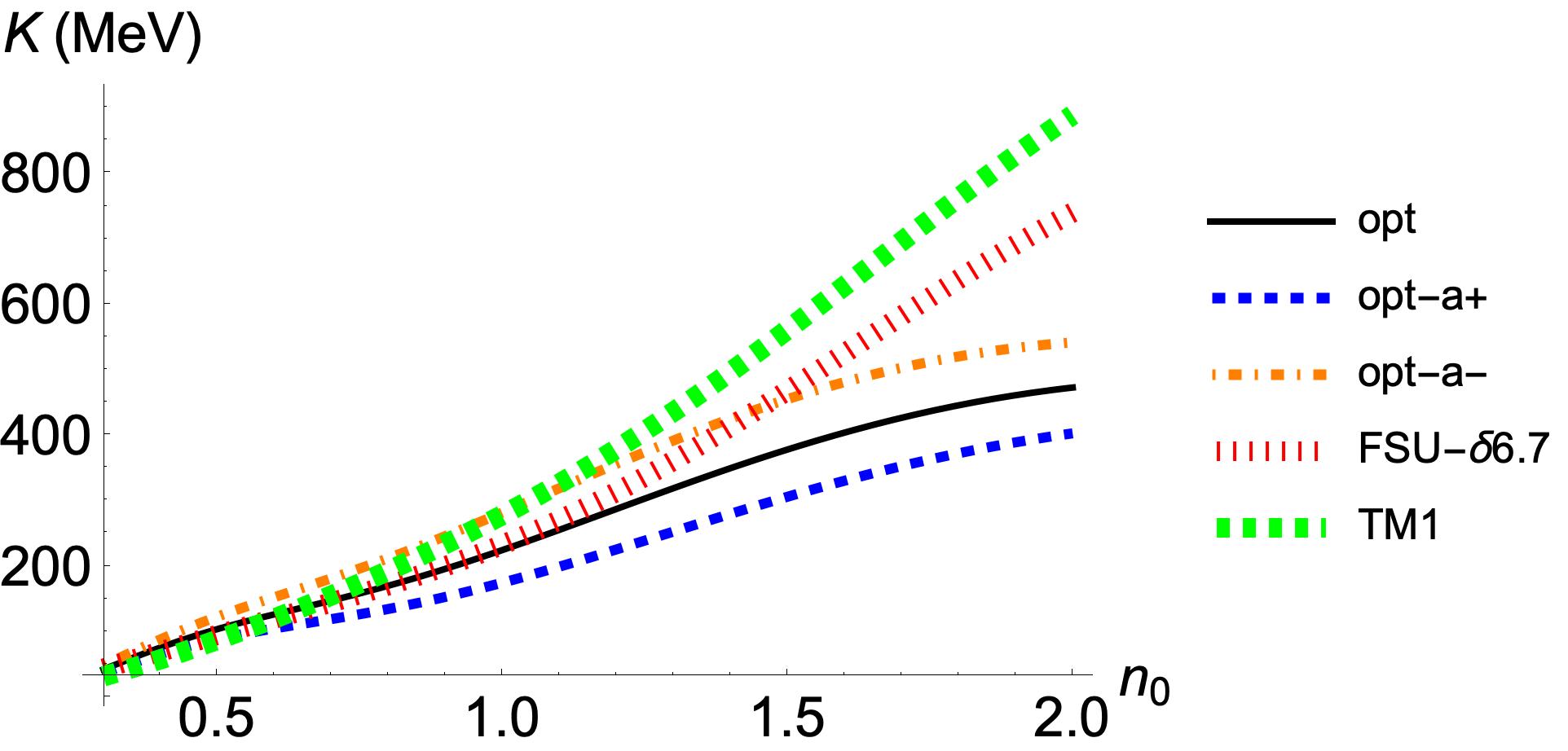}}
	\subfigure[\(\beta\) variarion.]{\includegraphics[scale=0.2]{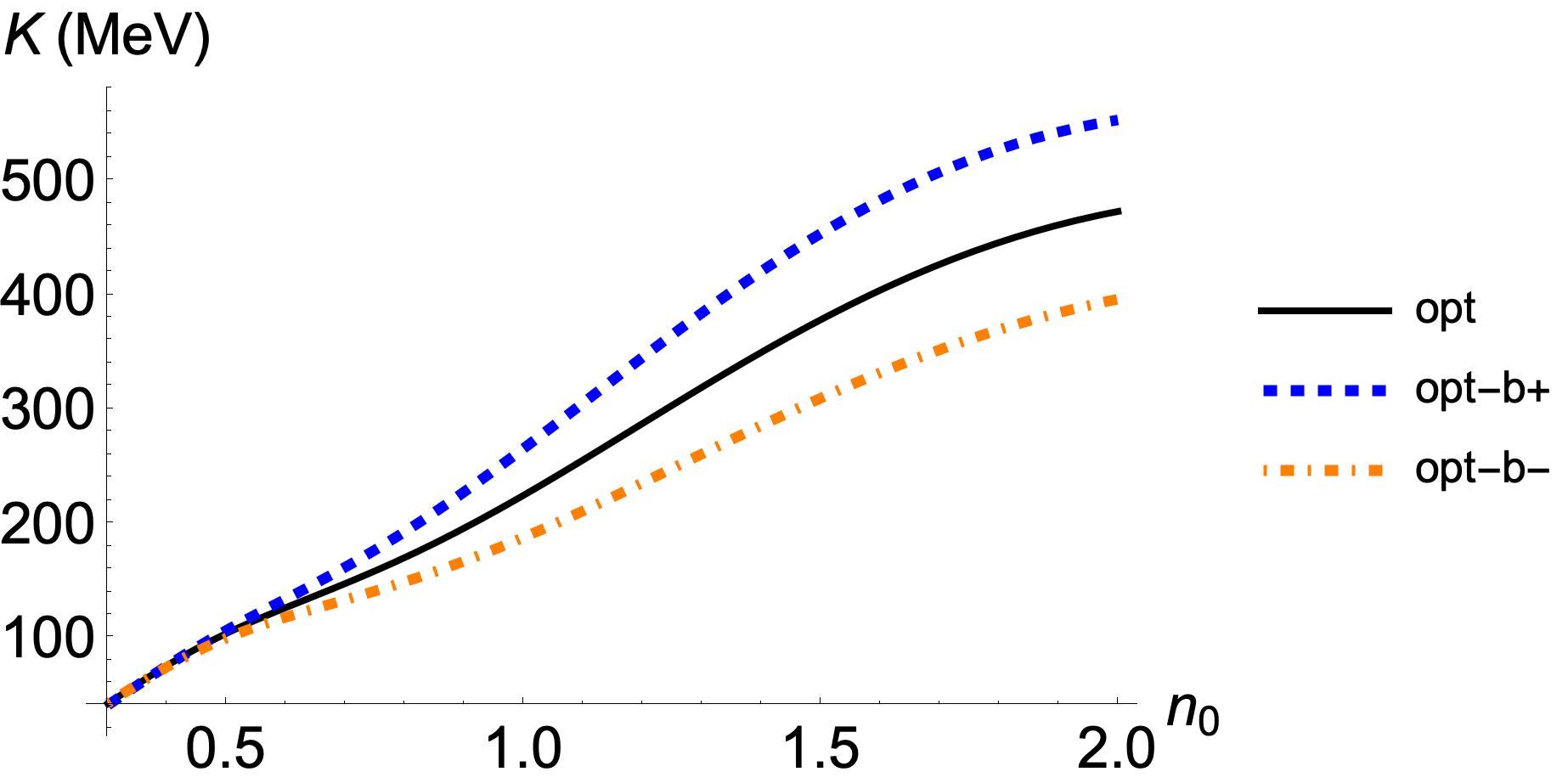}}
	\subfigure[\(c_4\) variarion.]{\includegraphics[scale=0.2]{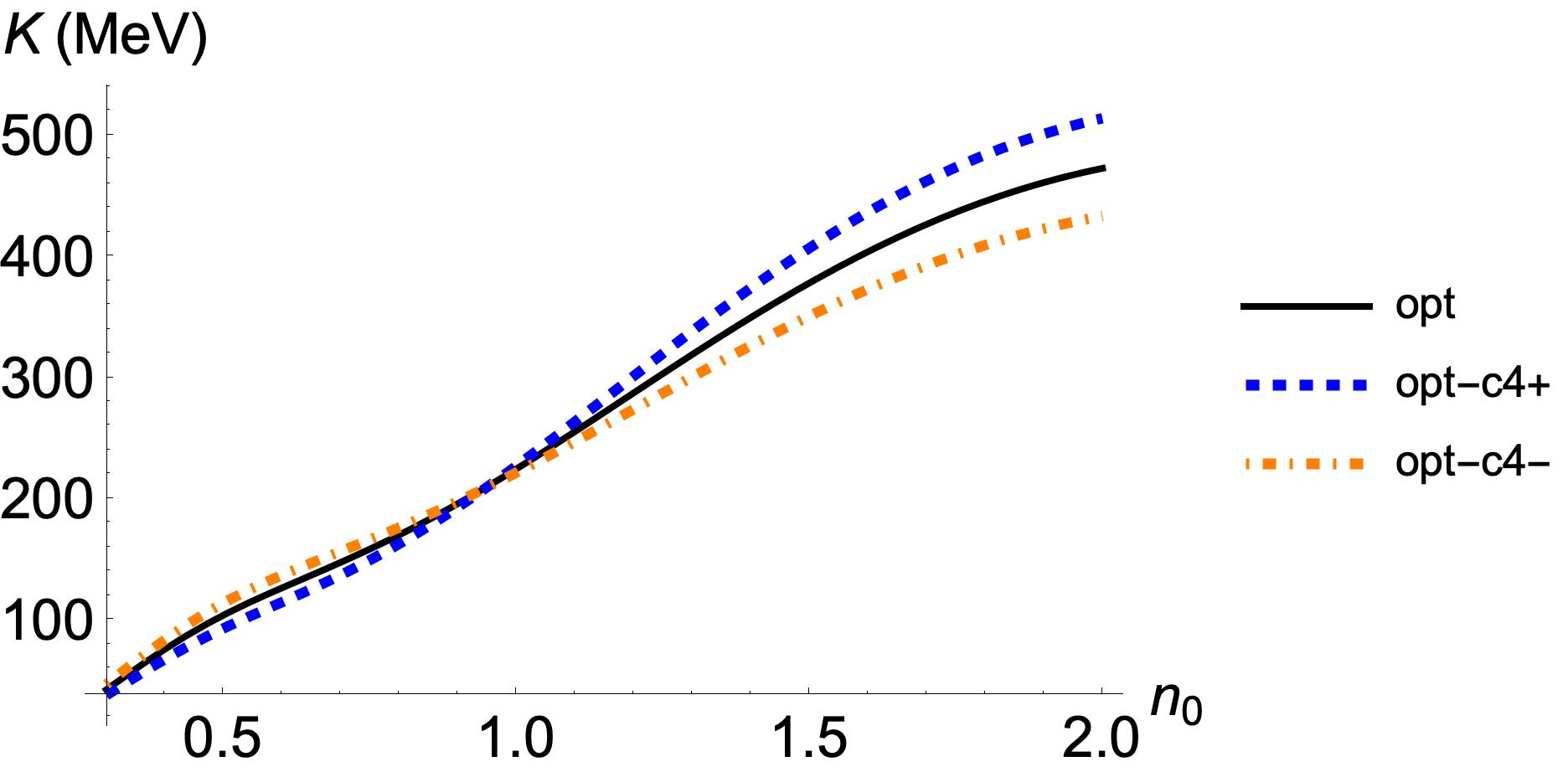}}
	\subfigure[\(e_3\) variarion.]{\includegraphics[scale=0.2]{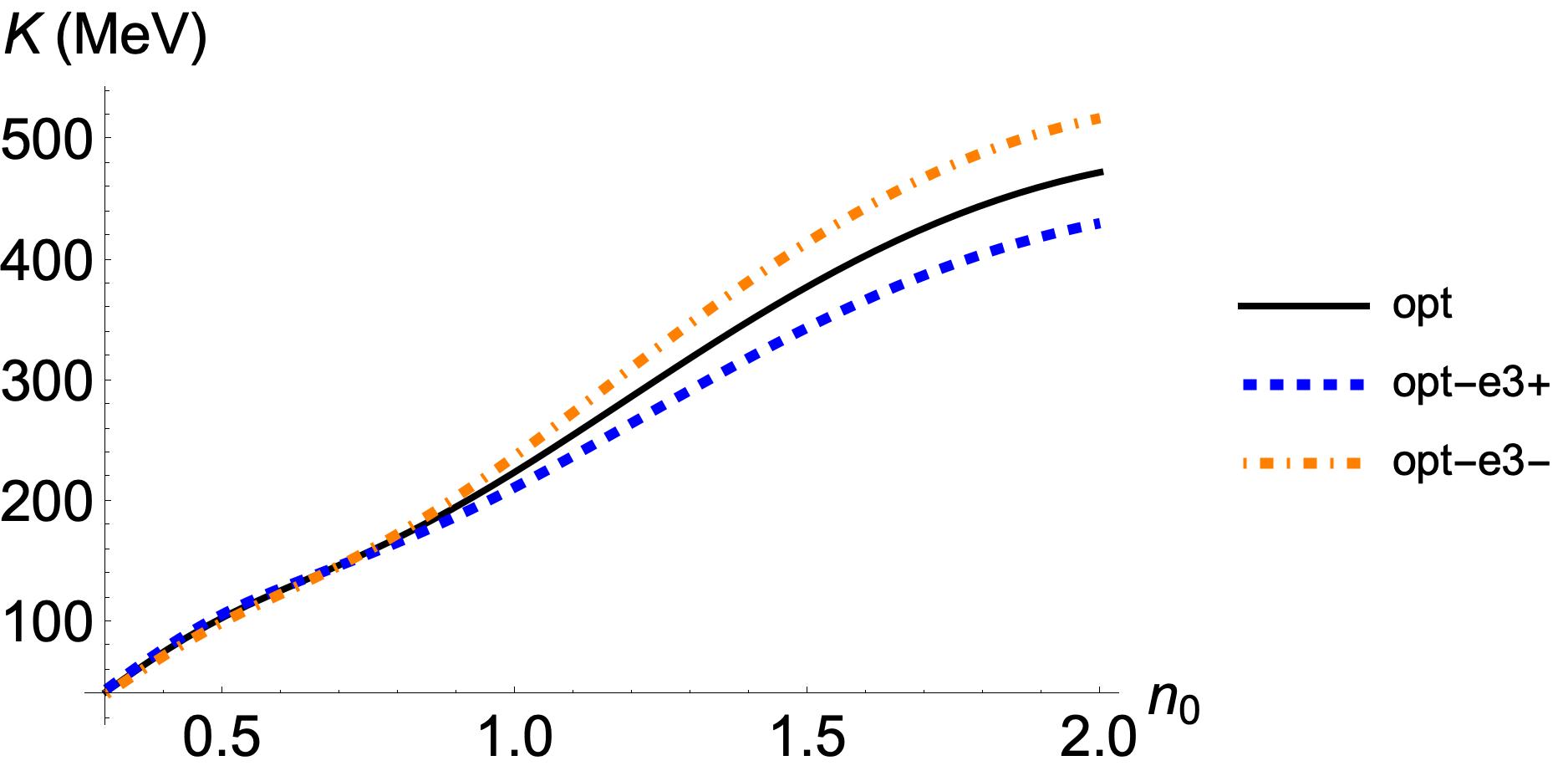}}
	\subfigure[\(c\) variarion.]{\includegraphics[scale=0.2]{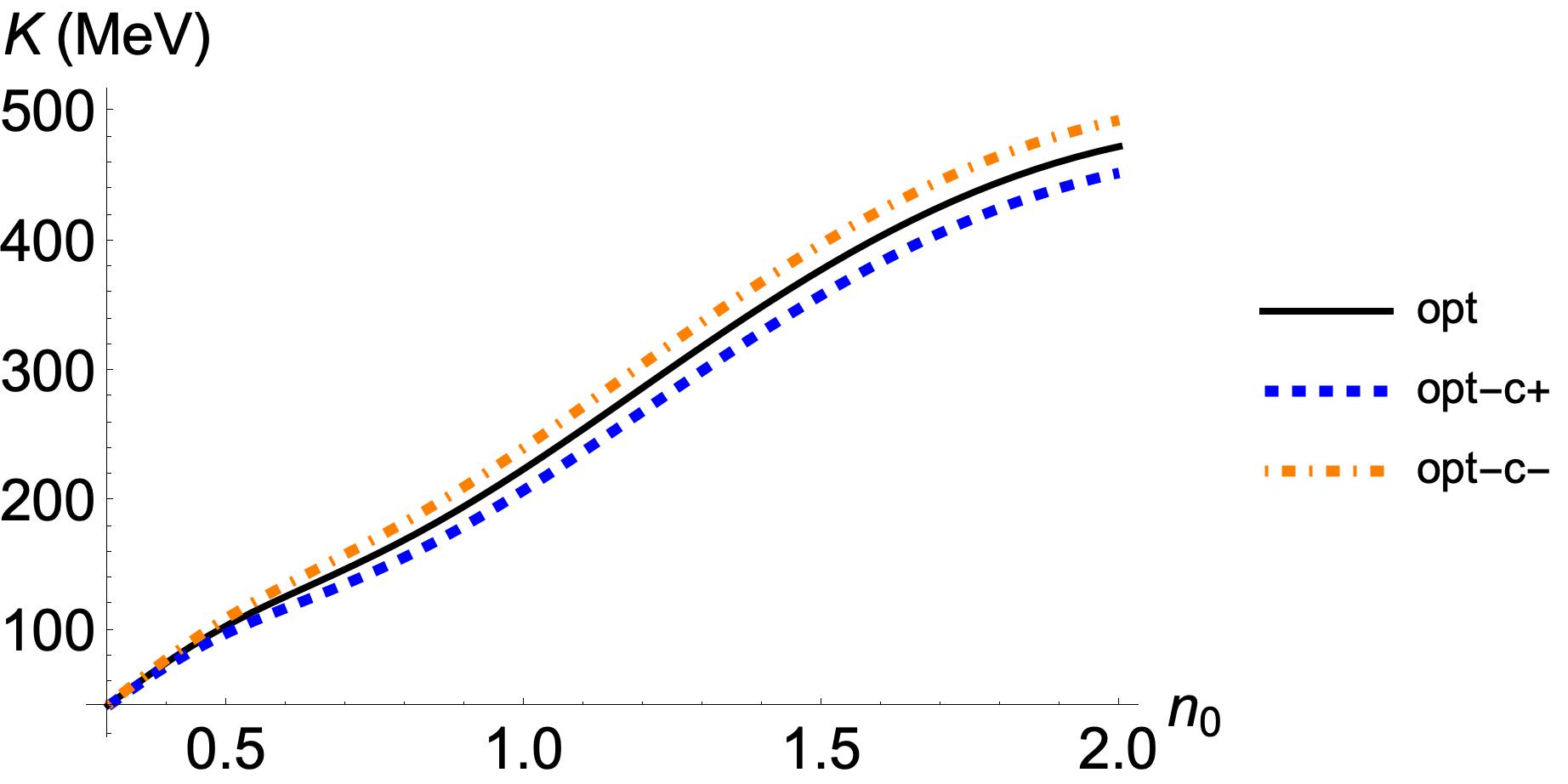}}
	\subfigure[\(g_3\) variarion.]{\includegraphics[scale=0.2]{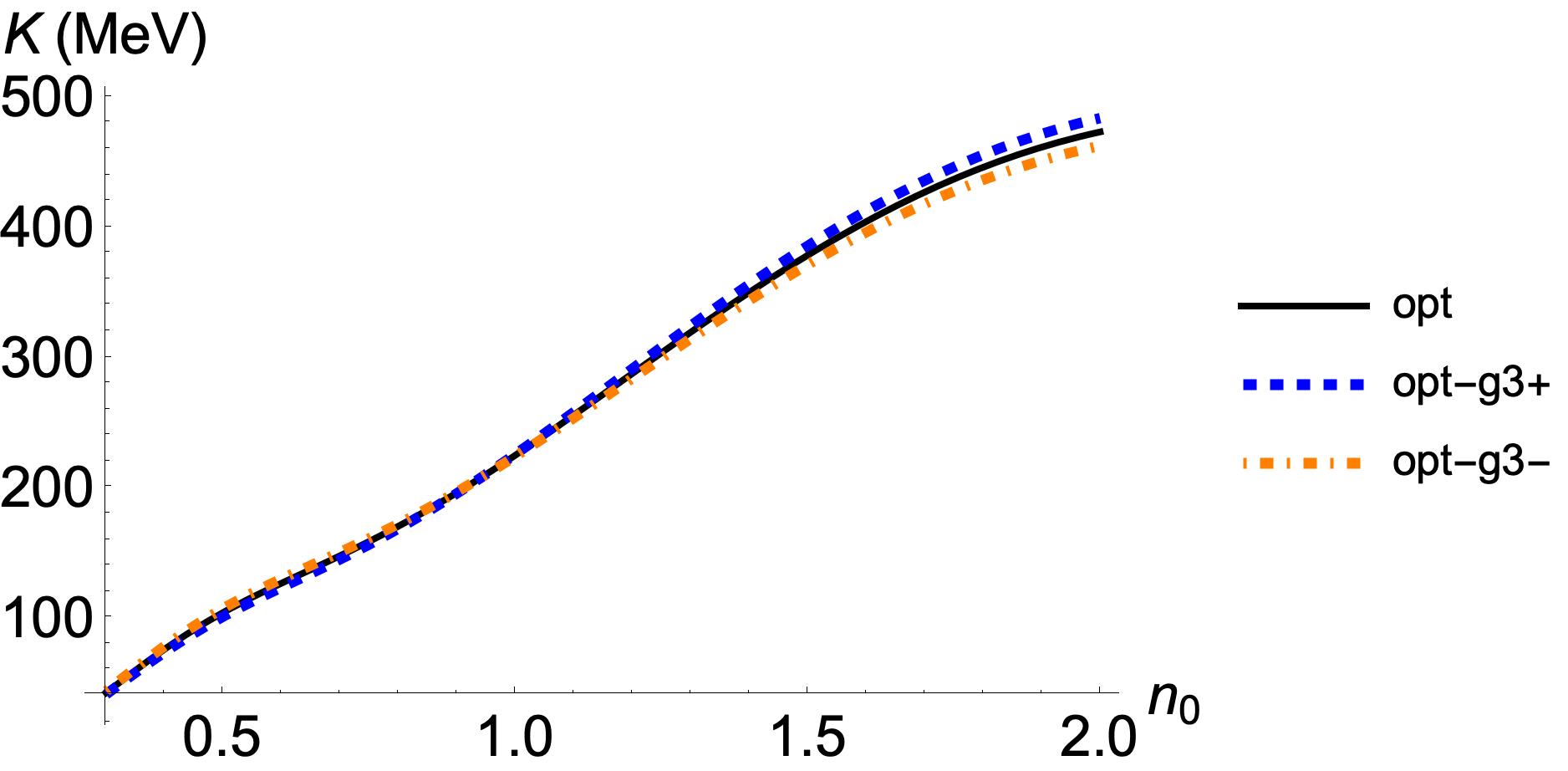}}
	\subfigure[\(h_2\) variarion.]{\includegraphics[scale=0.2]{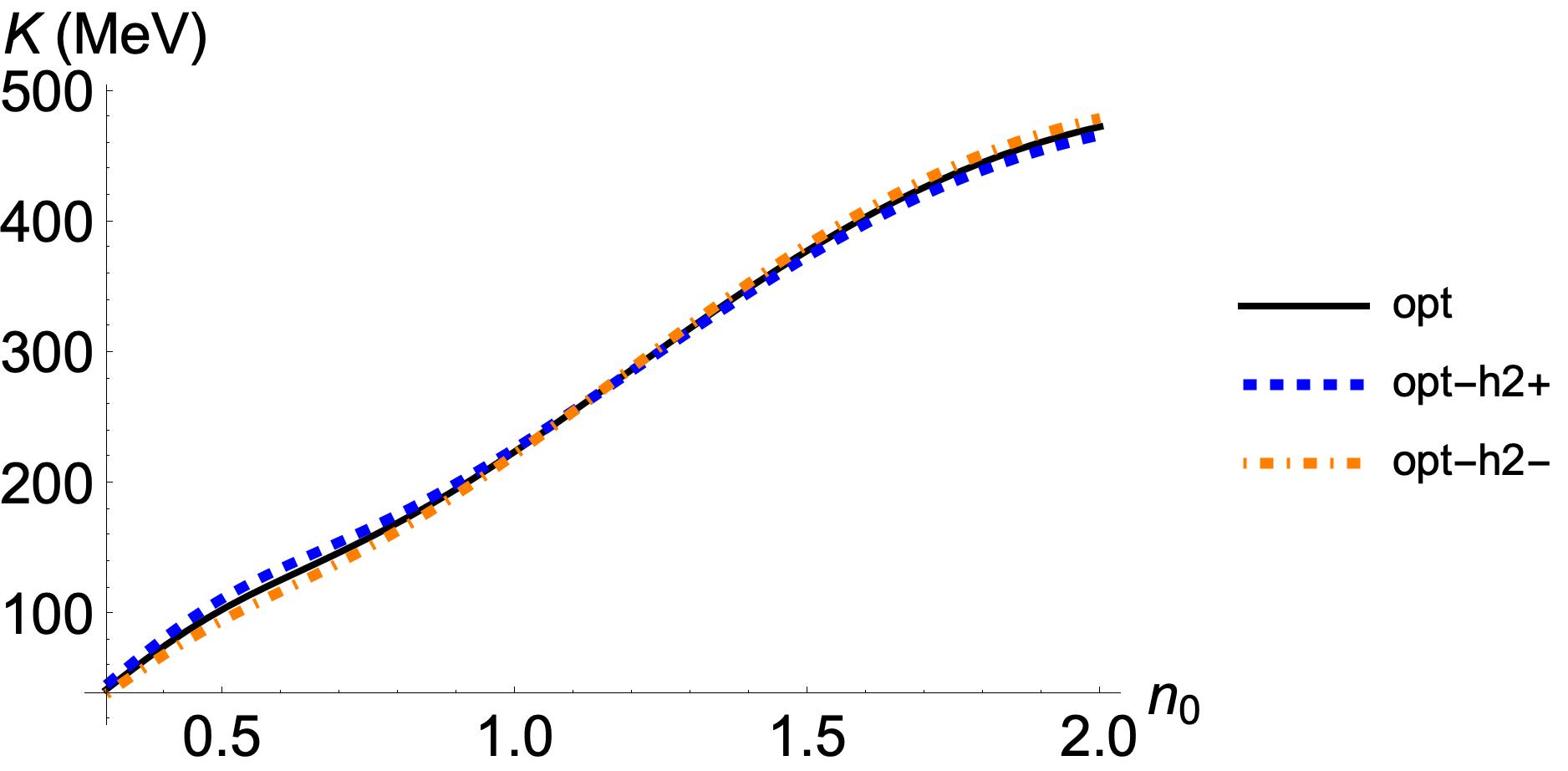}}
	\subfigure[\(\hat{h}_2\) variarion.]{\includegraphics[scale=0.2]{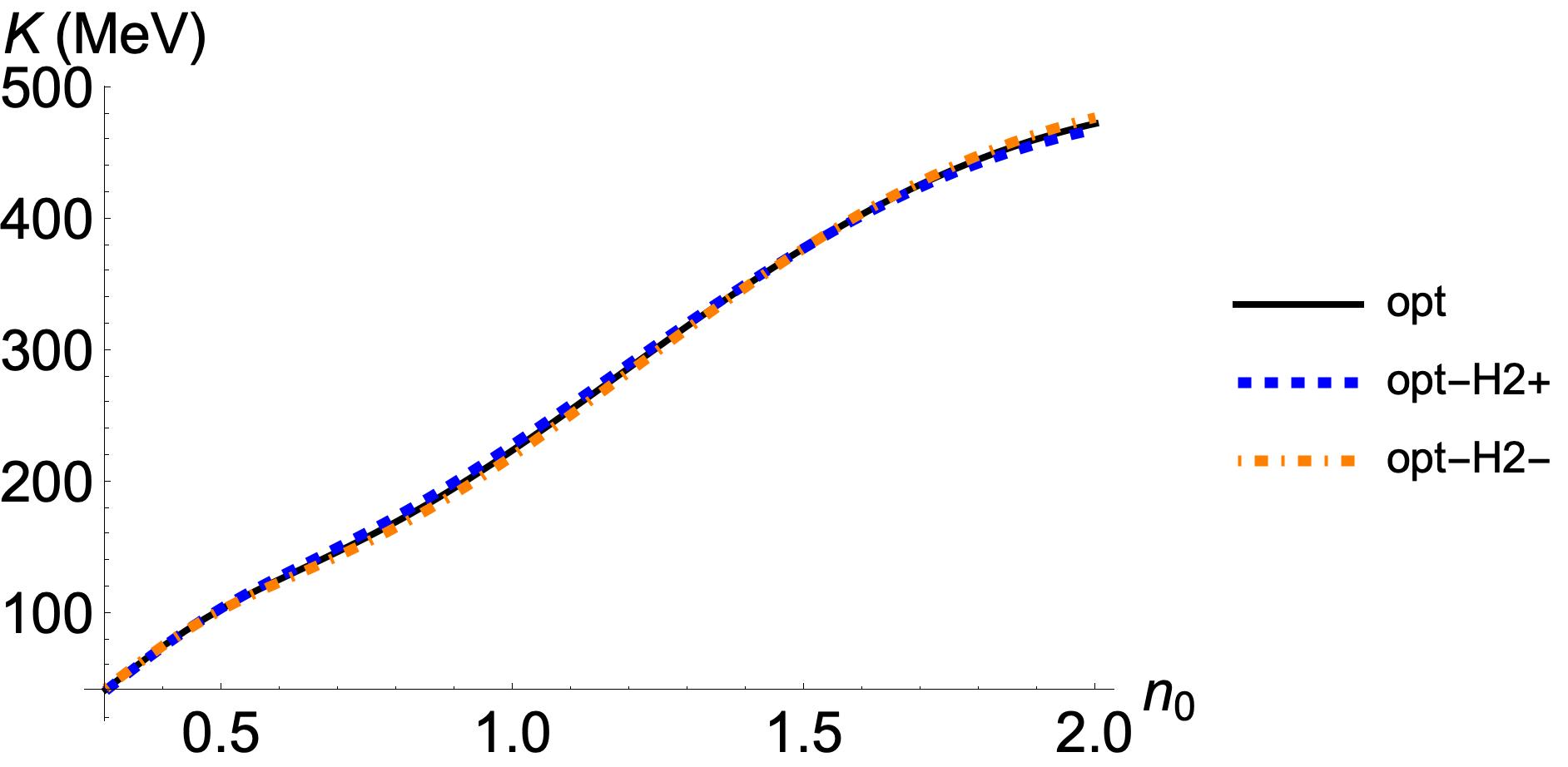}}
	\caption{Parameter effects on incompressibility coefficient $K$.}
	\label{fig:kvar}
\end{figure*}

\begin{figure*}[htb]
	\centering
	\subfigure[\(\alpha\) variation.]{\includegraphics[scale=0.2]{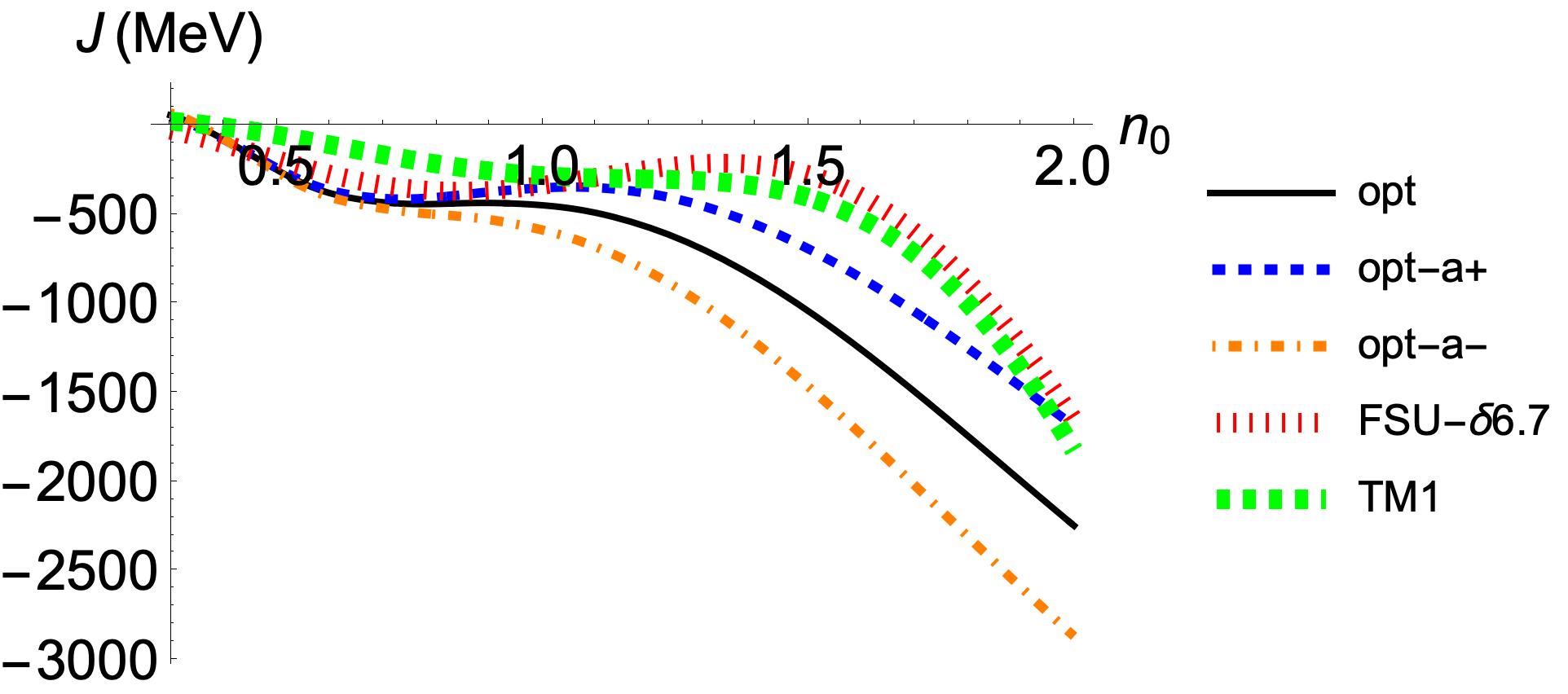}}
	\subfigure[\(\beta\) variation.]{\includegraphics[scale=0.2]{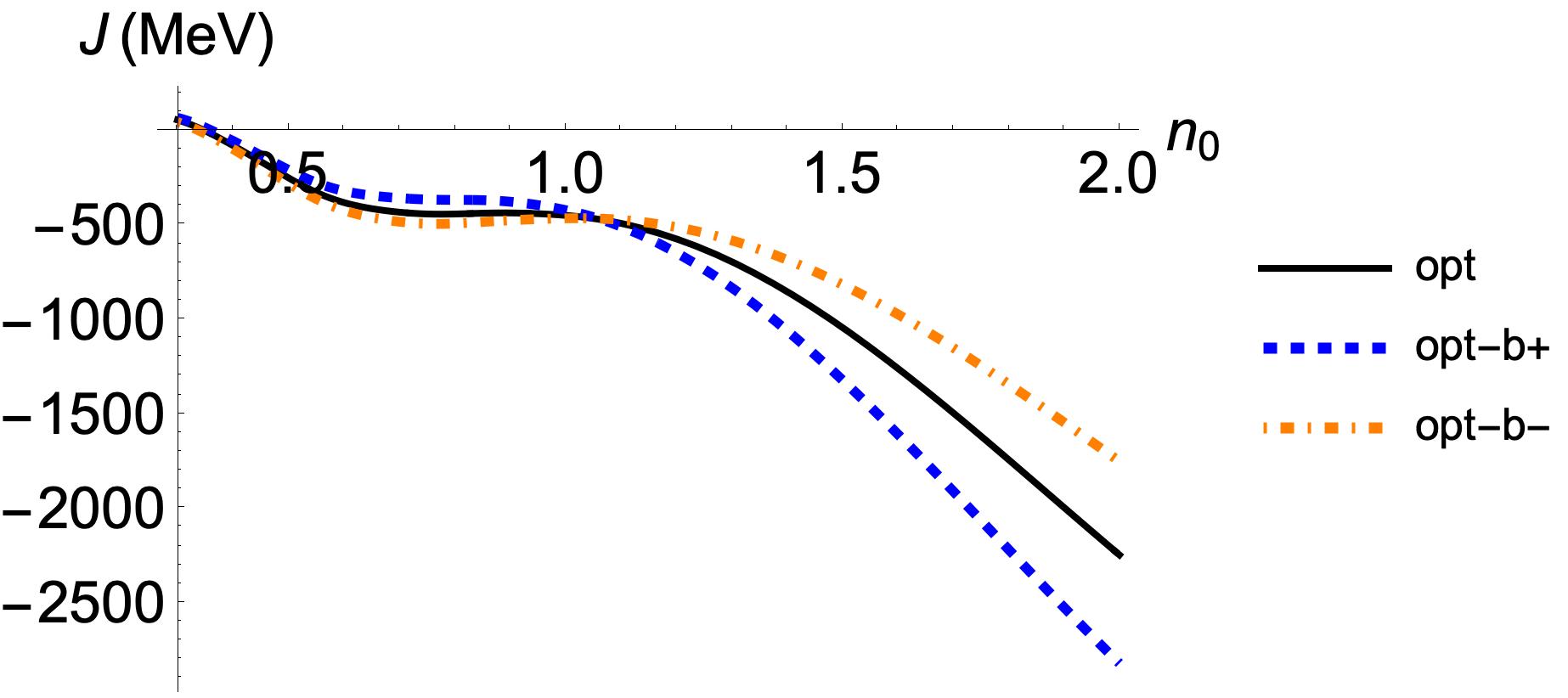}}
	\subfigure[\(c_4\) variation.]{\includegraphics[scale=0.2]{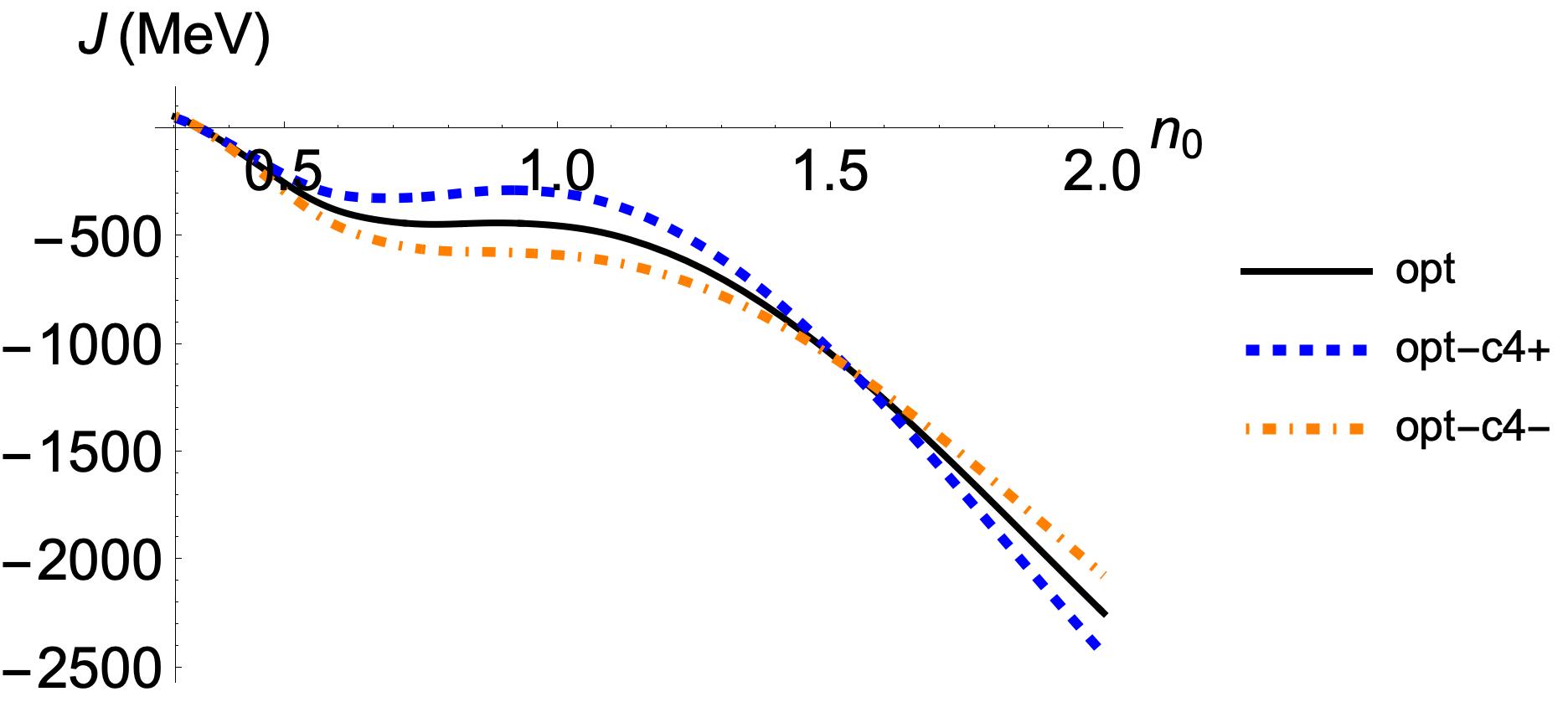}}
	\subfigure[\(e_3\) variation.]{\includegraphics[scale=0.2]{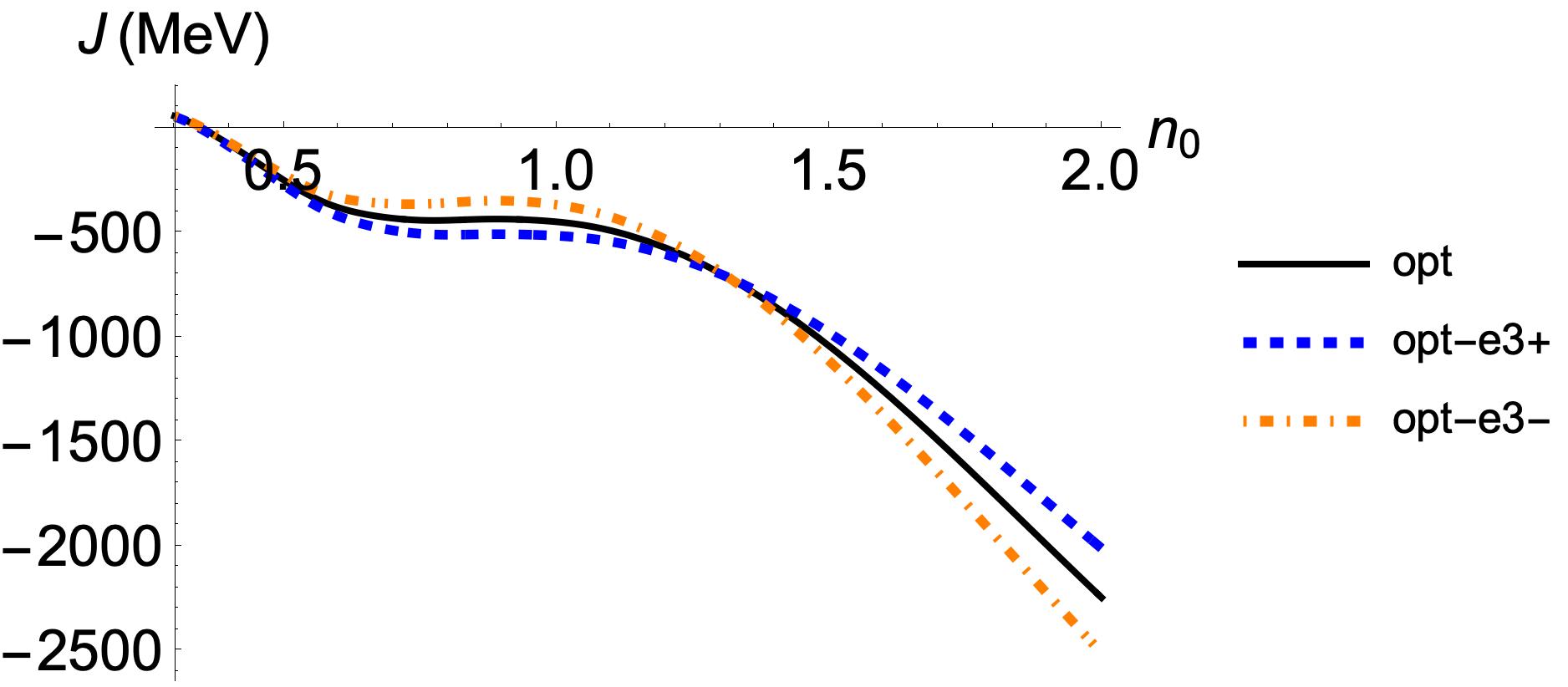}}
	\subfigure[\(c\) variation.]{\includegraphics[scale=0.2]{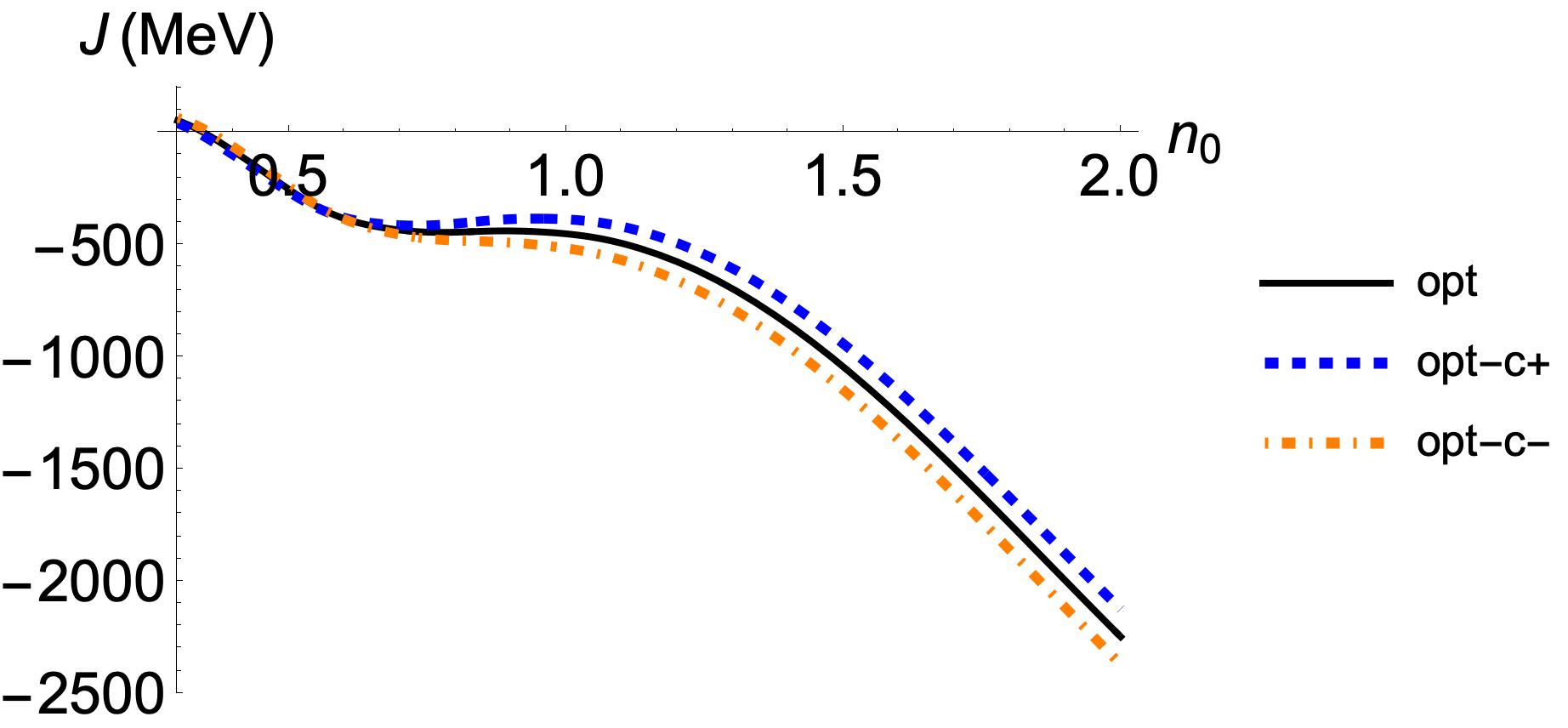}}
	\subfigure[\(g_3\) variation.]{\includegraphics[scale=0.2]{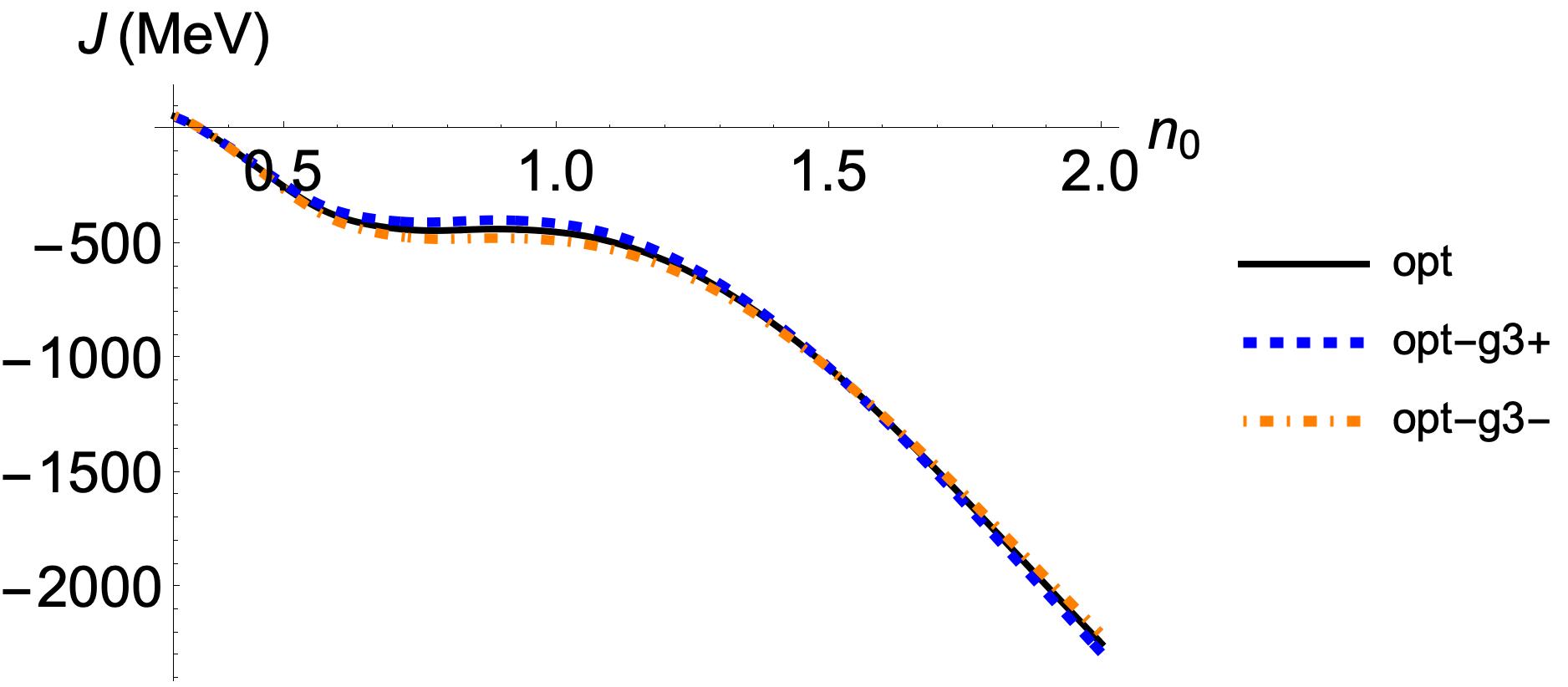}}
	\subfigure[\(h_2\) variation.]{\includegraphics[scale=0.2]{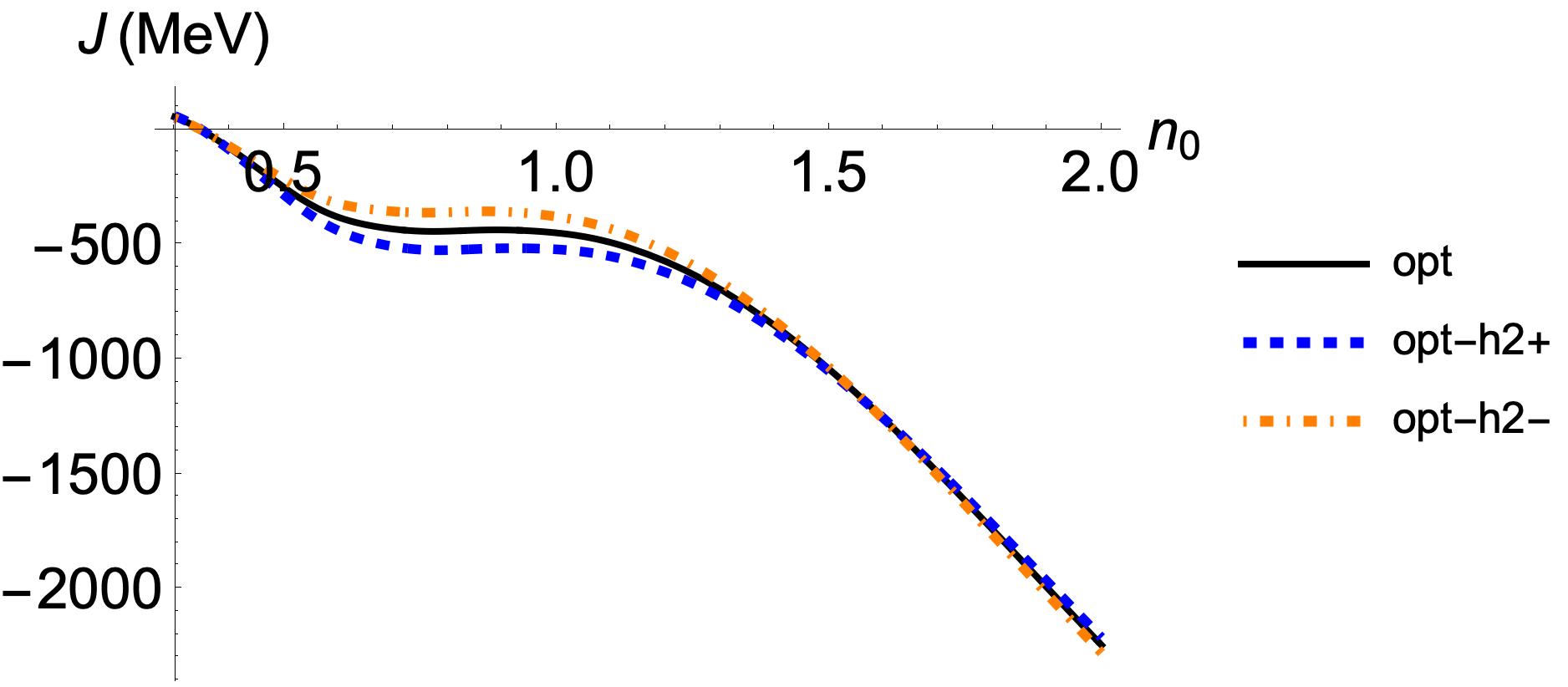}}
	\subfigure[\(\hat{h}_2\) variation.]{\includegraphics[scale=0.2]{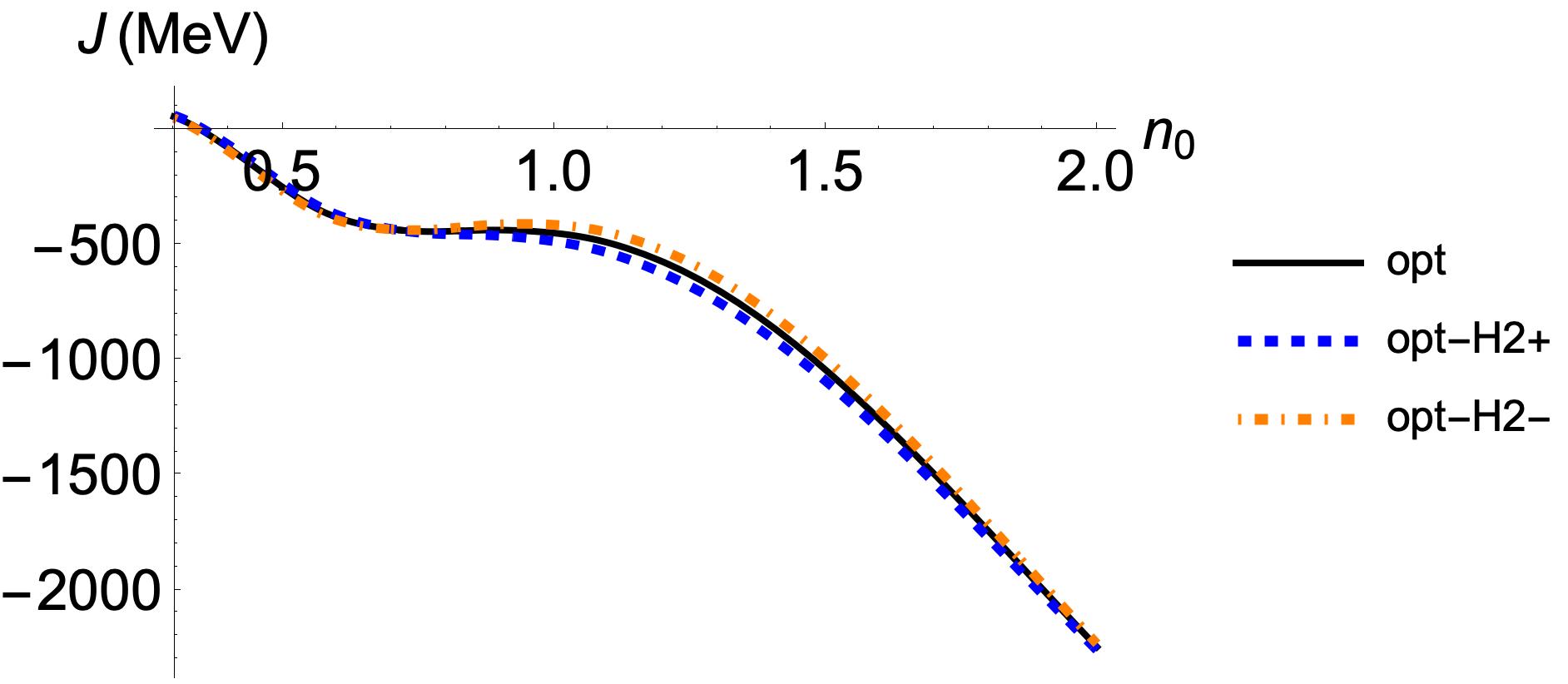}}
	\caption{Parameter effects on skewness coefficient $J$.}
	\label{fig:jvar}
\end{figure*}

We next investigate the effects of each parameter on the NM properties by changing $1\%$ of the opt values. The effects on the spectrum are shown in Tables~\ref{tab:para-var-scal} and \ref{tab:para-var-vec}, and those on NM properties are illustrated in Figs.~\ref{fig:bevar}--\ref{fig:jvar}.
\begin{widetext}
\begin{table*}[tbh]\small
	\centering
	\caption{
		The effects on spectra and two-quark configuration by parameter variation. opt-a$\pm$ refers to the parameter set with $\tilde{\alpha}=(1\pm0.01)\alpha$ with $\alpha$ being the value in opt. The same convention is used for other parameters with b for $\beta$, c4 for $c_4$, and e for $e_3$.
	}
	\label{tab:para-var-scal}
	\begin{threeparttable}
		\begin{tabular}{@{}lllllllllllllllll}
			\hline
			\hline
			&opt-a+&opt-a-&opt-b+&opt-b-&opt-c4+&opt-c4-&opt-e3+&opt-e3-\\
			\hline
			$m_{\sigma}$ (MeV)&715&687&696&706&706&696&699&703\\
			\hline
			$m_{a_0}$ (MeV)&974&956&969&960&973&956&961&968\\
			\hline
			$m_{\omega}$ (MeV)&789&770&777&782&779&779&779&779\\
			\hline
			\(\sigma\) two-quark (\(\%\))&56.5&57.2&57.9&55.8&57.6&56.2&56.2&57.6\\
			\hline
			\(a_0\) two-quark (\(\%\))&42.4&42.6&42.9&42.1&42.8&42.2&42.2&42.8\\
			\hline
			\hline
		\end{tabular}
	\end{threeparttable}
\end{table*}

\begin{table*}[tbh]\small
	\centering
	\caption{
		The effects on vector spectra.
		The notations are the same as Table~\ref{tab:para-var-scal}: h for $h_2$, H for $\hat{h}_2$, g for $g_3$ and c for $c$.
	}
	\label{tab:para-var-vec}
	\begin{threeparttable}
		\begin{tabular}{@{}lllllllll}
			\hline
			\hline
			&opt-c+&opt-c-&opt-h2+&opt-h2-&opt-g3+&opt-g3-&opt-H2+&opt-H2-\\
			\hline
			$m_{\omega}$ (MeV)&779&779&784&774&780&780&778&781\\
			\hline
			\hline
		\end{tabular}
	\end{threeparttable}
\end{table*}
\end{widetext}

From Fig.~\ref{fig:bevar}, it can be seen that the binding energy is sensitive to $\alpha,\ \beta,\ c_4,\ e_3,$ and $c$, since these parameters determine the spectra and couplings between nucleons and mesons. As expected, multimeson couplings, the $g_3, h_2$ and $\hat{h}_2$ terms, affect less because they represent higher-order density dependence. In general, the binding energy is consistent with TM1~\cite{Sugahara:1993wz} and FSU-$\rm \delta$6.7~\cite{Li:2022okx}. So, its behavior is similar to the previous conclusion which is determined by the competition between attractive and repulsive potentials.

The parameter dependence of the symmetry energy $E_{\rm sym}$ is shown in Fig.~\ref{fig:esymvar}.
A generic property which can be found is the existence of a plateaulike structure which is due to the possible multimeson couplings allowed by chiral symmetry and power-counting rules set up above.
Actually, this structure has also been found in FSU-$\rm \delta$6.7~\cite{Li:2022okx} to describe the neutron skin thickness of ${}^{208}\rm Pb$ and tidal deformability of neutron stars. Explicitly, the $^{208} \rm Pb$ measurement suggests that $L(2/3 n_0)$ ($E_{\rm sym}$ slope) should be larger than $49~\rm MeV$, and the neutron star tidal deformation (TD) $\Lambda_{1.4}$ is estimated as $642-955$~\cite{Reed:2021nqk}, if $E_{\rm sym}$ is stiff at intermediate densities. Such calculated value of $\Lambda_{1.4}$ is larger than the one extracted from GW170817 $\Lambda_{1.4}\leq 580$~\cite{LIGOScientific:2018cki}. In order to yield the neutron star TD consistent with both GW170817 and neutron skin thickness of $^{208} \rm Pb$, the symmetry energy should be stiff at subsaturation densities but soft at intermediate regions, leading to a plateaulike structure in symmetry energy. The difference is that the plateaulike structure in this work appears at a lower-order region than that in FSU-$\rm \delta$6.7. The reason is that the current bELSM is built with exact $\rm U(3)_V$ symmetry, which sets $g_{\rho NN}=g_{\omega NN}$ and also constrains other vector meson couplings. With these constraints, the $E_{\rm sym}$ will first increase to a large value when the binding energy is tuned to a reasonable value by adjusting $g_{\omega NN}$, but the growth rate of $E_{\rm sym}$ must be suppressed around $n_0$ to meet the requirement $E_{{\rm sym}}\approx 32\ \rm MeV$.

This reasoning based on the symmetry argument can also be seen in the analysis of the slope of the symmetry energy $L$. From Fig.~\ref{fig:lvar}, one can see that there is quantitative difference between opt and FSU-$\rm \delta$6.7 due to the exact $\rm U(3)_V$ symmetry applied here but the qualitative behavior is similar. The valley structure of the slope around $0.6 n_0$ yields the plateaulike structure of $E_{\rm sym}$ below $n_0$ and gives a realistic $E_{\rm sym}$. In addition, it is found that the slope is affected mainly by the quark condensates. It's reasonable to expect that the behavior of $L$ can be revised by breaking the $\rm U(3)_V$ symmetry into $\rm SU(3)_V\times U(1)_V$.

It is found in Fig.~\ref{fig:kvar} that the variations of $\alpha,\ \beta$, and $c$ increase or decrease $K$ at all regions below $2n_0$, while those multimeson coupling contributions are opposite between regions below and above $n_0$, especially those triple- and quadruple-scalar-meson couplings. The behavior of $K$ can affect the speed of sound $v_s$ NM. Explicitly, a large $K$ always leads to a large $v_s$ by definition. $v_s$ is at the center of the stage in the studies on strong interaction in dense systems, since its unique behavior always indicates interesting phenomena~\cite{Ma:2018qkg,Zhao:2020dvu,Kapusta:2021ney,Margueron:2021dtx,Hippert:2021gfs}. The analysis here may provide some hints about introducing multimeson couplings in a more logical way.

Different from other properties discussed above, the constraint on the skewness coefficient $J_0$ is not well understood, and it varies between $-1200$ and $400\ {\rm MeV}$ based on different targets or analysis methods~\cite{Farine:1997vuz,Xie:2019sqb,Choi:2020eun}. An interesting point is that a plateaulike structure is found in every model from Fig.~\ref{fig:jvar}. The parameter variation effects are similar to that in the incompressibility $K$; that is, the multimeson coupling contributions are opposite between high-density regions and low-density regions.

The above analysis shows that the stiffness of the EOS obtained with the parameter choice opt is different from other approaches, for example, FSU-$\rm \delta$6.7, due to the $\rm U(3)_V$ symmetry applied. Using the parameter choice opt we found that the maximum mass of neutron star is $1.98\ M_{\odot}$, and the radius of the neutron star with mass $1.4\ M_{\odot}$ neutron star is $14.5\ \rm{km}$ with the tidal deformability $\Lambda_{1.4}=1440$. Obviously, the neutron star predicted above is less compact than expected to describe the GW170817 event and the tidal deformability is beyond the constraint. The possible reasons are the following: (i) The $\rm U(3)_V$ symmetry should break into $\rm SU(3)_V\times U(1)_V$, especially in a dense environment~\cite{Paeng:2013xya}; (ii) The explicit chiral symmetry-breaking terms are important because too strong multimeson couplings are introduced to obtain reasonable spectrums without quark mass terms and it may cause too much noise at high-density regions.

\section {Discussion and perspective}

In this work, a bELSM is built to study the properties of nuclear matter with special interest in the effects of quark configurations of scalar mesons and vector resonances. The model is built with three flavors such that the splitting of scalar meson octet and singlet can be analyzed and the $a_0$ and $f_0$ mesons can be included with $\sigma$ meson in a unified framework. 

By pinning down the spectra of mesons and NM properties at saturation density with RMF approximation, the parameters are determined. It is found the NM properties are sensitive to the two-quark condensate $\alpha$, which determines the couplings between scalar mesons and nucleons, and $|g_{\sigma NN}|$ should be around $\simeq 10$. This result is consistent with previous works where $\sigma$ is parametrized as OBE form.

We find that there is a plateaulike structure in the $E_{\rm sym}$ curve because of the valley structure in its slope $L$. This behavior has already been addressed in Ref.~\cite{Li:2022okx} where $a_0(980)$, also known as $\delta$, is introduced to produce the plateaulike structure to describe neutron skin thickness and tidal deformation of neutron star (NS) constraints at the same time. But the plateaulike structure appears too early in current bELSM due to $\rm U(3)_V$ symmetry applied which sets $g_{\rho NN}=g_{\omega NN}$ making $E_{\rm sym}$ grow too rapidly below $n_0$. This implies that $\rm U(3)_V$ symmetry is necessarily broken into $\rm SU(3)_V\times U(1)_V$ in a dense environment. And, because of the early plateaulike structure, the star properties are slightly beyond the constraints of GW170817.

Besides the breaking of the $\rm U(3)_V$ symmetry, another factor that may remedy the drawback of the present model is the explicit chiral symmetry-breaking effect which is not considered in the present work. This is because the quark masses may apparently reduce the strength of multimeson couplings producing the meson spectra in current bELSM. This effect will be studied in a forthcoming work.

With more and more NS signals detected, the more stringent multisource constraints of nucleon interactions in the medium will become available. Therefore, a systematical investigation of the hadron spectrum, NM properties, and neutron star properties will bring insights into strong interactions and the nature of hadrons, such as the structure of light scalar mesons. The present model will be a good starting point to further explore in a wider density range of NM.

In the future, the bELSM will be extended to a more realistic one with a better symmetry pattern to describe the neutron stars with RMF approximation, and the strange freedoms will also be considered naturally. In addition, it is also interesting to introduce the quantum effects of hadrons using the bELSM with the Hartree-Fock method, so that the contributions of pseudo-scalar and axial-vector mesons can be included. All these will be clarified in forthcoming works.

\section*{ACKNOWLEDGMENTS}
The work of Y.-L. M. is supported in part by the National Key R\&D Program of China under Grant No. 2021YFC2202900 and the National Science Foundation of China (NSFC) under Grants No. 11875147 and No. 12147103.

\appendix

\section*{APPENDIX: THE TRIPLE-MESON AND QUADRUPLE-MESON COUPLINGS}

In this appendix, we explicitly list the triple-meson couplings and quadruple-meson couplings which contribute to the present calculation after the RMF.
\begin{widetext}
The triple-meson couplings that survive under RMF are the following:
\be
\mathcal{L}_3 & = &{}  -\frac{1}{6}\left(12 \sqrt{3} c_4 f_0^2 \alpha \sigma \cos \theta_0-4 \sqrt{3} h_2 \alpha \rho^2 \sigma \cos \theta_0+6 \sqrt{3} c_4 \alpha \sigma^3 \cos \theta_0-4 \sqrt{3} h_2 \alpha \sigma \omega^2 \cos \theta_0\right.\nonumber\\
& & \;\;\;\;\;\;\;\;\; {} + 2 \sqrt{3} c_4 \alpha \sigma^3 \cos (3 \theta_0)-6 \sqrt{3} c_4 f_0^2 \alpha \sigma \cos (\theta_0-2 \theta_8)-\sqrt{6} e_3 f_0^3 \cos \theta_8 \nonumber\\
& &\;\;\;\;\;\;\;\;\;{} +\sqrt{6} e_3 f_0^3 \cos (3 \theta_8) - 6 \sqrt{3} c_4 f_0^2 \alpha \sigma \cos (\theta_0+2 \theta_8)-2 \sqrt{3} e_3 f_0^2 \sigma \sin \theta_0 \nonumber\\
& &\;\;\;\;\;\;\;\;\;{} -4 \sqrt{3}\hat{h}_2\beta \rho^2 \sigma \sin \theta_0+2 \sqrt{3} e_3 \sigma^3 \sin \theta_0 - 4 \sqrt{3}\hat{h}_2 \beta \sigma \omega^2 \sin \theta_0+ 2 \sqrt{3} e_3\sigma^3 \sin(3 \theta_0) \nonumber\\
& &\;\;\;\;\;\;\;\;\;{} -\sqrt{3} e_3 f_0^2 \sigma \sin(\theta_0-2 \theta_8)+3 \sqrt{6} c_4 f_0^3 \alpha \sin(\theta_8)+ 2 \sqrt{6} h_2 f_0  \alpha \rho^2 \sin\theta_8+2 \sqrt{6} \hat{h}_2 \beta f_0\rho^2 \sin\theta_8 \nonumber\\
& &\;\;\;\;\;\;\;\;\;{} +12 \sqrt{2} h_2 \alpha a_0 \rho \omega \sin\theta_8+ 12 \sqrt{2} \hat{h}_2 \beta a_0\rho \omega \sin\theta_8 + 2 \sqrt{6} \hat{h}_2 \alpha f_0\omega^2 \sin \theta_8+2 \sqrt{6} \hat{h}_2 \beta f_0\omega^2 \sin \theta_8\nonumber\\
& &\;\;\;\;\;\;\;\;\;{} +8 \sqrt{3} e_3 a_0^2 \sigma \cos\theta_0 \cos\theta_8\sin\theta_8 + 24 \sqrt{3} c_4 \alpha \sigma a_0^2\cos\theta_0 \sin^2\theta_8+ 12 \sqrt{6} e_3f_0a_0^2 \cos\theta_8 \sin\theta_8^2\nonumber\\
& &\;\;\;\;\;\;\;\;\;{} -4 \sqrt{3} e_3 \sigma a_0^2\sin\theta_0 \sin^2\theta_8 - 12 \sqrt{6} c_4  \alpha f_0 a_0^2 \sin^3\theta_8-\sqrt{6} c_4 f_0^3 \alpha \sin(3 \theta_8)\nonumber\\
& &\;\;\;\;\;\;\;\;\;{} \left. +3 \sqrt{3} e_3 f_0^2 \sigma \sin(\theta_0+2 \theta_8)\right) .
\ee
The quadruple-meson couplings contributing to the present calculation are summarized as 
\be
\mathcal{L}_4 & = &{} -\frac{1}{48}\left(9 c_4 f_0^4-4  h_2f_0^2 \rho^2-4\hat{h}_2 f_0^2 \rho^2-12 g_3 \rho^4+24 c_4 f_0^2 \sigma^2-8 h_2 \rho^2 \sigma^2-8\hat{h}_2 \rho^2 \sigma^2+6 c_4 \sigma^4\right. \nonumber\\
& &\;\;\;\;\;\;\;\;\;\;\;{} - 4  h_2f_0^2 \omega^2-4 \hat{h}_2f_0^2 \omega^2-72 g_3 \rho^2 \omega^2-8 h_2 \sigma^2 \omega^2-8 \hat{h}_2 \sigma^2 \omega^2-12 g_3 \omega^4\nonumber\\
& &\;\;\;\;\;\;\;\;\;\;\;{} +24 c_4 f_0^2 \sigma^2 \cos(2 \theta_0) - 8 h_2 \rho^2 \sigma^2 \cos(2 \theta_0)+8 \hat{h}_2 \rho^2 \sigma^2 \cos (2 \theta_0)+8 c_4 \sigma^4 \cos (2 \theta_0)\nonumber\\
& &\;\;\;\;\;\;\;\;\;\;\;{} -8 h_2 \sigma^2 \omega^2 \cos(2 \theta_0) + 8 \hat{h}_2 \sigma^2 \omega^2 \cos(2 \theta_0)+2 c_4 \sigma^4 \cos (4 \theta_0)+8 \sqrt{2} \hat{h}_2 f_0  \rho^2 \sigma \cos (\theta_0-\theta_8)\nonumber\\
& &\;\;\;\;\;\;\;\;\;\;\;{} + 8 \sqrt{2} \hat{h}_2f_0\sigma \omega^2 \cos (\theta_0-\theta_8)-12 c_4 f_0^2 \sigma^2 \cos [2(\theta_0-\theta_8)]-12 c_4 f_0^4 \cos [2 \theta_8] \nonumber\\
& &\;\;\;\;\;\;\;\;\;\;\;{} +4 h_2f_0^2\rho^2 \cos(2 \theta_8) + 4 \hat{h}_2f_0^2  \rho^2 \cos\theta_8-24 c_4 f_0^2 \sigma^2 \cos\theta_8+4  h_2f_0^2 \omega^2 \cos\theta_8 \nonumber\\
& &\;\;\;\;\;\;\;\;\;\;\;{} +4 \hat{h}_2f_0^2 \omega^2 \cos\theta_8+3 c_4 f_0^4 \cos\theta_8 - 8 \sqrt{2} \hat{h}_2f_0 \rho^2 \sigma \cos(\theta_0+\theta_8)-8 \sqrt{2} \hat{h}_2f_0 \sigma \omega^2 \cos(\theta_0+\theta_8)\nonumber\\
& &\;\;\;\;\;\;\;\;\;\;\;{} -12 c_4 f_0^2 \sigma^2 \cos[2(\theta_0+\theta_8)] + 4 \sqrt{2} c_4 f_0^3 \sigma \sin(\theta_0-3 \theta_8)-12 \sqrt{2} c_4 f_0^3 \sigma \sin(\theta_0-\theta_8) \nonumber\\
& &\;\;\;\;\;\;\;\;\;\;\;{} -8 \sqrt{2} h_2f_0  \rho^2 \sigma \sin[\theta 0-\theta 8] - 8 \sqrt{2} h_2f_0  \sigma \omega^2 \sin(\theta_0-\theta_8)+32 \sqrt{6} h_2 a_0\rho \sigma \omega \cos\theta_0 \sin\theta_8 \nonumber\\
& &\;\;\;\;\;\;\;\;\;\;\;{} +32 \sqrt{6}  \hat{h}_2 a_0 \rho \sigma \omega \sin\theta_0 \sin\theta_8 - 24h_2a_0^2 \rho^2 \sin^2\theta_8-24  \hat{h}_2a_0^2 \rho^2 \sin^2\theta_8\nonumber\\
& &\;\;\;\;\;\;\;\;\;\;\;{} -32 \sqrt{3} h_2 a_0f_0 \rho \omega \sin^2\theta_8-32 \sqrt{3}  \hat{h}_2 a_0 f_0 \rho \omega \sin^2\theta_8 - 24h_2a_0^2 \omega^2 \sin^2\theta_8-24  \hat{h}_2a_0^2 \omega^2 \sin^2\theta_8 \nonumber\\
& &\;\;\;\;\;\;\;\;\;\;\;{} +96 c_4 a_0^2  \sigma^2 \cos^2\theta_0 \sin^2\theta_8-96 \sqrt{2}c_4 a_0^2  f_0 \sigma \cos\theta_0 \sin\theta_8^3 + 24 c_4 a_0^4 \sin\theta_8^4+48 c_4 a_0^2  f^2 \sin\theta_8^4 \nonumber\\
& &\;\;\;\;\;\;\;\;\;\;\;{} +12 \sqrt{2} c_4 f_0^3 \sigma \sin(\theta_0+\theta_8)+8 \sqrt{2} h_2f_0  \rho^2 \sigma \sin(\theta_0+\theta_8)\nonumber\\
& & \left.\;\;\;\;\;\;\;\;\;\;\;{} + 8 \sqrt{2} h_2f_0  \sigma \omega^2 \sin(\theta_0+\theta_8)-4 \sqrt{2} c_4 f_0^3 \sigma \sin(\theta_0+3 \theta_8)\right) ,
\ee
where the mixing angles have already been chosen based on the numerical results in Sec. \ref{sec:num}.
	

\end{widetext}

\bibliography{NuclMatTeraQ}

\end{document}